\documentclass[a4paper,usenatbib]{mnras}

\usepackage{graphicx}   
\usepackage{newtxtext,newtxmath}
\usepackage{url}
\usepackage{lscape}
\usepackage{multirow}
\usepackage{float}
\usepackage{array}
\usepackage{natbib}
\usepackage{amsmath, amsfonts}
\usepackage{amssymb}

\newcommand{\kms}          {\mbox{${\rm km~s^{-1}}$}}

\newcommand{\simgt}{\lower.5ex\hbox{$\; \buildrel > \over \sim \;$}}
\newcommand{\simlt}{\lower.5ex\hbox{$\; \buildrel < \over \sim \;$}}

\title[Discovering Large-Scale Structure]{Establishing a New Technique for Discovering Large-Scale Structure Using the ORELSE Survey}
\author[Hung et al.]{D. Hung$^{1}$, B.~C. Lemaux$^{2}$, R.~R. Gal$^{1}$, A.~R. Tomczak$^{2}$, L.~M. Lubin$^{2}$, O. Cucciati$^{3}$, 
\newauthor D. Pelliccia$^{2,4}$, L. Shen$^{2}$, O. Le F\`{e}vre$^{5}$, P-F. Wu$^{6,7}$, D.~D. Kocevski$^{8}$, S. Mei$^{9,10,11}$, 
\newauthor G.~K. Squires$^{12}$ \\
$^{1}$ University of Hawai'i, Institute for Astronomy, 2680 Woodlawn Drive, Honolulu, HI 96822, USA \\
$^{2}$ Department of Physics, University of California, Davis, One Shields Ave., Davis, CA 95616, USA \\
$^{3}$ INAF - Osservatorio di Astrofisica e Scienza dello Spazio diBologna, via Gobetti 93/3 - 40129 Bologna - Italy \\
$^{4}$ Department of Physics and Astronomy, University of California, Riverside, 900 University Ave, Riverside, CA 92521, USA \\
$^{5}$ Aix-Marseille Universit{\'e}, CNRS, LAM (Laboratoire d'Astrophysique de Marseille) UMR 7326, 13388 Marseille, France \\
$^{6}$ Max-Planck Institut f\"{u}r Astronomie, K\"{o}nigstuhl 17, D-69117, Heidelberg, Germany\\
$^{7}$ EACOA Fellow, National Astronomical Observatory of Japan, Osawa 2-21-1, Mitaka, Tokyo 181-8588, Japan \\
$^{8}$ Department of Physics and Astronomy, Colby College, Waterville, ME 04961, USA \\
$^{9}$ University of Paris Denis Diderot, University of Paris Sorbonne Cit{\'e} (PSC), 75205 Paris Cedex 13, France \\
$^{10}$ Sorbonne Universit{\'e}, Observatoire de Paris, Universit{\'e} PSL, CNRS, LERMA, F-75014, Paris, France \\
$^{11}$ Jet Propulsion Laboratory, Cahill Center for Astronomy \& Astrophysics, California Institute of Technology, 4800 Oak Grove Drive, Pasadena, California, USA \\
$^{12}$ Spitzer Science Center, California Institute of Technology, M/S 220-6, 1200 E. California Blvd., Pasadena, CA 91125, USA}

\date{Accepted XXX. Received YYY; in original form ZZZ}

\pubyear{2019}

\begin{document}
\label{firstpage}
\pagerange{\pageref{firstpage}--\pageref{lastpage}}
\maketitle

\begin{abstract}
The Observations of Redshift Evolution in Large-Scale Environments (ORELSE) survey is an ongoing imaging and spectroscopic campaign initially designed to study the effects of environment on galaxy evolution in high-redshift ($z\sim1$) large-scale structures. We use its rich data in combination with a powerful new technique, Voronoi tessellation Monte-Carlo (VMC) mapping, to search for serendipitous galaxy overdensities at $0.55<z<1.37$ within 15 ORELSE fields, a combined spectroscopic footprint of $\sim$1.4 square degrees. Through extensive tests with both observational data and our own mock galaxy catalogs, we optimize the method's many free parameters to maximize its efficacy for general overdensity searches. Our overdensity search yielded 402 new overdensity candidates with precisely measured redshifts and an unprecedented sensitivity down to low total overdensity masses ($\mathcal{M}_{tot}\ga5\times10^{13}$ $M_{\odot}$). Using the mock catalogs, we estimated the purity and completeness of our overdensity catalog as a function of redshift, total mass, and spectroscopic redshift fraction, finding impressive levels of both 0.92/0.83 and 0.60/0.49 for purity/completeness at $z=0.8$ and $z=1.2$, respectively, for all overdensity masses at spectroscopic fractions of $\sim$20\%. With VMC mapping, we are able to measure precise systemic redshifts, provide an estimate of the total gravitating mass, and maintain high levels of purity and completeness at $z\sim1$ even with only moderate levels of spectroscopy. Other methods (e.g., red-sequence overdensities and hot medium reliant detections) begin to fail at similar redshifts, which attests to VMC mapping's potential to be a powerful tool for current and future wide-field galaxy evolution surveys at $z\sim1$ and beyond.
\end{abstract}

\section{Introduction}

Galaxy groups and clusters define the extreme high-mass end of the large-scale structure in the Universe, and the study of such overdensities provides valuable clues to a variety of open questions in astrophysics. From a galaxy evolution perspective, it is thought that the environment in which a galaxy resides, both on kpc and Mpc scales, plays a significant role in shaping its physical characteristics and evolution \citep[e.g.,][]{muz12,muz14,balogh16,Owers17,Tomczak17,tomczak19,Lemaux17,Lemaux19}. Such effects are likely to be a function both of the dynamic range of group/cluster masses observed and cosmic epoch. As such, the overarching environment a galaxy experiences can change dramatically during the assembly of the overdensity. Large-scale structures present around high redshift ($z\sim1$) clusters allow us to observe the full range of environments and their effects on galaxies as they collect into the denser regions of already established clusters. In parallel, a census of a large number of overdensities over a large baseline in cosmic time allows to decrease the noise associated with assembly bias and dynamical maturity. From a cosmological perspective, the physical properties, characteristics, and number counts of overdensities at both low and especially high redshift are useful for providing constraints on cosmological models \citep[e.g.,][]{clerc12,arnaud17,planck16,ridl17}. Such a sample is, however, challenging to assemble as structures become increasingly difficult to detect at lower total masses and higher redshifts, and most detection methods are biased for or against certain types of overdensities.

Four broad classes of methods have been used to detect mass overdensities: two methods which rely on the presence of a hot medium, surveys in the X-ray focused on photons emitted via bremsstrahlung emission \citep[e.g.,][]{voges99,ebeling01,piffaretti11} and radio/sub-mm surveys searching for signatures of thermal Sunyaev-Zel'dovich \citep[SZ;][]{sz72} effect \citep[e.g.,][]{staniszewski09,menanteau10,planck16}, strong and weak gravitational lensing techniques \citep[e.g.,][]{tyson90,kubo09a,kubo09b,ford14}, and those employing optical/near-infrared (NIR) imaging/spectroscopy that use galaxies themselves as tracers of such overdensities \citep[e.g.,][]{abell58,Oke98,gladnyee00,gilbank11,milkeraitis10,sousbie11,ascaso12,rykoff14,rykoff16}. The former three methods, while overwhelmingly successful at finding at least some types of galaxy overdensities at $z\la1.5$, quickly begin to fail at higher redshifts due to a variety of effects. At such redshifts the time since formation of overdensities necessarily decreases, meaning the processes with the hot intracluster or intragroup medium (ICM/IGM) become less effective due to the limited time they have had to act on member galaxies. As such, X-ray and SZ surveys become increasingly ineffectual when exploring the high-redshift Universe as well as increasingly biased towards the most massive overdensities with the earliest formation times. Further, the increasing fraction of active galactic nuclei (AGN) activity \citep[e.g.,][]{martini13} and more prevalent and severe deviations from hydrostatic equilibrium at higher redshift \citep[e.g.,][]{burns08} mean that uncertainties and biases associated with mass estimates from such methods necessarily grow with redshift. Practical concerns also enter, such as X-ray surface brightness dimming ($\propto (1+z)^4$) and resolution effects, which constrain the highest redshift detections to $z\sim2$ in both types of surveys at least with current technology \citep[e.g.,][]{gobat11,taowang16,strazzullo19}. While strong and weak lensing surveys do not suffer similar astrophysical concerns, as they are ostensibly only sensitive to the total mass projected along the line of sight, practical concerns such as projection effects and the necessity of extremely deep imaging to effectively probe and measure the shapes of source populations at $z\ga2$ become increasingly overwhelming when attempting to detect overdensities at $z>1$ with current technologies. As such, the only broad class of method that is likely feasible for future large-scale structure surveys over a large redshift baseline (i.e., $0 \le z \la 8$) involves optical/NIR imaging and spectroscopy of the galaxies themselves \citep[or, alternatively, at least for $2\la z\la 5$, HI gas, e.g.,][]{lee16} to trace matter overdensities. 
 
However, this class of methodology carries with it a plethora of effects that have plagued searches since their inception more than 50 years ago. In the absence of well-measured photometric redshifts and/or extensive spectroscopy, finding overdensities typically requires one to focus on overdensities of a particular galaxy class. As it was found that local clusters contain both a fractional and absolute excess of quiescent, redder galaxies per comoving volume, searching for overdensities of such galaxies quickly became popular among cluster searches \citep[as was done, e.g., in the Red-Sequence Cluster Survey;][]{gladnyee00}. These searches were extremely successful and searches based on this methodology have recently culminated in the detection of statistically significant samples of clusters over large sky areas by looking for overdensities of red galaxies in the projected on-sky galaxy distribution \citep[e.g.,][]{gilbank11,rykoff16}. Despite their success, determining systemic redshifts and other properties such as total mass can be extremely challenging with such methods and require considerable effort to calibrate \citep[e.g.,][]{McClintock19}. The inclusion of high-quality photometric redshifts leads to improved cluster detection and allows detection to extend to higher redshifts where the number of red galaxies populating overdensities begins to decrease \citep{bo84}, though spectroscopy is still required for confirmation. The use of high-quality photometric redshifts for finding high-redshift cluster candidates was established by \citet{Stanford05}. In this study, a version of this technique was used to select candidate clusters over a 8.5 square degree Bo{\"o}tes field \citep{Brodwin06,Elston06}, one of which was spectroscopically confirmed to be what was then the highest redshift galaxy cluster to date at $z$ = 1.41. \citet{Eisenhardt08} reported the full candidate cluster sample from these data using this technique, which included 335 overdensity candidates, with 106 candidates at $z > 1$, twelve of which were spectroscopically confirmed at these redshifts. With photometric redshifts based on similar but deeper data, \citet{Stanford12,Zeimann13} were able to identify and eventually spectroscopically confirm clusters at even higher redshifts of $z$ = 1.75 and 1.89. In recent years, photometric redshift searches have expanded to covering greater breadths of the sky at similar redshifts, such as \citet{Radovich17,Bellagamba18}, who found nearly 2000 cluster candidates over an area of 114 square degrees, and the Massive and Distant Clusters of {\it WISE} Survey \citep[MaDCoWS;][]{Gonzalez19}, the first cluster survey capable of discovering massive clusters over the full extragalactic sky at $z\sim1$.

Detections using such methods are also complicated by the presence of background and foreground objects which can quickly overpower the density peaks at higher redshift if extreme care is not taken. This is especially true at higher redshift when the colors of galaxies populating overdensities begins to approach those galaxies in the field. In an attempt to mitigate such noise, photometric large-scale structure detection algorithms often use filters which make some assumptions about the properties of clusters they search for including, e.g., the shape or size of the overdensity profile or the extent of the overdensity in redshift space \citep[e.g.,][]{banerjee18}. Other searches, such as those mentioned in the above paragraph, generally focus on finding the most massive systems, and thus often see relatively low number densities over the search area. Including spectroscopic redshifts, with their greater than order of magnitude higher precision and accuracy, can also help mitigate such projection effects, but spectroscopy must be unbiased with respect to the underlying galaxy population in order to avoid biasing the overdensity search. To date only a few surveys at moderately high redshift ($z\sim1$) have achieved extensive, representative, wide-field spectroscopy including the Deep Extragalactic Evolutionary Probe 2 \citep[DEEP2;][]{Davis03,Newman13}, the VIMOS Very Deep Survey \citep[VVDS;][]{lefevre05,lefevre13}, zCOSMOS \citep{lilly07,lilly09}, and the VIMOS Public Extragalactic Redshift Survey \citep[VIPERS;][]{garilli14,guzzo14}. Such surveys are typically limited to fields which are broadly devoid of massive groups, clusters, and other large-scale structures \citep[e.g.,][]{Gerke12,Owers17}. Conversely, studies of large-scale structures (LSS) at these redshifts are typically limited to the cores of clusters and groups, have limited or severely biased spectroscopy, and/or are limited to the study of one or a few LSSs.

Unlike many past LSS surveys, the Observations of Redshift Evolution in Large-Scale Environments \citep[ORELSE;][]{Lubin09} survey has the advantage of having both unprecedentedly deep, representative spectroscopy, with hundreds to thousands of spectra per field, as well as deep imaging over a broad baseline in wavelength across a large number of fields. Multi-wavelength observations are able to probe the properties of overdensities from a variety of perspectives and allow for the measurements of a wide range of spectroscopic features. In this paper we use the rich ORELSE dataset, which provides high-quality spectroscopic and photometric redshifts across 15 LSS fields, to develop and test a new method of overdensity finding which makes limited assumptions on the underlying galaxy populations and the overdensities which house them. Though this method, known as Voronoi tessellation Monte Carlo (VMC) mapping, has already been used in a variety of studies that probe overdensities over the broad redshift range $0.6 < z < 4.6$ \citep[e.g.,][]{Tomczak17,tomczak19,Lemaux17,Lemaux18,Lemaux19,shen17,shen19,Rumbaugh17,Cucciati18,pelliccia19}, here we expand and fully establish the methodology, as well as extensively test, both observationally and through mock catalogs, the precision of the method in recovering the properties of overdensities (e.g., systemic redshift, redshift extent, total mass). Additionally, we quantify through the use of mock galaxy catalogs the purity and completeness of our VMC overdensity search with ORELSE-like data properties as a function of systemtic redshift, fraction of objects with spectroscopic redshifts, and total mass, finding, e.g., purity/completeness values of $\ge$0.5/0.8 for all overdensities ($\mathcal{M}_{tot}\ga5\times10^{13} M_{\odot}$) at $z \sim 0.8$ for spectroscopic redshift fractions $\ge$5\%. This high level of completeness allows us to blindly recover essentially all of the known ORELSE clusters and groups and detect $\sim$400 new overdensity candidates across the 1.4 square degrees searched, as well as to assign precise redshifts and total masses to each candidate.

This paper is organized as follows: In \S\ref{data}, we discuss the photometric and spectroscopic data used as input to our overdensity candidate detection. We also describe tests used to establish the minimum requirements for photometric data to be useful in our overdensity candidate detection method. In \S\ref{methodology}, we outline the VMC method for overdensity candidate detection, and its application here using redshift slices. We then describe in general the overdensity candidate detection using SExtractor to detect overdensity peaks in each redshift slice, followed by a linking algorithm to identify unique overdensities and estimate their redshifts. In \S\ref{optimize}, we describe extensive testing of various parameters in the overdensity candidate detection  process. In \S\ref{mocks}, we examine the purity and completeness of our catalog as a function of total mass and redshift. We present the overdensity candidate catalog in \S\ref{final}. We adopt a flat $\Lambda$CDM cosmology throughout this paper, with $H_{0}$ = 70 km s$^{-1}$ Mpc$^{-1}$, $\Omega_{m}$ = 0.27, and $\Omega_{\Lambda}$ = 0.73. All distances reported are in proper units.

\section{Data}\label{data}

\subsection{The ORELSE Survey}

This study makes use of data taken from the Observations of Redshift Evolution in Large-Scale Environments \citep[ORELSE;][]{Lubin09} survey. ORELSE is a large multi-wavelength photometric and spectroscopic campaign designed to map out large-scale structures in 15 fields over the redshift range of $0.6 < z<1.3$. Imaging covers an area of $\sim5$ square degrees across a wide range of wavelengths, from optical ($BVriz$) to near-infrared  ($JK$, {\it Spitzer}/IRAC). The spectroscopic footprint, defined by first assigning circles of 0.5 Mpc radii to all spectroscopic objects in each field at all redshifts of our interest and then summing the total projected area all those circles, has an area of $\sim1.4$ square degrees. In this work, we restrict our study  to the spectroscopic footprint only (see \S\ref{subsec:spectroscopy}) of all the 15 ORELSE fields (Table \ref{tab.fields}). ORELSE distinguishes itself from similar competing studies thanks to its unprecedented spectroscopic coverage \citep{Lubin09}, which includes  $\sim11,000$ high quality spectroscopic objects, with 100-500 confirmed members per structures. This extensive dataset has already been shown to contain many possible high-redshift structures beyond those initially targeted \citep[e.g.,][]{Gal08,Lemaux19}.

In this study, we use the fully-processed photometric and spectroscopic catalogs available for all the 15 ORELSE fields to detect overdensity candidates and to determine the detection efficiency as a function of spectroscopic completeness, redshift, mass, and other properties. For known structures, those which have been identified in the ORELSE fields through other overdensity detection methods, the spectroscopic completeness ranges from 25\% to 80\%. Moreover, for all analysis presented in this paper, we cut the catalogs at 18 mag $\leq i \leq$ 24.5 mag\footnote{The particular type of $i$ filter curve will differ from field to field, e.g., $I^{+}$ (equivalent to SDSS $i^{\prime}$) or Cousins $I$, and some fields have multiple $i$-bands available. This is also true for the $r$- and $z$-bands used. For the sake of simplicity in this paper, we will refer to all variants of these bands by their generalized $riz$ names. Refer to Table \ref{tab.imaging} for details on the exact photometry bands used for each field.} (or the equivalent 18 mag $\leq z \leq$ 24.5 mag when the redshift of the targeted large-scale structure was greater than 0.95), a magnitude range that encompasses nearly all high-quality ORELSE objects. Every field's detection limit is fainter than 24.5, so our magnitude cut homogenizes the completeness statistics for all fields. This magnitude cut essentially produces a stellar mass-limited sample at $10^{9} - 10^{10} M_{\odot}$, depending on the redshift and field (see \citet{Tomczak17} for further details on the galaxy stellar mass function of our sample).

\begin{table}
\caption{ORELSE Fields}
\label{tab.fields}
\begin{tabular}{ccccc}
\hline
Name & RA (J2000) & Dec (J2000) & Redshift & Area$^{a}$ \\
\hline
SG0023 & 00 23 52.2 & +04 23 07 & 0.845 & 0.077 \\
RCS0224 & 02 24 34.0 & +00 02 30 & 0.772 & 0.058 \\
XLSS005 & 02 27 09.7 & --04 18 05 & 1.000 & 0.422 \\
SC0849 & 08 48 56.3 & +44 52 16 & 1.261 & 0.049 \\
RXJ0910 & 09 10 44.9 & +54 22 09 & 1.110 & 0.061 \\
RXJ1053 & 10 53 39.8 & +57 35 18 & 1.140 & 0.063 \\
Cl1137 & 11 37 33.4 & +30 07 36 & 0.959 & 0.066 \\
RXJ1221 & 12 21 24.5 & +49 18 13 & 0.700 & 0.067 \\
SC1324 & 13 24 52.0 & +30 35 43 & 0.756 & 0.142 \\
Cl1350 & 13 50 48.5 & +60 07 07 & 0.804 & 0.054 \\
Cl1429 & 14 29 06.4 & +42 41 10 & 0.920 & 0.084 \\
SC1604 & 16 04 25.5 & +43 13 25 & 0.910 & 0.089 \\
RXJ1716 & 17 16 49.6 & +67 08 30 & 0.813 & 0.057 \\
RXJ1757 & 17 57 19.4 & +66 31 31 & 0.691 & 0.063 \\
RXJ1821 & 18 21 32.9 & +68 27 55 & 0.811 & 0.048 \\
\hline
\end{tabular}
\begin{flushleft}
$a$: Effective area of the spectroscopic footprint of each field, where the overdensity search is performed, in square degrees. This is estimated with assigning 0.5 Mpc radii circles to all spectroscopic objects in the redshift range $0.55 < z<1.37$ and summing their total projected area.

ORELSE fields with complete photometric redshift and spectroscopic catalogs used in this cluster search study, adapted from \citet{Lubin09}. The redshift for each field is that of the targeted known structures in the field. The original two Cl1604 and two Cl1324 fields were combined to the single SC1604 and SC1324 supercluster fields, respectively.
\end{flushleft}
\end{table}

\subsection{Optical/Near-infrared Imaging and Photometry}

Initial optical $riz$ imaging for most ORELSE fields was obtained with Suprime-Cam \citep{Miyazaki02} on Subaru and the Large Format Camera \citep[LFC;][]{Simcoe00} on the Palomar 200-inch Hale telescope. For XLSS005, the initial optical imaging was instead acquired with MegaCam \citep{Boulade03} on the Canada France Hawaii Telescope (CFHT) as part of the ``Deep'' portion of the CFHT Legacy Survey (CFHTLS). Additional $B$- and $V$-band imaging was taken for all ORELSE fields with Suprime-Cam with the exception of XLSS005 which had $u^{*}$- and $g^{\prime}$-band imaging available. The optical imaging has typical depths ranging from $m_{AB}$ = 26.4 in the $B$-band to $m_{AB}$ = 24.6 in the $z$-bands using the estimation methods described in \citet{Tomczak17}. Table \ref{tab.imaging} shows the available photometry with depth estimates for every ORELSE field and the facilities and telescopes used to acquire the data.

All LFC data were reduced with a suite of image processing scripts\footnote{\url{http://www.ifa.hawaii.edu/~rgal/science/lfcred/lfc_red.html}} written in Image Reduction and Analysis Facility \citep[IRAF;][]{Tody93} and following the methods of \citet{Gal08}. Suprime-Cam data were reduced using the \texttt{SDFRED} pipeline \citep{Ouchi04} and several Traitement \'El\'ementaire R\'eduction et Analyse des PIXels (\texttt{TERAPIX})\footnote{\url{http://terapix.iap.fr/}} software packages. We performed photometric calibration from same-night observations of standard star fields from the \citet{Landolt92} catalogs. The optical CFHTLS observations were reduced and photometrically calibrated using \texttt{TERAPIX} routines following the methods described in \citet{Ilbert06} and the T0006 CFHTLS handbook\footnote{\url{http://terapix.iap.fr/cplt/T0006-doc.pdf}}. For further details on the reduction of these data, see \citet{Tomczak17}.

Near-infrared (NIR) $J$ and $K$/$K_{s}$ imaging was taken for every ORELSE field but Cl1350. These observations were conducted with the Wide-field InfraRed Camera \citep[WIRCam;][]{Puget04} on the CFHT and the Wide Field Camera \citep[WFCAM;][]{Casali07} on the United Kingdom InfraRed Telescope (UKIRT). The $J$ and $K$/$K_{s}$ bands reached a typical depth of $m_{AB}$ = 21.9 and 21.7 respectively. Both facilities implement automated data reduction pipelines that output fully-reduced mosaics and weight maps. The UKIRT data were reduced through the standard UKIRT processing pipeline provided courtesy of the Cambridge Astronomy Survey Unit\footnote{\url{http://casu.ast.cam.ac.uk/surveys-projects/wfcam/technical}}. The CFHT data were ran through  the I'iwi pre-processing routines and \texttt{TERAPIX}. We photometrically calibrate these mosaics using bright ($<$15 mag), non-saturated objects with existing photometry from the Two Micron All Sky Survey \citep[2MASS;][]{Skrutskie06} in each field.

Additional imaging in the NIR was taken with the {\it Spitzer} \citep{Werner04} Space Observatory using the InfraRed Array Camera \citep[IRAC;][]{Fazio04}. All 15 ORELSE fields were observed in the two non-cryogenic channels ([3.6]/[4.5]). Four fields (SC1604, RXJ1716, RXJ1053, and XLSS005) were additionally observed in the two cryogenic channels ([5.8]/[8.0]) to an average respective depths of 24.0, 23.8, 22.4 and 22.3 magnitudes. These data were provided by the {\it Spitzer} Heritage Archive in the form of basic calibrated data (cBCD) images and were reduced using the MOsaicker and Point source EXtractor \citep[\texttt{MOPEX};][]{Makovoz05} package and several custom Interactive Data Language (\texttt{IDL}) scripts written by J. Surace. Further details on these data reduction can be found in \citet{Tomczak17}.

For each field, all optical and non-{\it Spitzer} images were registered to a common grid of plate scale 0.2$\arcsec$ pixel$^{-1}$ and then convolved to the field's worst point spread function (PSF) using the methods described in \citet{Tomczak17}. The worst PSF for each field was between $\sim$1.00$\arcsec$-1.96$\arcsec$, with Cl1350 being the only field with an image that had a PSF greater than 1.4$\arcsec$. Source detection and photometry for each field were obtained by running Source Extractor \citep[SExtractor;][]{Bertin96} in dual-image mode using either a stacked $\chi^{2}$ optical image or a single-band image as a detection image. For details on the specific image used for each field, see \citet{Tomczak17,Rumbaugh18,Lemaux19}. Photometry is extracted from PSF-matched images with SExtractor using fixed circular apertures with diameters 1.3 times the full-width half-maximum (FWHM) of the largest PSF. The total magnitudes are obtained through using the ratio of aperture and SExtractor AUTO flux densities as measured in the detection image. Magnitude uncertainties were calculated by adding the SExtractor uncertainties and background noise in quadrature. The background noise was estimated by 1$\sigma$ root mean square (RMS) scatter of measurements in hundreds of blank sky regions for each band. We incorporated {\it Spitzer}/IRAC magnitudes by running the software \texttt{T-PHOT} \citep{Merlin15} on the fully reduced mosaic images. This took the segmentation maps from the ground-based detection images as the input, where flux density uncertainties were estimated from the scaled best fit model for each object. For more details on the reduction and measurements of ORELSE imaging data, see \citet{Tomczak17}.

\subsection{Spectroscopy} \label{subsec:spectroscopy}

The majority of spectroscopic data were obtained as part of a 300 hour Keck II/DEep Imaging Multi-Object Spectrograph \citep[DEIMOS;][]{Faber03} campaign. The number of slitmasks per field varied between 4 (for RCS0224) and 18 (for SC1604), as more extensive coverage was given to the larger and more complex large-scale structures, as well as those at higher redshift.  These observations were taken using the 1200 line mm$^{-1}$ grating with 1$\arcsec$ slit widths. Central wavelengths were chosen to be between 7200\AA\ to 8700\AA\ depending on the redshift of the field. Average exposure times were between $\sim$7000s to $\sim$10500s, chosen to roughly obtain an identical distribution in continuum S/N across all masks independent of conditions and the median faintness of the target population. This configuration produced spectra with a pixel scale of 0.33\AA\ pix$^{-1}$, a resolution of R $\sim$ 5000 ($\lambda$/$\theta_{\textrm{FWHM}}$, where $\theta_{\textrm{FWHM}}$ is the full-width half-maximum  spectral resolution), and a wavelength range of $\Delta\lambda \sim$ 2600\AA.

The selection for the DEIMOS targets was based on color and magnitude cuts to maximize the number of objects with a high likelihood of being on the cluster/group red sequence at the presumed redshift of the large-scale structure in each field using methods described in \citet{Lubin09}. These targets were the highest priority (priority 1), and we assigned progressively lower priority to progressively bluer objects. Though our selection scheme heavily favored redder objects, the majority of our spectroscopic targets had colors bluer than the highest priority objects, due to the relative rarity of objects at these red colors and the strictness of our cuts, as discussed in depth in \citet{Tomczak17}. The fraction of priority 1 targets in our final sample ranged from $\sim$1\% to $\sim$45\% across all ORELSE fields. This fraction generally varied strongly with the density of spectroscopic sampling in each field. We also assigned additional priority to a very small number of special interest targets such as X-ray or radio detected objects for use in other ORELSE studies that primarily focused on AGN activity \citep[e.g.,][]{Rumbaugh12,Rumbaugh17,shen17,shen19}. We generally restricted targets to a magnitude limit of $i < 24.5$, though we also had  2-5\% targets per field fainter than this limit. As shown in \citet{shen17} and \citet{Lemaux19}, the resultant ORELSE spectral sample is found to be broadly representative of the underlying galaxy population at $i/z<24.5$ for all but the bluest galaxy types. 
 
Spectroscopic data were reduced using the Deep Evolutionary Extragalactic Probe 2 \citep[DEEP2;][]{Davis03,Newman13} \texttt{spec2d} pipeline, which generates processed two-dimensional and one-dimensional spectra for each slit. The version used to reduce our data additionally had several modifications to improve the response correction precision, perform absolute spectrophotometic flux calibration, and improve the method of joining the blue and red ends of the spectra over the $\sim$5\AA\ gap separating the two CCD arrays. See \citet{Lemaux19} for greater discussion on the reduction of our spectroscopic data. Additionally, every two-dimensional spectrum was inspected to identify serendipitous detections (see \citealt{lem09} for details on these types of detections and the method used for finding them).

\subsection{Spectroscopic and Photometric Redshifts}

The DEEP2 \texttt{spec1d} pipeline is run on all the one-dimensional DEIMOS spectra to find 10 first-guess redshifts, by cross-correlating a suite of galactic and stellar templates. These redshifts are then used to inform a visual  inspection process performed  using the publicly available DEEP2 redshift measurement program, \texttt{zspec} \citep{Newman13} to determine, if possible, the redshift of each target. All targeted and serendipitously observed objects were visually inspected and assigned a spectroscopic redshift $z_{spec}$ and a redshift quality code $Q$ according to the DEEP2 convention, with secure stellar ($Q$ = -1) and extragalactic ($Q$ = 3, 4) redshifts scientifically usable at the $\geq$95\% confidence level \citep{Newman13}. $Q$ = -1 objects were identified securely as stars, which required either the presence of multiple significant narrow photospheric absorption features (e.g., H$\alpha$ and the Ca 2 triplet) or broad continuum features indicative of a late-type star (primarily TiO). $Q$ = 3,4 were objects identified as secure galaxies because they had two or more emission or absorption features, with $Q$ = 3 objects having one or more of the features slightly questionable in S/N. The presence of the unblended [O\,II] $\lambda$3726, 3729\AA\ doublet emission line was sufficient to assign a $Q$ = 4 code if both components were significantly detected. If the doublet was moderately blended by velocity effects and there were no other features, a $Q$ = 3 code was assigned. Further discussion on these quality codes and their accuracies can be found in \citet{Newman13}. For additional details on the quality codes as they pertain to ORELSE data, see \citet{Lemaux19}. For our work, we only use spectroscopic redshifts if they have a quality code of $Q$ = -1, 3, or 4. $Q$ = -1 objects were used to exclude stellar redshifts in the analysis.

In addition to our DEIMOS data, we use spectroscopic redshifts from a few previous studies using various telescopes and instruments \citep{Oke98,Gal04,Tanaka08,Mei12}, which comprised $\simlt$3\% of all spectroscopic redshifts for all fields except XLSS005, where the majority of high-quality spectroscopic redshifts (92\%)  were drawn from the VIMOS Very Deep Survey \citep[VVDS;][]{lefevre13}. For redshifts coming from these surveys we required that they have quality codes that correspond to a high probability ($\simgt$75\%) of being correct.

Photometric redshifts were derived through broadband spectral energy distribution (SED) fitting of optical to mid-IR photometry of each object. These redshifts were estimated using the code Easy and Accurate Redshifts from Yale \citep[\texttt{EAZY};][]{Brammer08}, and the methods are described in depth in \citet{Tomczak17}. To summarize, \texttt{EAZY} performs $\chi^2$ minimization for a grid of user-defined redshifts using linear combinations of a default set of six basis template SEDs. It then calculates a probability density function (PDF) from the minimized $\chi^2$ values in the form of $\mathrm{P}(z) \propto e^{-\chi^2 / 2}$. The PDF is finally modulated by a magnitude prior, for which we use $r$-band, that is designed to mimic the intrinsic redshift distribution for galaxies of a given apparent magnitude. Throughout this paper, we take \texttt{EAZY}'s ``$z_{peak}$'' to be the photometric redshift of an object, which is obtained by marginalizing over the final PDF. In the cases where an object has multiple peaks in its PDF, \texttt{EAZY} will only marginalize over the peak with the largest integrated probability.

We assess the accuracy of our photometric redshifts, $z_{phot}$, by comparing them with our spectroscopic redshifts, $z_{spec}$. To achieve this, we fit a Gaussian to the distributions of the residual $(z_{phot}-z_{spec})/(1+z_{spec})$ for all objects. The best-fit $\sigma_{\Delta z/(1+z_{spec})}$ is taken as the $z_{phot}$ uncertainty. The $z_{phot}$ uncertainties across all ORELSE fields with no magnitude restriction typically ranged between $\sigma_{\Delta z/(1+z_{spec})}$ = 2.2-3.2\%. The fraction of catastrophic outliers, or objects with SED fits with reduced $\chi^{2}_{\textrm{galaxy}} > 10$ from fitting with \texttt{EAZY}, is around 6\% on average for all fields. We also imposed a use flag criterion throughout this work such that the $z_{phot}$ values that we used were more likely to be reliable. This use flag required a $z_{phot}$ object to have a signal-to-noise of at least 3 in its detection image and to have coverage in at least five images. The use flag also excluded any $z_{phot}$ detections that were identified as a star, had over 20\% of its pixels saturated, or was in the worst 1\% of reduced chi-squared values of all objects in that field.

\begin{figure}
\includegraphics[width=\columnwidth]{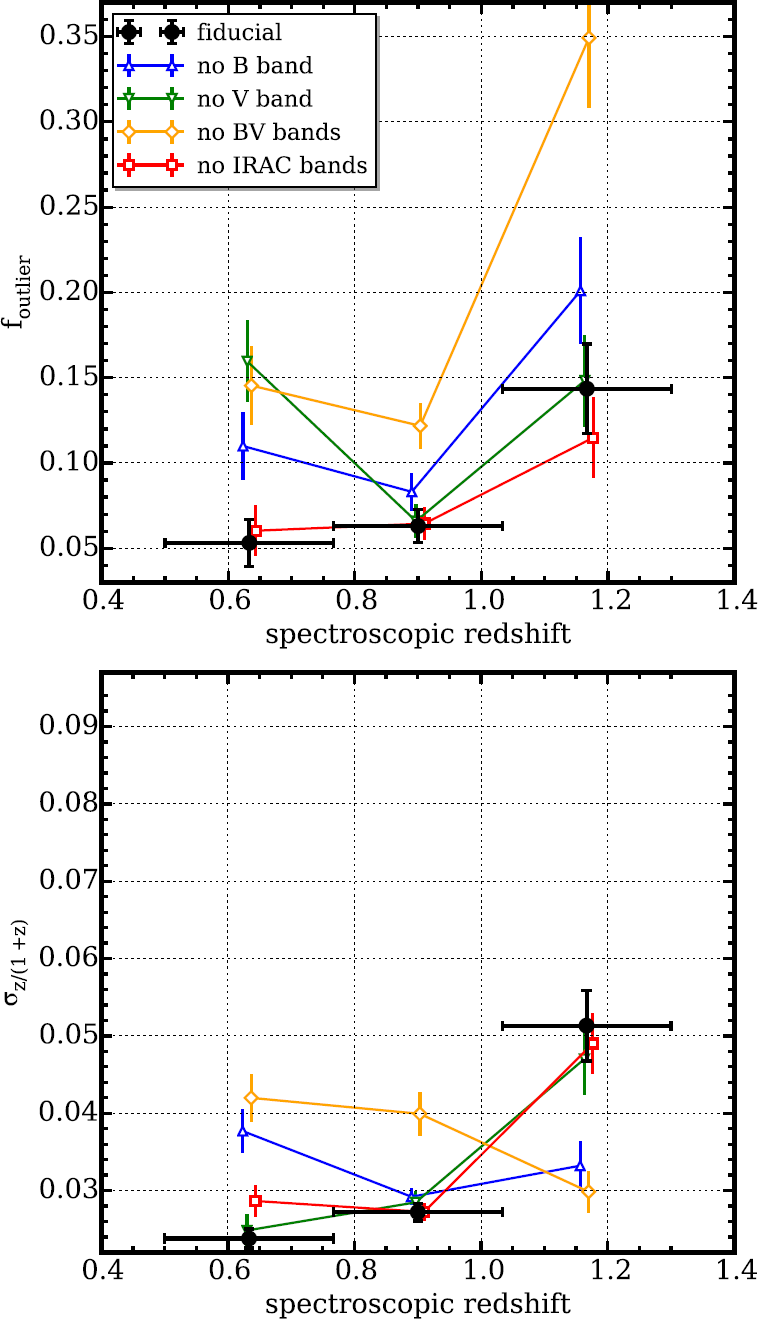}
\caption{The top panel shows the outlier fraction $f_{outlier}$ in three neighboring redshift regions in SC1604 after removing various bands. The fiducial points refer to no elimination of any band. $f_{outlier}$ here is defined as the fraction of galaxies with $|z_{phot}-z_{spec}|/(1+z_{spec}) > 0.15$. The bottom panel plots the size of the Gaussian's standard deviation, $\sigma_{\Delta z}/(1+z_{spec})$. Eliminating both the $B$- and $V$-bands produces the least constrained photometric redshift sample, indicating the necessity of including them.}
\label{fig.bandcuts}
\end{figure}

Since we will eventually be including objects with $z_{phot}$ values in our analysis, we attempted to test the consequence of varying the number of bands and the specific bands in which an object is detected on the accuracy and precision of the recovered $z_{phot}$. This is done in order to limit the final catalog to those objects with higher quality $z_{phot}$ values so as to maximize the purity and completeness of the eventual overdensity candidates that we find. To test this, we compared $z_{phot}$ and $z_{spec}$ values for $\sim$1400 galaxies in one of the ORELSE fields (SC1604) with secure spectral redshifts running \texttt{EAZY} fitting on a variety of different combinations of photometry. In total we tested five cases, i) all photometric data included (fiducial), ii) $B$-band imaging removed, iii) $V$-band imaging removed, iv) both $B$- and $V$-band imaging removed, and v) all IRAC imaging removed. In each case, the $z_{phot}$ values generated from that set of photometry are compared to the $z_{spec}$ values by $\sigma_{\Delta z/(1+z_{spec})}$ and $f_{outlier}$. These results are shown in Fig. \ref{fig.bandcuts}. We found that removing $B$-band information increases $\sigma_{\Delta z/(1+z_{spec})}$ by over 50\% at low redshift as well as drastically increasing $f_{outlier}$. Additionally, $f_{outlier}$ is significantly higher at high redshift when both the $B$- and $V$-band information are excluded. Interestingly, excluding the information from the IRAC bands from our fitting had very little effect on the $z_{phot}$ precision or accuracy relative to the fiducial setup at least for the galaxy population studied here. We chose the IRAC bands for this exercise rather than in the $J/K$ bands because the latter are relatively shallow and cutting on these bands results in fewer total objects remaining to perform this test on. Additionally, we tested the effect of requiring different significance detections in the $B$- and $V$- bands, finding that requiring magnitude errors of $\le0.3$ in both bands gave the best combination of precision and accuracy while still allowing us to include most photometric objects in our final sample. This criteria was imposed on all photometric objects to generate our final sample (with the exception of the XLSS005 field, see below) and corresponds to detection significance of $\ge$3.6$\sigma$ in both bands. 

\begin{figure}
\includegraphics[width=\columnwidth]{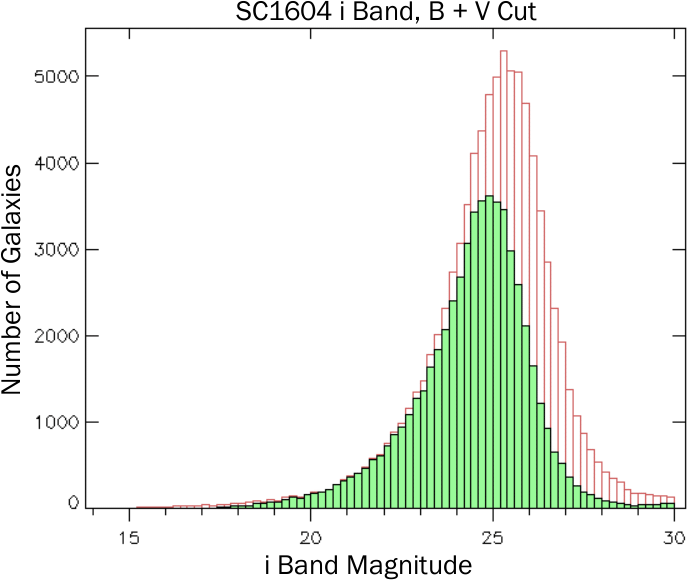}
\caption{The $i$-band magnitude distribution for all objects in the SC1604 field compared with the population of objects with $B$- and $V$-band photometric errors less than 0.3 mag in green. The $B$ and $V$ cut still contains the majority of objects in the magnitude range of interest, 18 $\leq i \leq$ 24.5.}
\label{fig.cuts}
\end{figure}

In essentially every field, because the $B$- and $V$-band images are deep (see Table \ref{tab.imaging}), and because we include only those objects brighter than $i/z<24.5$ in our final sample, the above criteria essentially amounts to only including those areas where $B$- and $V$-band coverage is available, which is the case for essentially every spectroscopic object. The $B$- and $V$-band requirement additionally included most of the photometric objects in our redshift catalog in the range of 18 $\leq i/z \leq$ 24.5 (Fig. \ref{fig.cuts}). These cuts were used for all spectroscopic and photometric objects in the 14 ORELSE fields that have similar imaging depth in the $B$- and $V$-bands. The one remaining field, XLSS005, has CFHT Legacy Survey (CFHTLS\footnote{\url{ftp://cdsarc.u-strasbg.fr/cats/II/317/T0007-doc.pdf}}) $u^{*}$/$g^{\prime}$ imaging that acted in place of our typical $B$- and $V$-band requirement (Table \ref{tab.imaging}).

\begin{figure}
\includegraphics[width=\columnwidth]{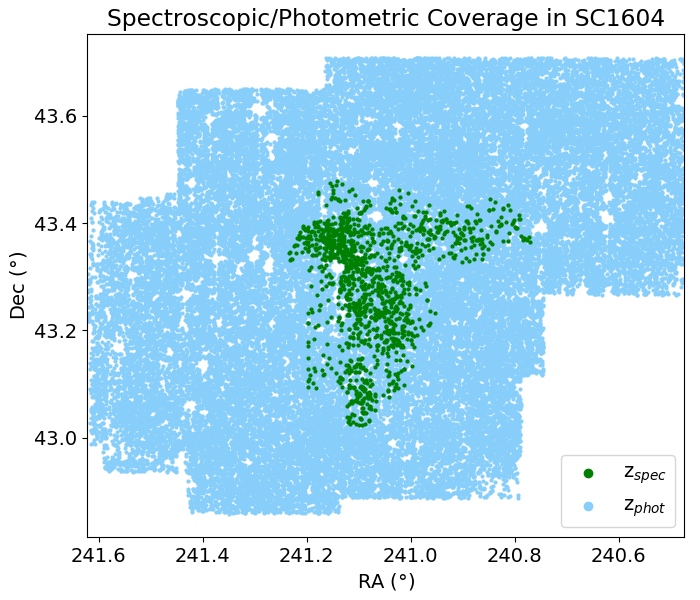}
\caption{All photometric and high-quality spectroscopic members in the ORELSE field SC1604. The photometric and spectroscopic redshifts in this field cover the ranges $0.03 < z_{phot} < 5.76$ and $0.11 < z_{spec} < 2.76$, respectively. Of all the ORELSE fields, SC1604 is the most well sampled field, with high spectroscopic coverage and superb photometric redshift accuracy and precision. The photometric coverage far exceeds the spectroscopic range, which is primarily limited to the regions around the known structures in the field. Regions with no imaging data or with severe issues with bright stars or other imaging artifacts (depicted in white) are masked and excluded from the overdensity calculations of each ORELSE field. We limit our search range to the spectroscopic footprint as overdensity candidates detected outside the spectroscopic range of coverage will likely have more uncertain redshifts than those inside due to the higher uncertainty in the photometric redshifts.}
\label{fig.specprint}
\end{figure}

For the purpose of this work, spectroscopic redshifts are extremely helpful, since they provide highly accurate information on the position of the galaxies along the line of sight and are therefore extremely important in identifying and mapping the large-scale structures. However, obtaining spectra for a large and contiguous field is difficult, and often the spectroscopic coverage is not evenly distributed in the sky (see an example in Fig. \ref{fig.specprint}). This is why our approach in detecting overdensity candidates (see \S\ref{methodology}) includes the use of both spectroscopic and photometric redshifts. The latter, although less accurate, generally has a more uniform spatial distribution. In conjunction with the spectroscopic redshifts, $z_{phot}$ values, if treated properly, are able to provide a more complete mapping of the density field. As a reminder, however, we limited our sample to areas in and near the spectroscopic footprint, as the effectiveness of our methodology degrades considerably in the complete absence of spectroscopic redshifts.

\section{Methodology}\label{methodology}

Our goal is to discover and characterize new overdensity candidates in the ORELSE fields. Once we identify these candidates, we can translate each of their overdensities derived from the VMC overdensity maps to their total gravitating halo mass. This translates the observed spatial clustering of galaxies into a mass distribution, which can be used to trace the underlying dark matter distribution as described in \citet{Cucciati14}. For overdensity candidates with sufficient spectroscopy, we can also estimate masses from their measured velocity dispersions \citep{Gal08,Lemaux12}.

To find new overdensity candidates, we look for overdensities that subtend a large angular distance and are coherent over some redshift range. We apply the standard photometry software package Source Extractor \citep[SExtractor;][]{Bertin96} to the VMC overdensity maps, which are divided into several small redshift slices, to identify overdensities candidates in each slice.

In this section, we describe the methods used in our overdensity candidate detection algorithm. We discuss our optimization schemes and how we set our parameters in \S\ref{optimize}.

\subsection{Voronoi Tessellation}\label{slices.intro}

Mapping the density field of galaxies requires a large, homogeneous, and unbiased sample of galaxies with accurate redshifts \citep[spectroscopic and/or photometric,][]{Darvish15}. 2D surface density estimates are made using a series of narrow redshift slices, where the widths of the z-slices are set at first-order by some characteristic of the data, for example the photometric redshift precision or the redshift extent of structures in the field. A too narrow width might miss galaxies belonging to a structure extended along the line of sight, while a too broad width risks contamination from foreground and background galaxies. We construct our overdensity maps with what is known as the Voronoi tessellation Monte-Carlo (VMC) method. Optimizing our VMC overdensity map code is critical for accurately determining the redshifts of all overdensity candidates in a field. 

A Voronoi tessellation is the division of a 2D plane into a number of polygonal regions equal to the number of objects in that plane. The Voronoi cell of each object is defined as the region closer to it than to any other object in the plane. Objects in high density regions therefore have small Voronoi cells, while objects in lower density regions have larger cells. The inverse area of the cell sizes can thus be used to measure the local density at the position of the object bounded by the cell.

When we apply the Voronoi tessellation to our data, the redshift slices are our 2D planes and the galaxies are the objects in the planes. Voronoi tessellation is advantageous to use over other density field estimators as it is scale-independent and can be used over large physical lengths. Most importantly for the detection of often irregularly shaped overdensity candidates, it makes no assumptions about the geometry or morphology of structures in the field \citep{Darvish15}.

Not all galaxies have equally well-determined redshifts. We must take into account the high uncertainties in using the photometric redshifts. To do so, we use a VMC technique broadly following the weighted Voronoi tessellation estimator method outlined in \citet{Darvish15} and described in \citep{Lemaux18}. For the galaxies in our sample with only photometric redshifts, $z_{phot}$, we use a Monte-Carlo acceptance-rejection process to treat these redshifts and their uncertainties from \texttt{EAZY} as statistically asymmetric Gaussians.

For each Monte-Carlo realization, we assign a new $z_{phot,MC}$ to each $z_{phot}$ galaxy. This $z_{phot,MC}$ is randomly sampled from a simplified version of the $z_{phot}$ PDF, where we assume the PDF is a Gaussian centered on the original $z_{phot}$ PDF. The $\sigma$ of the $z_{phot,MC}$ is either the upper or lower $z_{phot}$ error depending on whether the sampled random number was above or below the mean of the Gaussian peak. If the sample point is lower than the mean of the Gaussian peak, it is multiplied by the lower 1$\sigma$ on the galaxy's $z_{phot}$ and subtracted from the original $z_{phot}$. If the sample point is higher than the mean of the Gaussian peak, it is multiplied by the upper 1$\sigma$ on the galaxy's $z_{phot}$ and added to the original $z_{phot}$.

\begin{figure*}
\centering
\includegraphics[width=2\columnwidth]{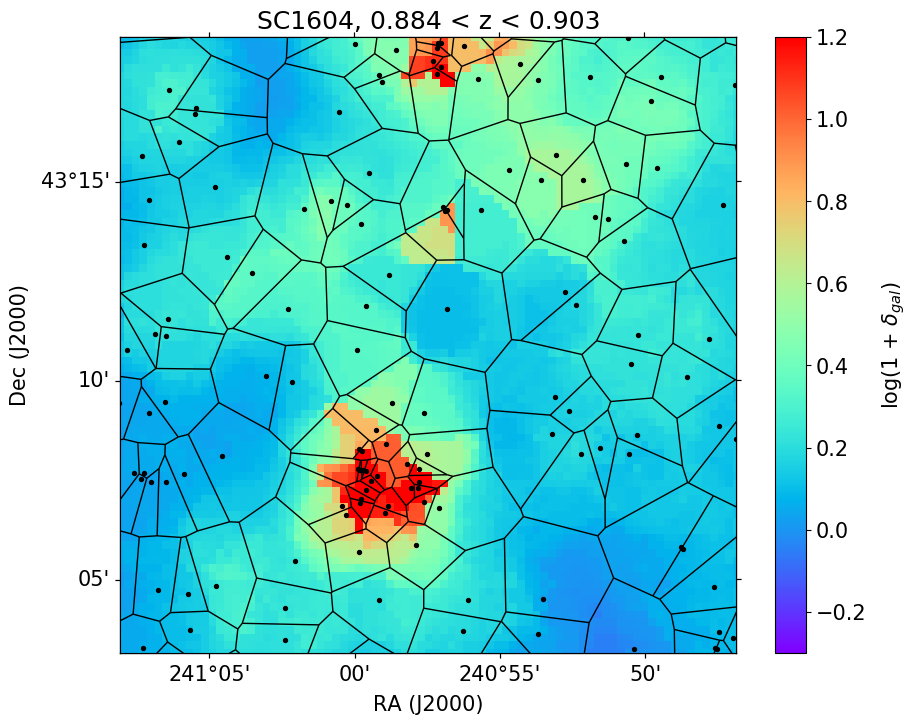}
\caption{Depicted is a portion of one redshift slice in the Voronoi Monte-Carlo overdensity map around the SC1604 A overdensity with one of the Voronoi tessellation realizations overlaid. Each ORELSE field is sliced into redshift bins of $\Delta \mathrm{v} = \pm 1500$ km s$^{-1}$ wide across the redshift range of $0.55 < z<1.37$. For each slice, the photometric redshifts are randomly sampled 100 times based on their estimated $z_{phot}$ uncertainties. The black points, or galaxies, in the slice are partitioned into polygonal cells with a single realization of the Voronoi tessellation, where each cell represents the projected area that is closest to the galaxy in it than any other galaxy. The underlying overdensity map, shown in the background, is the median among the 100 VMC realizations, on a grid of 75$\times$75 kpc pixels, as described in the text. The color code is shown in the color bar on the right.}
\label{fig.voronoimap}
\end{figure*}

These $z_{phot,MC}$ and $z_{spec}$ galaxies are sliced into bins of approximately $\pm$1500 km s$^{-1}$ in velocity space over $0.55 < z<1.37$. We discuss how we set our number and width of slices in \S\ref{slices.opt}. The Voronoi tessellation is applied on 100 realizations of each bin. For each realization, a grid of 75$\times$75 proper kpc pixels is used to sample the local density distribution for each slice. The local density of each grid point for each realization is set equal to the inverse of the Voronoi cell area that encloses the grid point, multiplied by the square of the angular diameter distance. As the slices go to higher redshift, the projected size of the sky covers a larger proper area. Because the pixel scale is fixed, this means the image size for each redshift slice will increase with higher redshift for the same field. The final local overdensities for each grid point in the redshift slice are computed by median combining the values from the 100 Monte-Carlo realizations. The local overdensity in a pixel ($i,j$) is approximated with

\begin{equation} \label{eq.overdense}
\mathrm{log}(1 + \delta_{\mathrm{gal}}) = \mathrm{log}(1+(\Sigma_{i,j}-\tilde{\Sigma})/\tilde{\Sigma})
\end{equation}

where $\delta_{\mathrm{gal}}$ is the density of galaxies, $\Sigma_{i,j}$ is the given pixel's density, and $\tilde{\Sigma}$ is the median density of all pixels in the slice (Fig. \ref{fig.voronoimap}). As discussed in \citet{Tomczak17,Lemaux19}, these local overdensities have been shown through tests to correlate well with other density metrics and, as we will show later, trace out the known structures extremely well. 

\subsection{Source Extractor}

We used SExtractor to find the overdensity candidates in our VMC overdensity maps. As clusters and groups are not necessarily regular in shape, we use the isophotal fluxes rather than any of SExtractor's elliptical or circular apertures. When running SExtractor over a VMC overdensity map, SExtractor outputs isophotal flux values, which are its measure of a given region's overdensity. Thus, higher fluxes indicate higher densities. SExtractor identifies pixels as significant if their overdensities are above some given detection threshold. The pixels are then identified as detections if their groupings are larger than some given minimum area. The higher the detection threshold and the larger the minimum area, the fewer detections SExtractor will find. Therefore, carefully choosing the optimal parameters is key for finding as many overdensity candidates in the map as possible without being overwhelmed by astrophysical and random noise. We discuss the optimization of these parameters in greater detail in \S\ref{optimize}.

\subsubsection{Linking SExtractor Detections}\label{masks.linking}

In order to find significant overdensity candidates in our fields, we must first properly assess the background galaxy density, which is calculated by SExtractor. The outer edges of the imaging footprint in the VMC overdensity maps have higher galaxy incompleteness and thus artificially low overdensities. Including such regions in our SExtractor analysis would skew the average background galaxy overdensity low. When the VMC overdensity maps are made, we compute the densities after masking out regions without imaging data or which have been severely corrupted by bright stars or image artifacts, and then calculate the overdensities. In the final overdensity maps, we still have low density regions around the boundaries of the imaging footprint. To exclude these low density regions, we constructed a mask for every redshift slice in each field to remove the areas of the maps with overdensities less than log(1+$\delta_{gal}$) = -0.35 and passed them into SExtractor. Over all ORELSE fields, this masked roughly 5.8\% of the spectroscopic footprint, but 0\% of the spectroscopic footprint of five fields: SG0023, SC0849, RXJ1053, Cl1350, and SC1604.

For every redshift slice, SExtractor outputs a position and total isophotal flux for each detection it finds. To find coherent overdensity candidates across separate redshift slices, we calculate the distances between all SExtractor detections in one slice with all SExtractor detections in the immediate next redshift slice. The distance calculated is the angular diameter distance evaluated at the redshift which is the average of the two slices' central redshifts.

If two detections are within a certain linking radius, we consider them as part of the same overdensity candidate. We then use their flux weighted position to attempt to link the pair with a third detection in the next redshift slice, where the link is successful if the third detection is also within the same linking radius as before. This process is repeated across redshift slices until no further links are found. The final centroided position is the flux weighted average of all linked detections in that overdensity candidate. Further details of this search and our tests with different linking radii can be found in \S\ref{slices.opt}.

\begin{figure}
\includegraphics[width=\columnwidth]{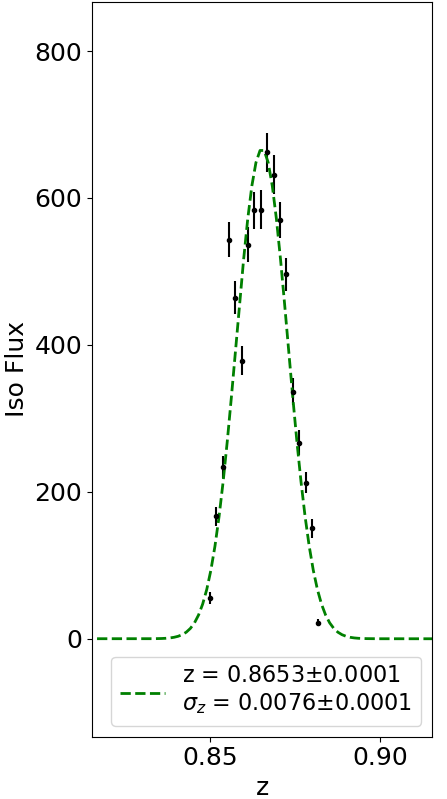}
\caption{The panel shows an example of a Gaussian fit of a linked overdensity candidate. The individual points are the isophotal flux values of individual SExtractor detections in neighboring redshift slices. The position of the overall overdensity candidate is an average weighted by the isophotal flux of all linked detections in the overdensity candidate. $\sigma_{z}$ is the redshift dispersion, describing the width of the Gaussian fit and therefore the extent of the overdensity candidate across redshift.}
\label{fig.linking}
\end{figure}

The redshift of the overdensity candidate is then determined by fitting a Gaussian to the isophotal fluxes of all the linked detections as a function of redshift, where the isophotal flux and error for each detection are calculated by SExtractor. We use the standard deviation of the Gaussian, $\sigma_z$, to describe the redshift dispersion. We expect the redshift of a overdensity candidate to be where the density of galaxies is highest, and we take the mean of the Gaussian fit to be the redshift of the overdensity candidate (Fig. \ref{fig.linking}). To avoid cases where the Gaussian is largely extrapolated and fitted only to a few data points near one tail, we require the amplitude of the Gaussian to be no more than 20\% of the highest value data point in the fit and remove all candidates that do not meet this criterion.

Because we attempt to link all possible SExtractor detection chains starting at each redshift slice, there will be some linked overdensities which are subsets of links that begin at earlier redshifts. However, these overdensities will likely have similar redshifts and centroided positions. We control for these duplicate detections by iterating over all the detections in a field, starting from the largest Gaussian fits by amplitude, and removing any other detections within both 0.7 Mpc, a distance which is the average extent of a group or cluster, and $\Delta z < 0.02$. There will likely be a few duplicate detections of the more irregularly shaped overdensities remaining after this removal process, but we expect these to be few in number. We go over the results of setting this separation threshold in \S\ref{final}.

\section{Optimization and Choice of Parameters}\label{optimize}

\subsection{Detection Parameter Optimization}\label{optimize.1}

SExtractor's object identification strongly depends on the choice of DETECT\textunderscore THRESH, the detection threshold significance above the median overdensity, and DETECT\textunderscore MINAREA, the minimum area of an object in square pixels. The isophotal fluxes calculated by SExtractor are the overdensities above the detection floor. A higher floor in other words translates to smaller isophotal fluxes. Too restrictive parameters means we lose detection of structures, but too inclusive parameters inundates our identified overdensity candidates with noise and false detections. Larger minimum areas require the detection of more of an overdensity candidate's subtended angular size, lowering the chance of a false positive detection, but can miss detecting lower mass clusters. Smaller areas only require detecting overdensity cores but are more susceptible to noise contamination.

We tested a grid of DETECT\textunderscore THRESH of 3, 4, 5, and 6$\sigma$ and DETECT\textunderscore MINAREA of 10, 20, 40, 80, and 160 square pixels. The $\sigma$ value is what SExtractor calculates as the RMS noise in the background of a given slice of the VMC overdensity map. For a single detection, all the pixels must be above the detection threshold and adjacent to each other, and the total area of the pixels must be at least as big as the minimum area. Ideally, we should set our detection threshold low enough to pick up groups and low mass clusters but not so low that we are overwhelmed by small fluctuations of noise. The results of these tests are detailed in \S\ref{recovering}.

We did not use any smoothing Gaussian filter in SExtractor as filters are best suited for recovering regularly shaped large-scale structures, and real structures are not necessarily all regularly shaped. We did test how various filtering schemes performed when attempting to recover injected mock structures in \S\ref{mock.compur} and found only a very modest improvement when using a filter versus not using a filter at all.

We found the background RMS values in our fields were generally around $\log(1+\delta_{gal})$ = 0.09-0.15. Our grid of detection threshold $\sigma$s probe below and above a local overdensity of log(1+$\delta_{gal}$) = 0.5, which is the typical high end of the $\log(1+\delta_{gal})$ distribution for field surveys and likely corresponds to group-like environments \citep{Pelliccia17}. The minimum area is essentially a measure of the velocity dispersion of a structure. Smaller minimum areas are sensitive to smaller velocity dispersions. The velocity dispersions of clusters are typically calculated over a 0.5 or 1 Mpc radius. With our 75$\times$75 proper kpc pixel scale, our minimum areas cover the lowest end of this range, translating to circles with areas of 0.06 to 0.9 square Mpc, which allows us to more easily identify groups and low mass clusters.

\subsection{Linking Detections Across Redshift Slices}\label{slices.opt}

As first introduced in \S\ref{slices.intro}, we tested using VMC overdensity maps divided into redshift slices of different spacings. With more overlap between neighboring slices, we have more total redshift slices and thus a higher number of detections of a overdensity candidate in the field. We expect the overdensity candidates will be easier to detect with more detections, though increasing the total number of redshift slices can greatly increase the total computation time in constructing the VMC overdensity maps.

When constructing our VMC overdensity maps, we set our redshift slice size at 0.01(1+z), corresponding to an approximately $\pm$1500 $\kms$ velocity dispersion. This value is roughly twice the typical velocity dispersions for known structures in the ORELSE fields and rivals the velocity dispersions observed for the most massive galaxy clusters \citep[e.g.,][]{Ruel14,Owers17}. With narrower redshift slices, we run the risk of subsampling structures, missing massive cohesive structures because they could become separated over different bins. The same is true for galaxies with only photometric redshifts, as these redshifts have much coarser resolution than z $\sim$ 0.01. We tested using redshift slice sizes two and four times bigger than 0.01(1+z) and found that these wider slices placed a majority of distinct redshift structures into the same redshift slice, reducing the accuracy of their measured redshift. Additionally, instituting wider bins had the effect of several of the known ORELSE structures being missed entirely since their full redshift extent were fully contained in only one slice.

The VMC overdensity maps are made such that each redshift slice deliberately overlaps with the slice before it. This avoids scenarios where a single overdensity candidate gets separated over different slices. We tested maps where the overlapping redshift slices were centered at $\Delta z$ step sizes of 1/3, 1/4... 1/10 of the total width of each slice. In other words, at a 1/10 step size, we would for example have neighboring redshift slices covering $z$ = 0.600 to 0.620 and $z$ = 0.602 to 0.622, or 90\% overlap between the two slices. The width of the redshift slice remains the same regardless of the step size.

\begin{figure}
\includegraphics[width=\columnwidth]{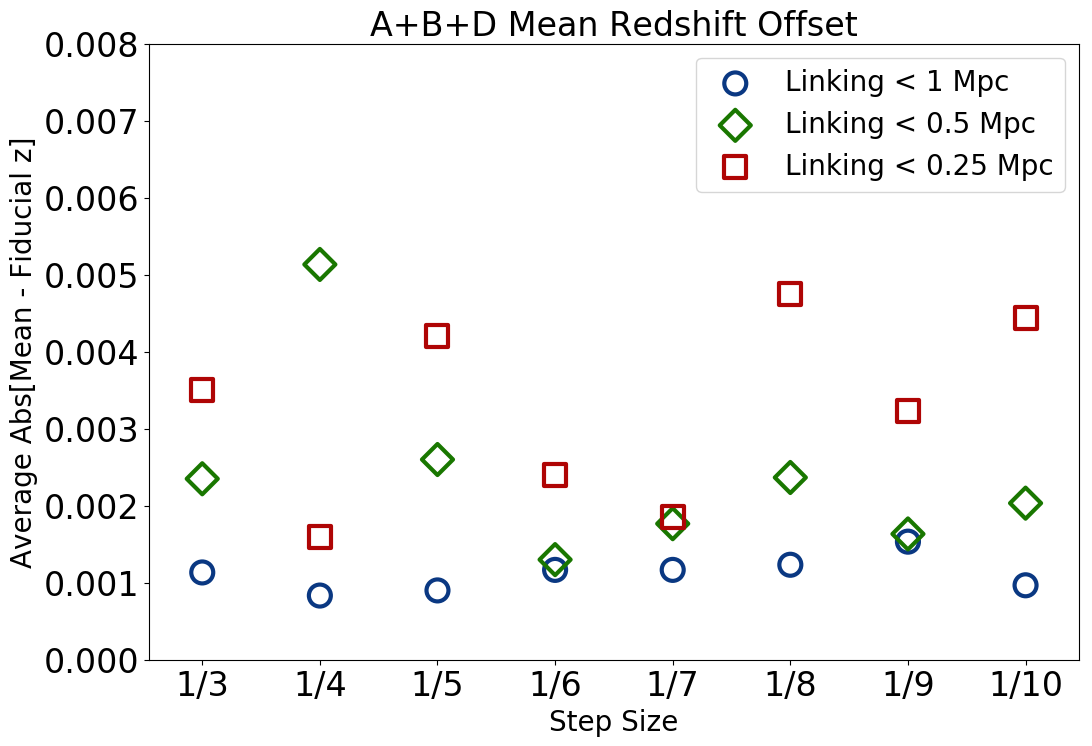}
\includegraphics[width=\columnwidth]{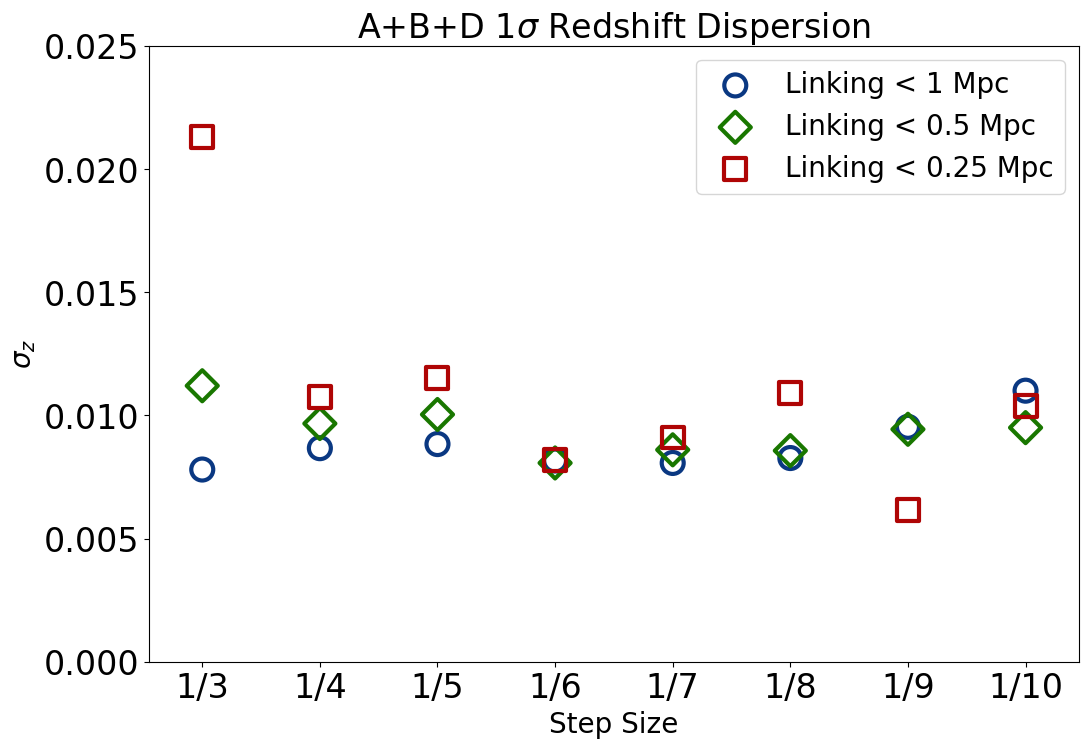}
\caption{The average absolute offset between the fiducial and fitted mean redshifts and 1$\sigma$ redshift dispersions $\sigma_{z}$ for known Clusters A, B, and D in SC1604. The step size is the spacing between redshift slices set as a fraction of the total width of each slice. The largest linking radius of 1 Mpc performs the best in recovering the fiducial redshifts. The redshift dispersions are all on the order of 0.01 for every linking radius, showing that we consistently recover the shape of each structure well. We find no significant dependence on step size for either the determined mean redshift or redshift dispersion.}
\label{fig.steps}
\end{figure}

We tested how well linking radius values, as described in \S\ref{masks.linking}, of 0.25, 0.5 and 1 Mpc performed in recovering three known structures in SC1604 at around $z \sim 0.9$, Clusters A, B, and D. We chose the SC1604 field because, of all the fields in ORELSE, SC1604 is the closest to being ideal for such a test in several regards. It is the field in ORELSE with the most dense spectroscopic coverage, and as well as some of the deepest. In addition, the large number of bands and the depth of the imaging have resulted in  excellent $z_{phot}$ accuracy and precision. We selected Clusters A, B, and D in this field for our tests because they are large and isolated, and they spanned a narrow redshift range, so we could construct VMC overdensity maps for these tests in a relatively short time-frame. These structures additionally span a range of different morphologies, where Cluster A appears as a typical cluster, Cluster B is close to dynamically relaxed, and Cluster D is elongated and irregular in shape (see \citealt{Lemaux12,Rumbaugh18} for more details). We found that a 1 Mpc linking radius was the only value large enough to link detections over the full Gaussian profile for Cluster D. The 1 Mpc linking radius also performed best overall in recovering the fiducial redshifts (Fig. \ref{fig.steps}). The fiducial redshifts were taken from the biweight mean of the known spectral members, which were within 3$\sigma$ of the LSS's velocity dispersion and a 1 Mpc projected radius.

The spectroscopic coverage in each ORELSE field is not uniform. Our densest spectroscopic coverage is around known structure targets. Many of the overdensity candidates in the field are likely to be less spectroscopically sampled. As a cursory estimation on how well our large-scale structure detection algorithm performs for such cases, we also test how well we can recover the known structures using smaller fractions of the spectroscopic data available. We thus repeated the same step size tests with SC1604 using 50\% and 25\% of the available spectroscopic redshifts in constructing the VMC overdensity maps. Decreasing the number of spectroscopic members is a good approximation for how our large-scale structure detection algorithm would fare with the overdensity candidates we are trying to find, as they generally will have lower levels of spectroscopic completeness than the nearly complete ($i \leq$ 24.5) SC1604 field.

\begin{figure}
\includegraphics[width=\columnwidth]{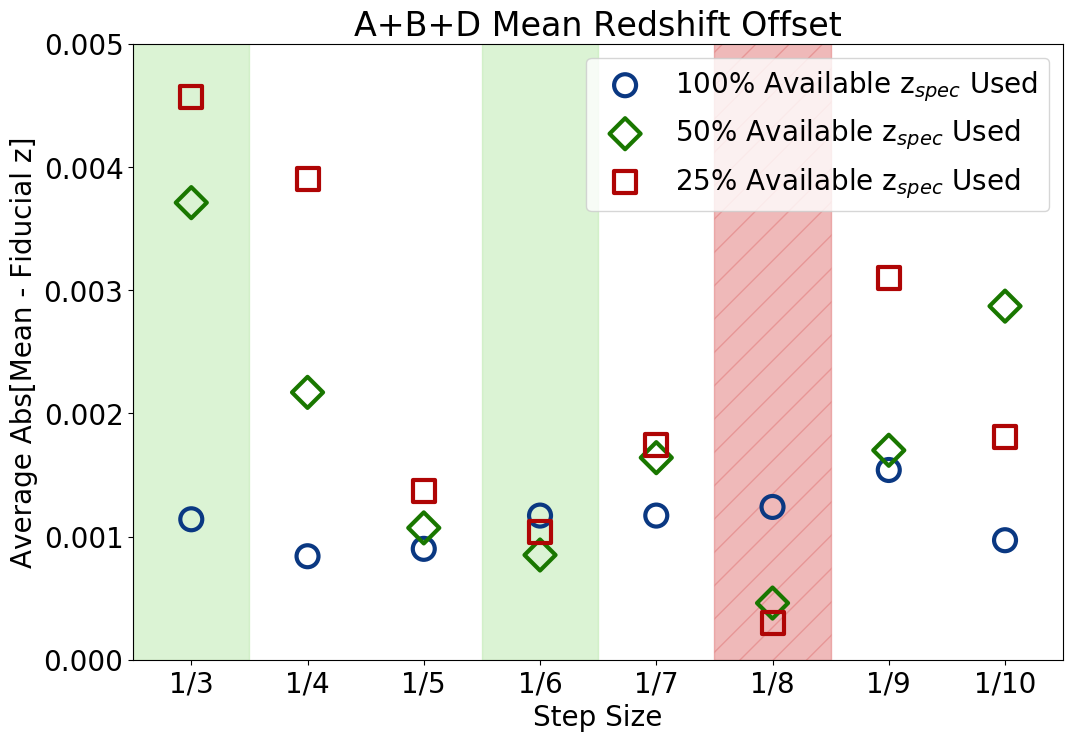}
\includegraphics[width=\columnwidth]{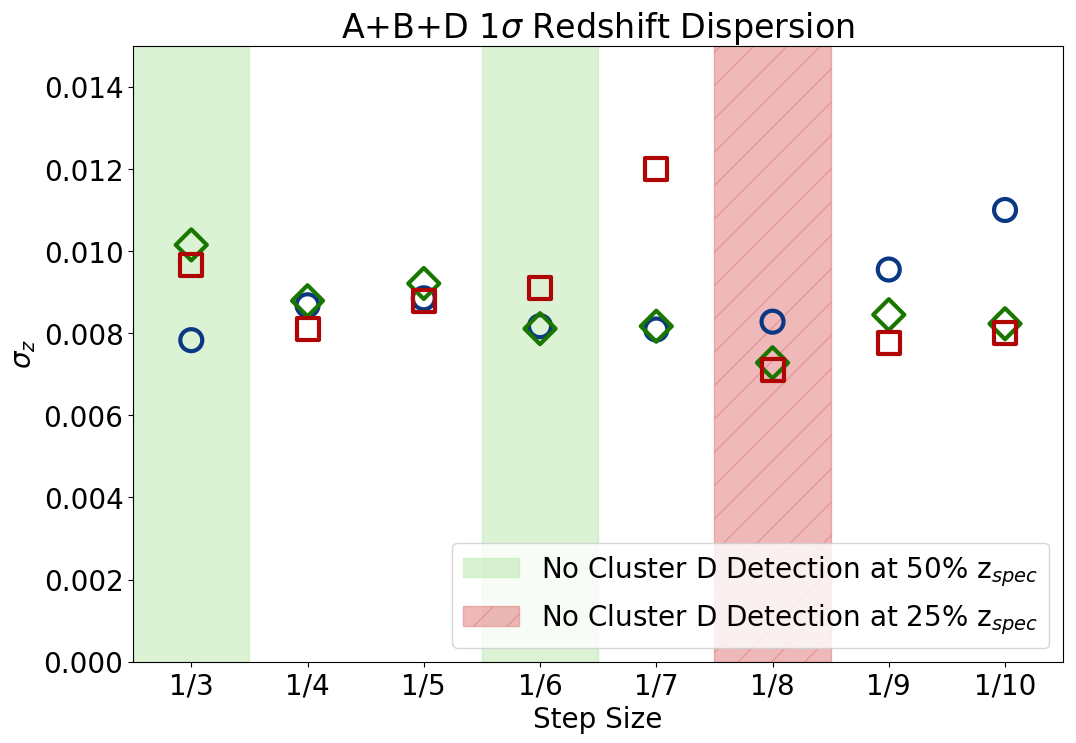}
\caption{Average absolute mean redshift offset and 1$\sigma$ redshift dispersions for Clusters A, B, and D using varying fractions of the available spectroscopic redshifts using a linking radius of 1 Mpc. The offset generally worsens with smaller fractions of spectroscopic data used, though even the maximum offset is still very small. The dispersions are consistently on the order of $\sigma_{z} \backsim$ 0.01, demonstrating how well the VMC method can recover the shape of the structures when only those objects with photometric redshifts that include significant detections in the $B$ and $V$ band even when a large percentage of the spectral redshifts are removed. The detection of Cluster D is lost at 1/3 and 1/6 step sizes for 50\% of the spectroscopic data used, and at 1/8 for 25\%. Though these tests operate at the mercy of small number statistics, having missed detections at all suggests that smaller step sizes are necessary to maximize overdensity candidate detections for less spectroscopically complete fields.}
\label{fig.fracsteps}
\end{figure}

When we dropped the fraction we used of our available spectroscopic data, we found that there was a greater chance of losing overdensity candidate detections with a step size larger than 1/10, as we missed the detection of Cluster D in our tests at 50\% (where the detection is lost at 1/3 and 1/6 step sizes) and 25\%  (where the detection is lost at 1/8 step size) of the spectroscopic data used. That we still successfully detect Cluster D with some smaller step sizes can be attributed to the small number statistics in these tests. However, having any missed detections at all implies that smaller step sizes are necessary for maximizing overdensity candidate detection completeness for less spectroscopically complete fields, especially for those of irregular shapes. The difference between the measured mean redshift and the fiducial redshift also generally increased with smaller fractions of spectroscopic data used, though even the largest difference we found was still very small. The redshift dispersions are consistently on the order of $\sigma_{z} \backsim$ 0.01 regardless of step size, demonstrating how well we can recover the shape of structures when we do detect them (Fig. \ref{fig.fracsteps}).

We thus elected to adopt a step size of 1/10, which means that adjacent redshift slices have 90\% overlap, for constructing the VMC overdensity map for the entire redshift range, so as to maximize our chances of successful overdensity candidate detections. To cover our entire redshift range of $z$ = 0.55 to 1.37 for our serendipitous overdensity candidate search, this translates to using 420 redshift slices for each field.

\subsubsection{Determining the Background RMS}

The background RMS value is critical for setting the detection threshold. When we pass our VMC overdensity maps to SExtractor, it estimates the background of the image and the RMS noise in that background. SExtractor computes the mean and standard deviation of the pixel value distribution in a 64 square pixel area. It then discards the most deviant and computes the mean and standard deviation again. This process is repeated until all the remaining pixel values are within 3 standard deviations of the mean. 

As mentioned in \S\ref{masks.linking}, we masked regions of the VMC overdensity maps with overdensities less than log(1+$\delta_{gal}$) = -0.35 in order to facilitate the accuracy of SExtractor's background RMS calculation. SExtractor excels in cases where the detections it finds are much smaller than the image containing them. This is not the case for many of our fields. We found that large-scale structures present in a field will often vary the field's background RMS as a function of redshift by more than 50\% higher than its mean value, clearly indicating that SExtractor had confused overdensity candidates for background. This effect is most prominent in RXJ1821, where a single large structure at $z$ = 0.8168 quadrupled the mean background RMS value.

There was additionally a persistent stochasticity in the background RMS as a function of redshift. The difference in the background RMS measured in neighboring redshift slices often exceeded the mean background RMS over the entire redshift range of the field. This variation in the background RMS is minimal for SC1604 and XLSS005, which are the targets with the largest imaging fields of view relative to the spectral footprint and the sizes of the structures, in contrast to smaller fields such as Cl1429 or RXJ1821. 

Due to the background RMS variations between fields, we would find structures of similar velocity dispersions with systematically smaller or larger isophotal flux peaks between fields of similar imaging depths, such as structures in SC1324 having much smaller peaks than SC1604. Because the detection threshold is set as a multiple of the background RMS, the higher background RMS in SC1324 compared to SC1604 meant that the former had an effectively higher threshold for the same relative DETECT\textunderscore THRESH value. For example, using a DETECT\textunderscore THRESH parameter of 4$\sigma$ would measure a smaller isophotal flux for an overdensity candidate of the same mass, velocity dispersion, and overdensity in SC1324 than it would in SC1604. This implied that the discrepancies we were finding arose due to how the data analysis was performed rather than an inherent characteristic of the structures in different fields.

\begin{figure*}
\centering
\includegraphics[width=2\columnwidth]{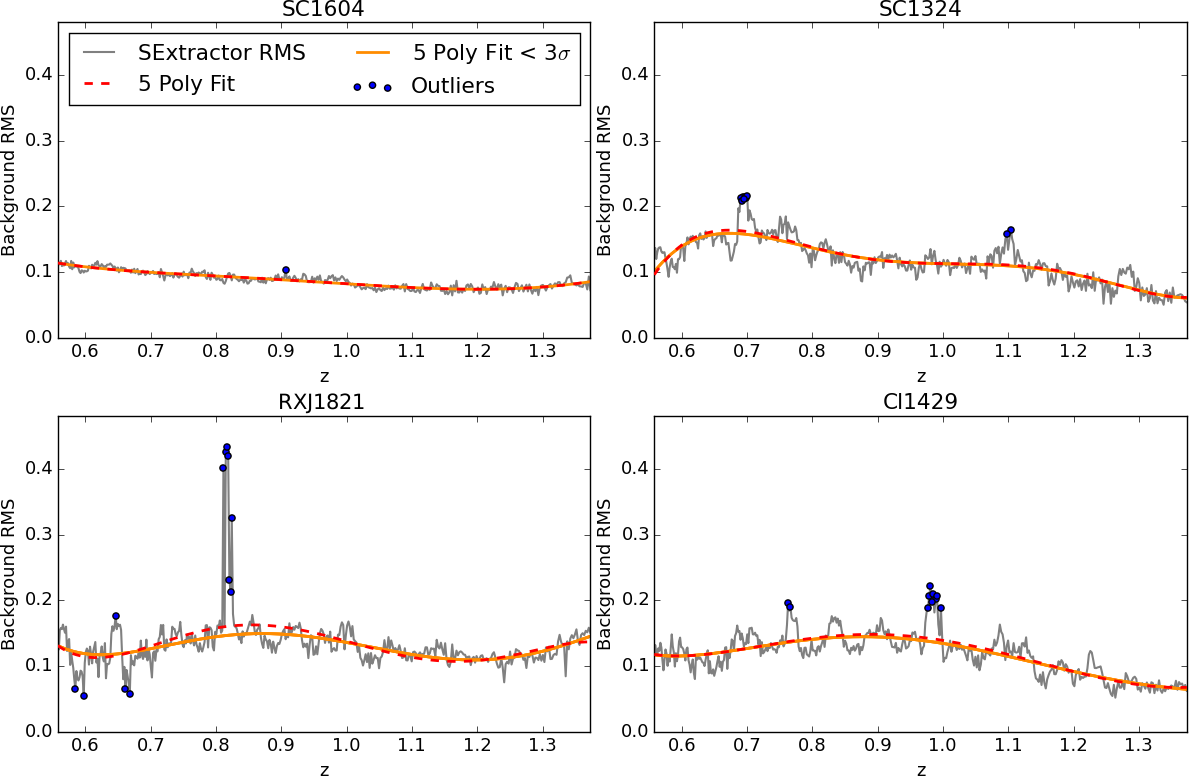}
\caption{The background RMS as a function of redshift in four fields, SC1604, SC1324, RXJ1821, and Cl1429. The gray lines are the background RMS as calculated by SExtractor. The dashed red line represents a fifth order polynomial fit without outlier rejection. The solid orange line represents a fifth order polynomial fit with iterative outlier rejection at above 3$\sigma$, based on the RMS of the preceding polynomial fit. The difference between the two fits with and without outlier rejection is largely minimal outside of areas with large peaks in the background RMS values. SC1604 is a very large field with deep imaging and shows a very flat RMS curve while Cl1429 sees a much more variation in its RMS as a function of redshift. SC1324 and RXJ1821 show noticeable peaks at redshifts of 0.7 and 0.8 respectively, which correspond to known large structures in each field.}
\label{fig.rms}
\end{figure*}

To allay the stochasticity of the background RMS between neighboring redshift slices, we fitted each field's RMS as a function of redshift with a fifth order polynomial. We used the RMS of that fit to identify outliers greater than 3$\sigma$ away from the value predicted by our polynomial fit and then fitted the background RMS again without the outliers. We performed this outlier rejection iteratively, repeating until the polynomial fit found no more outliers at above 3$\sigma$, which is similar to the methods adopted by \citet{Cucciati18}. In the majority of fields, the fit with outlier rejection was largely unchanged from the first fit, only noticeably deviating in fields with large peaks in the background RMS values (Fig. \ref{fig.rms}). We use these polynomial fits to set the RMS as a function of redshift in each ORELSE field in SExtractor.

\subsection{Recovering Known Structures}\label{recovering}

As a preliminary test of overdensity candidate completeness for different levels of spectroscopic coverage, we looked to see how well our choice of SExtractor parameters would recover 22 different known clusters and groups in ten ORELSE fields, which visually appeared to be isolated from other systems in the fields (Table \ref{tab.iso}). We refer to these hereafter as ``isolated structures.'' This additionally tests our sensitivity to low mass structures as some of these structures have virial masses as low as 3-4$\times$10$^{13}$ M$_{\odot}$, on par with that of the Local Group. Virial masses are computed based off the velocity dispersions $\sigma_{v}$ that we measured using the spectroscopic data, using the formula in \citet{Lemaux12}:

\begin{equation}
M_{vir} = \frac{3\sqrt{3}\sigma_{v}^3}{11.4 \mathrm{G} \mathrm{H(z)}}
\label{eq.vir}
\end{equation}

where G is the gravitational constant and $H(z)$ is the Hubble parameter. A structure's velocity dispersion is calculated using number of spectroscopic members in the structure, following the methods described in \citet{Rumbaugh13} and \citet{Ascaso14}. The velocity dispersion is calculated from the galaxies which make up the structure, and the galaxy membership is determined using an iterative process. Galaxies are initially identified as part of a structure if they fall within a 1 Mpc radius of the structure's density peak and then iteratively clipped for 3$\sigma$ outliers. Detections in the VMC overdensity maps were linked with known structures if their determined redshift was within $\Delta z < 0.02$ of their previously reported redshift and was centroided within 1 Mpc of their reported coordinates. We discarded all detections we found which had $\sigma_{z} > 0.05$, or a velocity dispersion of around $\pm$6000 km s$^{-1}$, as such systems would be unphysically large. 

\begin{table*}
\centering
\caption{Selected Isolated ORELSE Galaxy Clusters and Groups}
\label{tab.iso}
\begin{tabular}{lcccccc}
\hline
Structure & Redshift & RA (J2000) & Dec (J2000) & Members$^{a}$ & $\sigma_{v}$$^{b}$ & log(M$_{vir}$)$^{c}$ \\ 
\hline
SC1604 Lz & 0.5995 & 241.03282 & 43.2057 & 21 & 771.9 $\pm$ 110.0 & 14.711 $\pm$ 0.186 \\
SC1604 A & 0.8984 & 241.09311 & 43.0821 & 35 & 722.4 $\pm$ 134.5 & 14.551 $\pm$ 0.243 \\
SC1604 B & 0.8648 & 241.10796 & 43.2397 & 49 & 818.4 $\pm$ 74.2 & 14.722 $\pm$ 0.118 \\
SC1604 G & 0.9019 & 240.92745 & 43.403 & 18 & 539.3 $\pm$ 124.0 & 14.169 $\pm$ 0.300 \\
SC1604 H & 0.8528 & 240.89890 & 43.3669 & 10 & 287.0 $\pm$ 68.3 & 13.359 $\pm$ 0.310 \\
SC1604 Hz & 1.1815 & 241.07967 & 43.3215 & 15 & 661.5 $\pm$ 80.2 & 14.367 $\pm$ 0.158 \\
SC1324 A & 0.7566 & 201.20129 & 30.1924 & 43 & 873.4 $\pm$ 110.8 & 14.833 $\pm$ 0.165 \\
SC1324 H & 0.6990 & 201.2204 & 30.8408 & 19 & 346.4 $\pm$ 109.8 & 13.708 $\pm$ 0.393 \\
SC1324 I & 0.6956 & 201.2055 & 30.9665 & 35 & 847.1 $\pm$ 96.4 & 14.808 $\pm$ 0.148 \\
SC0849 A & 1.2637 & 132.23463 & 44.76178 & 13 & 714.4 $\pm$ 171.6 & 14.448 $\pm$ 0.313 \\
SC0849 D & 1.2703 & 132.14184 & 44.896338 & 23 & 697.2 $\pm$ 111.2 & 14.415 $\pm$ 0.208 \\
SC0849 E & 1.2601 & 132.27496 & 44.959253 & 14 & 445.1 $\pm$ 71.9 & 13.833 $\pm$ 0.210 \\
RCS0224 A & 0.7780 & 36.15714 & -0.0949 & 34 & 825.4 $\pm$ 193.2 & 14.754 $\pm$ 0.305 \\
RCS0224 B & 0.7781 & 36.14123 & -0.0394 & 52 & 710.7 $\pm$ 58.8 & 14.559 $\pm$ 0.108 \\
RXJ1221 B & 0.7000 & 185.34103 & 49.3138 & 18 & 426.6 $\pm$ 71.3 & 14.654 $\pm$ 0.222 \\
RXJ1053 & 1.1285 & 163.43097 & 57.591476 & 28 & 898.0 $\pm$ 142.0 & 14.778 $\pm$ 0.206 \\
RXJ1053 Hz & 1.2000 & 163.20387 & 57.58400 & 11 & 916.3 $\pm$  194.8 & 14.786 $\pm$ 0.277 \\
RXJ1821 & 0.8168 & 275.38451 & 68.465768 & 52 & 1119.6 $\pm$ 99.6 & 15.227 $\pm$ 0.218 \\
RXJ1757 & 0.6931 & 269.33196 & 66.525991 & 34 & 862.3 $\pm$ 107.9 & 14.832 $\pm$ 0.250 \\
Cl1137 & 0.9553 & 174.39786 & 30.008930 & 28 & 534.6 $\pm$ 81.1 & 14.144 $\pm$ 0.197 \\
RXJ1716 B & 0.8092 & 259.21686 & 67.139647 & 83 & 1120.6 $\pm$ 101.5 & 15.145 $\pm$ 0.118 \\
RXJ1716 C & 0.8146 & 259.25725 & 67.152497 & 39 & 678.4 $\pm$ 57.8 & 14.489 $\pm$ 0.111 \\
\hline
\end{tabular}
\begin{flushleft}
$a$: Number of galaxy members used for the velocity dispersion calculation within a 1 Mpc radius.

$b$: Velocity dispersion in $\kms$ within a 1 Mpc radius.

$c$: Virial mass in units of solar mass, calculated from the formula given in \citet{Lemaux12}.

Selected isolated known structures across 10 ORELSE fields used to test our detection threshold parameters. Though this table lists 22 structures, we effectively have 20 as we treat Clusters A and B in RCS0224 and B and C in RXJ1716 as single structures, with their redshifts, positions, and velocity dispersions member-weighted between each pair of clusters. The number of galaxy members, central positions, and velocity dispersions are calculated using the methods described in \citet{Rumbaugh13} and \citet{Ascaso14}.
\end{flushleft}
\end{table*}

As stated in \S\ref{optimize.1}, we tested a grid of parameters in SExtractor of DETECT\textunderscore THRESH of 3, 4, 5, and 6$\sigma$ and DETECT\textunderscore MINAREA of 10, 20, 40, 80, and 160 square pixels. We tested these parameters in conjunction with using 25\%, 50\%, 75\%, and 100\% of the available spectroscopic redshifts similar to what we did earlier in \S\ref{slices.opt}. For any percentage of $z_{spec}$ members, we looked for which pair of parameters would recover the highest percentage of our subset of known isolated structures. We found that generally the differences due to the choice of different minimum areas were small compared to those due to the choice of different detection thresholds, and similar minimum areas for a given detection threshold produced identical results.

\begin{figure*}
\centering
\includegraphics[width=2\columnwidth]{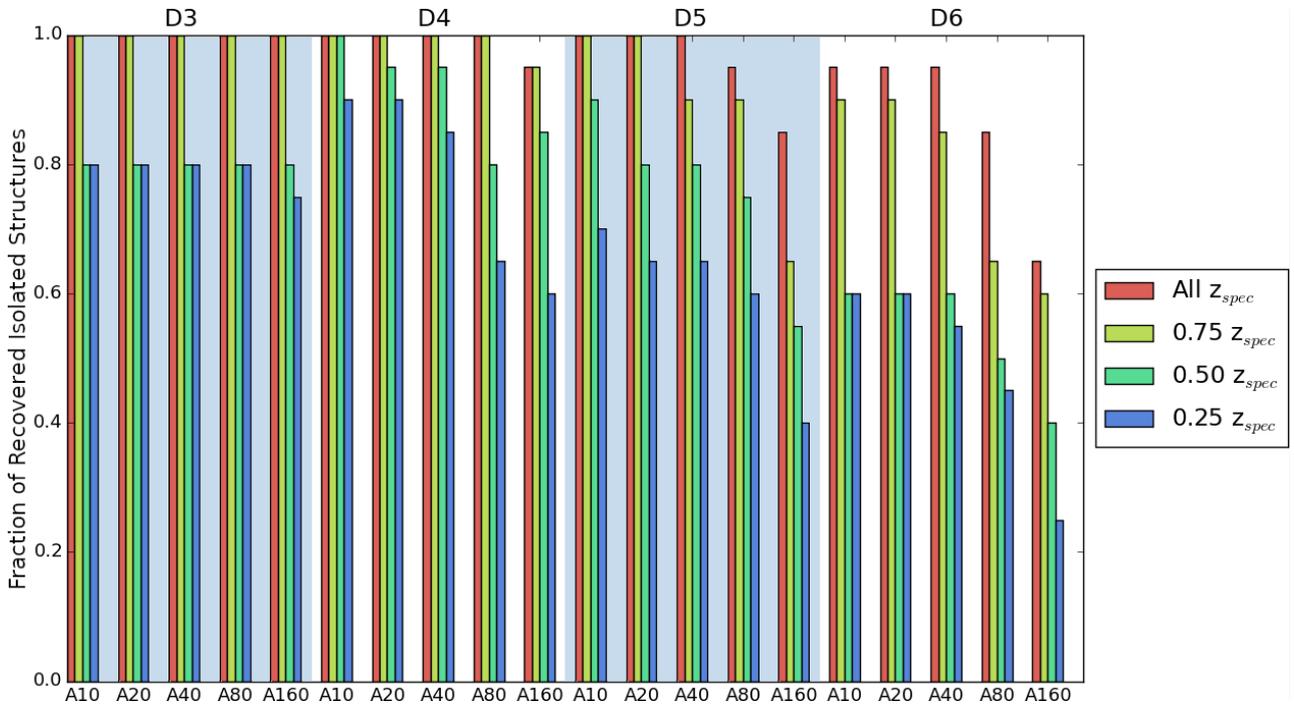}
\caption{The recovered fraction of known structures in Table \ref{tab.iso} based on different detection threshold (D, in units of $\sigma$, given on the top axis) and minimum area (A, in pixels, given on the bottom axis) parameters as well as fraction of available spectroscopic redshifts used in constructing the VMC maps. A known structures is considered successfully recovered if it lies within $\Delta z$ = 0.02 of the nominal redshift and 1 Mpc of the nominal central coordinates. We found that a 4$\sigma$ detection threshold was able to recover more structures at lower spectroscopic completeness.}
\label{fig.detectfrac}
\end{figure*}

Detection thresholds of 5 and 6$\sigma$ often were not able to find any of the known isolated structures nor produce any other overdensity candidates. This effect was more pronounced at smaller fractions of spectroscopic redshifts used. The 3$\sigma$ detection threshold also lost several detections of our lower mass structures at smaller spectroscopic fractions that were found by the 4$\sigma$ threshold, indicating the 3$\sigma$ threshold did not sufficiently filter out noise in the VMC maps. The 3$\sigma$ threshold additionally more poorly centroided the positions of the known structures it found compared with the 4$\sigma$ threshold. As the majority of these structures have high $z_{spec}$ fractions, we expect their fiducial positions to be highly accurate, so the centroid offsets are generally meaningful. From these results, we concluded it was best to use a 4$\sigma$ detection threshold (Fig. \ref{fig.detectfrac}).

\begin{figure*}
\centering
\includegraphics[width=2\columnwidth]{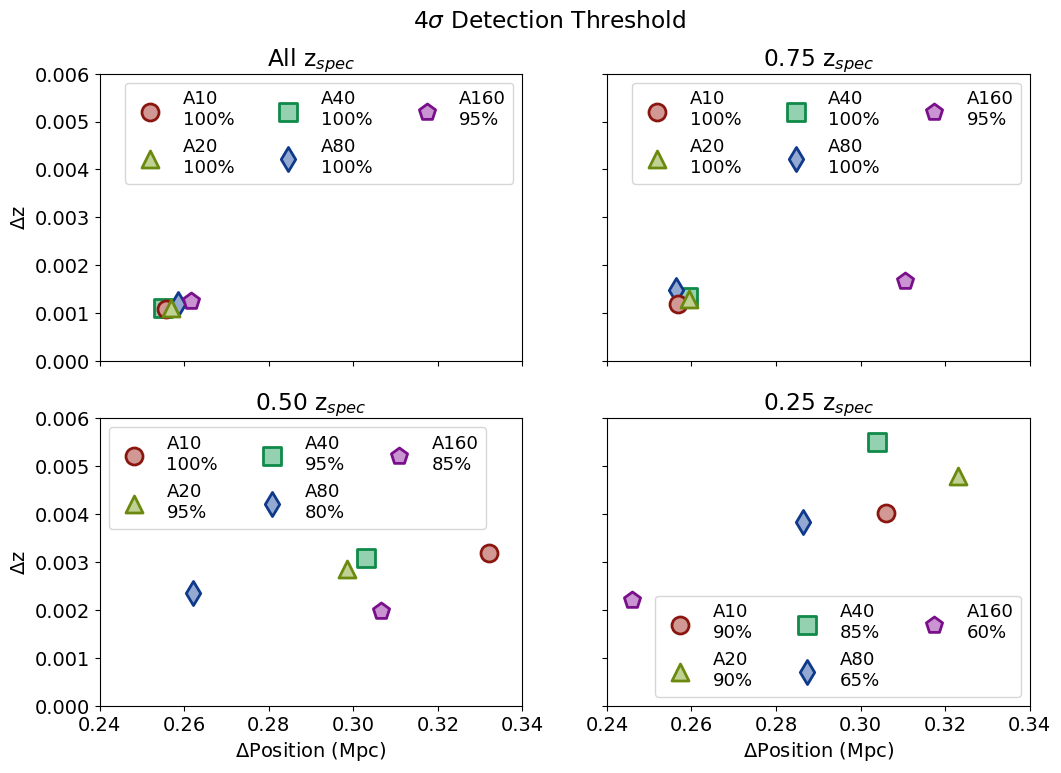}
\caption{The average absolute offset in redshift and position for all detected isolated structures using a 4$\sigma$ detection threshold with minimum areas (A) of 10, 20, 40, 80, and 160 square pixels. The annotated number next to each point represents the total fraction of the 22 known structures detected. In the top right panel, the 20 and 40 pixel points are overlapping. The 10 and 20 pixel areas recover the most known structures, reaching a minimum of 90\% at 25\% available $z_{spec}$ used. The difference in the redshift offset between 10 and 20 pixels is within z $<$ 0.001, but the 10 pixel area generally recovers more accurate structure positions while finding as many or more structures than the 20 pixel area.}
\label{fig.d4mean}
\end{figure*}

For a 4$\sigma$ detection threshold, minimum areas of 10 and 20 pixels were able to recover the most isolated structures for all fractions of spectroscopic redshifts, with a difference in redshift offset within z $<$ 0.001. The 10 pixel area had a better positional centroid in every case but at 50\% available $z_{spec}$ used, where the 10 pixel area has the advantage of recovering one structure that the 20 pixel area did not find (Fig. \ref{fig.d4mean}). The redshifts we determine for the known structures we recover agree very well with their fiducial values, with differences on the order of $\sigma <$ 0.001. This is an order of magnitude more precise than the $z_{phot}$ errors, which are typically on the order of $\sigma \backsim 0.02(1+z)$.

\subsubsection{Deblending Parameters}

We often find large overdensities that are in actuality multiple structures in close proximity. To split such overdensities into their component objects, SExtractor uses two deblending parameters, the number of deblending sub-thresholds DEBLEND\textunderscore NTHRESH and the minimum contrast DEBLEND\textunderscore MINCONT. SExtractor first defines a number of levels from the detection floor to the peak of the detection. This number is set by the DEBLEND\textunderscore NTHRESH parameter and the levels are spaced exponentially.  SExtractor builds a tree out of a detection, checking each level from the bottom up and branching every time it finds pixels above a threshold separated by pixels below it, similar to using cross-sections of a mountain to identify its peaks.

DEBLEND\textunderscore MINCONT is the fraction of the overdensities in a peak over the total overdensities in the entire structure. The smaller the minimum contrast, the smaller the peaks can be to be treated as single objects. Setting the deblending too coarse will lose out on detecting overdensity candidates, but too fine parameters will split individual overdensity candidates apart.

Once we settled on the detection parameters to use on the real data, we moved on to selecting the choice of deblending parameters. To do so, we qualitatively assessed how well we recovered structures close in proximity for five fields: Cl1350 at $z$ = 0.80, RCS0224 at $z$ = 0.78, RXJ1716 at $z$ = 0.81, SC1604 at $z$ = 0.90 and 0.93, and SG0023 at $z$ = 0.84. At these redshifts, there were the following numbers of known structures in each field: 3 in Cl1350, 2 in RCS0224, 3 in RXJ1716, 6 in SC1604, and 5 in SG0023.

We tested DEBLEND\textunderscore NTHRESH values of 16, 32, and 64, and DEBLEND\textunderscore MINCONT of 0.01, 0.001, 0.0001, and 0.00001. We found that a DEBLEND\textunderscore NTHRESH of 32 and MINCONT of 0.01 performed the best overall, missing only SG0023 A (a group with mass log(M$_{vir}$) = 13.836) and failing to deblend RXJ1716 B and C (clusters with centers 32 Mpc apart). Finer deblending parameters that could recover these clusters split the individual structures in other fields into multiple objects. Though our choice of deblending parameters cannot separate extremely close systems like RXJ1716 B and C, we are able to distinguish structures such as the components of the SC1604 supercluster and even closer systems like RCS0224 A and B. Our technique is able to find overdensity candidates, but for a more rigorous extraction of individual components, we encourage the reader to seek another more specialized technique, e.g., \citet{Golovich19}.

\section{Tests with Mock Catalogs}\label{mocks}

There are limitations to what we can test with the real data. Even with the availability of highly precise spectroscopic redshifts, projection effects can still complicate overdensity detection \citep{Lucey80}. Many of our observational tests lack a definite truth to compare with, which is an advantage using mock data can provide. In order to assess the purity and completeness of the new overdensity candidates we found using our detection algorithm, we tested how well it performed on mock structures across a range of numbers of members and velocity dispersions and at varying levels of spectroscopic coverage.

\subsection{Mock Galaxy Generation}

We generate the mocks by first populating a given volume with a population of field galaxies and then injecting galaxy clusters and groups. To simplify the distribution of the field galaxies in our mocks, we drew them from ORELSE fields where we did not find any structure candidates at the redshift of interest and also had deep enough imaging in the relevant bands such that it included a complete sample of galaxies to the magnitude limit given below (see \citealt{Tomczak17} for more details on the completeness limits of our imaging).

We drew the field galaxies for our mocks from the galaxies in SC0849 for $z$ = 0.8 and in RXJ1716 for $z$ = 1.2, using their $z_{phot}$ values to set the line-of-sight dimension and their positions to set the transverse dimension. We chose these two fields as they did not have any overdensity detections in our earlier findings, and being pseudo-realistic, they include an inherently more accurate distribution of galaxies that follows the two-point correlation function. The galaxies we included in these fields were within $\Delta z_{phot} \leq$ 0.025 of the target redshift. When selecting what field galaxies to use, we limited the magnitude range to between 18 and 24.5, using the Subaru $i$-band for SC0849 and the LFC $z$-band for RXJ1716. The field galaxies we use at each redshift cover similarly sized areas of 0.168 and 0.174 square degrees respectively, which are typical of the sizes of the fields in the ORELSE data.

We additionally attempted an alternative arrangement of the field galaxies to see if it would meaningfully change our results. For this method, we use a random distribution to populate the mocks with the same number of field galaxies over the same transverse area as in the pseudo-realistic fields. The line-of-sight dimension was covered with a random uniform distribution within $\Delta z_{phot} < 0.025$ of the central redshift of the mock. We compare this random distribution of field galaxies to the pseudo-realistic distribution drawn from SC0849 and RXJ1716. We found no significant difference between the random and pseudo-realistic fields but elected to use the latter in the mocks due to the more representative galaxy distribution.

\subsubsection{Field Galaxy Generation}

We generate the magnitude distributions of the field galaxies in our mock catalogs according to what is predicted by the Schechter function, using the rest-frame $M_{1700}$ luminosity function parameters from \citet{Hathi10} for both z $<$ 1 and z$>$1, which are modulated based on the average $M_B - M_{FUV}$ colors of galaxies at this redshift. The resulting $M_{*}$ values are consistent with the rest-frame $B$-band luminosity function parameters of \citet{Giallongo05}, and thus our results would be unchanged were we to adopt their parameters. We set a floor for the Schechter function such that we do not sample at luminosities $<0.1L_{*}$.

We allow samples from $0.1L_{*}$ to $10L_{*}$ at redshifts $z \sim 0.8$ and 1.2, as the rest-frame $B$-band is approximately the observed-frame $i$-band at $z \sim 0.8$ and very close to the $z$-band at $z \sim 1.2$. This matches the magnitude range 18 $< i <$ 24.5 we cover with the galaxies in our real data. The redshift range is limited to $\Delta z$ = 0.05 around each target redshift, and this small range is to limit the effect of k-corrections of the galaxies when transforming the $B$-band rest-frame luminosities to the observed-frame $i$-band apparent magnitudes at $z \sim 0.8$ and $z$-band at $z \sim 1.2$. For the mocks at $z\sim1.2$, we modify the $M_{*}$ parameter to be 1.35 magnitudes brighter than the \citet{Hathi10} value in order to transform it to the observed $i$-band. This value is supported by the average colors at this redshift range, where the average $M_B - M_{FUV}$ color is 1.1 from ORELSE photometric catalogs and the average $M_{FUV} - M_{NUV}$ color is 0.25 as determined by fits to the COSMOS $z_{spec}$ catalogs for galaxies in this redshift range \citep{Lemaux19}.

\subsubsection{Mock Structure Makeup}\label{mockmakeup}

We inject groups and clusters by drawing galaxies from a Gaussian distribution with a $\sigma$ equal to a  velocity dispersion chosen randomly to fit in the range of velocity dispersions we see in known structures. We use the same distribution of field galaxies in each mock for each redshift, and we inject different arrangements of mock groups and clusters over the field galaxies. We inject the mock groups and clusters at the central redshift for each of our two fields, with their centers forced to be within the central 50\% region of the mock field. We impose this constraint so that we can mask the outer 20\% region of the field when running our detection algorithm. This is to avoid picking up high overdensities due to edge effects from our field galaxy population while avoiding masking out mock cluster and group galaxies. This is effectively already done in the real data because each field is targeted such that the structures are in the center of the imaging footprints. We construct a corresponding VMC overdensity map for each arrangement of mock groups and clusters, and we then use the same detection and identification techniques as in \S\ref{recovering}.

\subsubsection{Using Real Data to Set Structure Membership}

We would like the mock groups and clusters to have similar numbers of members as our known groups and clusters had in all of the ORELSE fields. In an attempt to constrain the number of spectroscopic and photometric members in the ORELSE groups and clusters, we begin by estimating from the data the true number of members within a virial radius, $R_{vir}$, for each known cluster and group. $R_{vir}$ is defined as:

\begin{equation}
R_{vir} = \frac{\sqrt{3} \sigma_{v}}{11.4H(z)}
\end{equation}

where z is the systemic redshift of the cluster, $\sigma$ is the line-of-sight galaxy velocity dispersion for all galaxies within a projected radius of 1 Mpc of the luminosity-weighted spectral member center, and H(z) is the Hubble parameter. See \citet{Lemaux12} and references therein for details on this definition of $R_{vir}$, the measurement of $\sigma$, and the measurement of luminosity-weighted spectral member centers for ORELSE groups and clusters. Though other definitions of $R_{vir}$ are likely more well-motivated from theory (see, e.g., discussion in \S4.1 \citealt{Cucciati18} where $R_{vir}$ is defined to be $\sim$20\% larger), we adopt this value of $R_{vir}$ for consistency with previous ORELSE studies. In practice, since we do an aperture correction later in this section, and because we do not use $R_{vir}$ elsewhere, our results are unchanged if we instead adopt another definition of $R_{vir}$. 

The initial pool of possible $z_{spec}$ members have redshifts corresponding to peculiar velocities which are at most three times the velocity dispersion of the parent cluster or group, and their projected distances are within the virial radius. For every object in the magnitude range without a secure spectral redshift, we assign the $z_{phot}$ as measured by prior \texttt{EAZY} fitting. Objects within $R_{proj} \le R_{vir}$ and which have a $z_{phot}$ in the range $z_{min}-\sigma_{\Delta z/1+z}(1+z) < z < z_{max}+\sigma_{\Delta z/1+z}(1+z)$, where $z_{min}$ and $z_{max}$ are the minimum and maximum redshift bounds for spectral membership set by the criterion above, are considered $z_{phot}$ members.

The number of galaxies counted above still may contain contamination from foreground and background galaxies. We thus need to estimate the number of these interlopers and remove them. For every cluster and group for which we performed this estimate for, we chose an area of the imaging which did not to the best of our knowledge contain any large-scale structure or considerable photometric masking. The estimate of the number of contaminating objects, hereafter called $z_{phot,background}$, was performed by measuring the the number of objects within the same photometric redshift and projected spatial range as $z_{phot}$ members at a location on the sky where no cluster or group was detected. In order to futher mitigate any chance at contamination of $z_{phot,background}$ by large-scale structure features surrounding known clusters and groups, the number of objects was estimated at a redshift slightly higher ($\Delta z = 0.03$) than the systemic redshift of the cluster or group being measured. The number $z_{phot,background}$ objects for each cluster/group was estimated from estimates in the corresponding field in which it was observed to compensate for field-to-field variance in $z_{phot}$ accuracy/precision. 

We then apply the magnitude cut to both of these pools in the relevant band for the particular field, limiting the galaxy samples to objects brighter than 24.5. The total members where the projected radius is smaller than $R_{vir}$ are then the members with the background objects subtracted out, i.e., $N_{mem,R_{vir}} = z_{spec,members} + z_{phot,members} - z_{phot,background}$. 

Finally, to approximate the true number of members, $N$, for each group/cluster, the above numbers are aperture corrected in an attempt to include those real members that lie at $R>R_{vir}$. This aperture correction is estimated by multiplying the number of members calculated above by the average ratio of $z_{spec}$ members at $R_{vir}$ to those at 1.5$R_{vir}$ for all of the ORELSE clusters presented in \citet{Rumbaugh18}, where the definition of $z_{spec}$ members is the same as that stated earlier in the section. This ratio is computed to be 1/0.68. While an aperture correction to a projected radius of 1.5$R_{vir}$ is somewhat arbitrary, the number of interlopers within $\pm3\sigma$ increases severely at $R_{proj} > 1.5R_{vir}$ \citep{Wojtak07,Saro13}. Since we have no way to determine which galaxies are interlopers in our actual data, we limit our aperture correction to this radius. In practice, our results change very little if we instead apply an aperture correction to a larger radius, e.g., a correction to 2$R_{vir}$ results in a 17\% increase in the number of members, which would only serve to increase the completeness of the mock groups and clusters.

\begin{figure}
\includegraphics[width=\columnwidth]{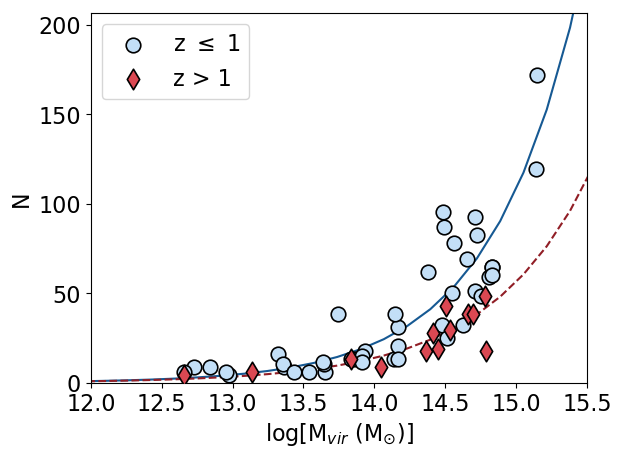}
\caption{Exponential function fits relating the known structures' virial masses, $M_{vir}$, to their magnitude cut member numbers $N$. The solid blue line is Equation \ref{eq.nmem1}, fitting the $z \leq 1$ structures denoted in blue, and the dashed red line is Equation \ref{eq.nmem2}, fitting the $z>1$ structures denoted in red.}
\label{fig.nmem}
\end{figure}

In order to populate the number of members in each mock structure, we require an analytic expression that provides the number of members of given structure at all overdensity masses simulated in the mocks. To that end, we perform a non-linear least squares fit of an exponential function that relates the virial masses of the known structures, $M_{vir}$, to the final aperture-corrected estimate of the true number of members brighter than the adopted magnitude cut calculated above. This function is broken up into two domains, one for $z \leq 1$ and one for $z>1$, such that:

\begin{equation}
N_{z \leq 1} = (4.25 \pm 0.10) \times 10^{-9} e^{(1.5972 \pm 0.0017) \textup{log}(M_{vir})}
\label{eq.nmem1}
\end{equation}
\begin{equation}
N_{z>1} = (4.01 \pm 0.14) \times 10^{-8} e^{(1.4041 \pm 0.0025) \textup{log}(M_{vir})}
\label{eq.nmem2}
\end{equation}

where $M_{vir}$ is in units of solar mass (Fig. \ref{fig.nmem}). Errors on the fit parameters are determined by the covariance matrix, though for the remainder of this exercise, we ignored their effect as they are negligibly small.

This function gives the number of members brighter than the $m_{AB}<24.5$ magnitude cutoff in each redshift domain that we simulate. When we generate our mock groups and clusters, all of the mock member galaxies have magnitudes that are brighter than the magnitude cutoff at $z$ = 0.8. However, this is not the case at $z$ = 1.2, where many of the galaxies that are generated are fainter than this limit. Thus, in order to match the number of mock members at $m_{AB}<24.5$ with the number of true members at $m_{AB}<24.5$ as estimated by Equation \ref{eq.nmem2}, we need to inject a larger number of members into the mocks to account for the eventual loss of fainter members. To determine the number of members that we must inject into our mock catalogs at a given structure mass at $z>1$, we first take the number of members predicted by Equation \ref{eq.nmem2} and then correct that number by dividing by the fraction of members in $z>1$ mock clusters and groups that are brighter than the magnitude limit cutoff. This value is calculated to be 0.743 and 0.583 for cluster and groups, respectively, on average, meaning that, in the mocks at $z>1$, the $N_{mem}$ recoverd by Equation \ref{eq.nmem2} must be multiplied by 1/0.743 and 1/0.583 for cluster and group members to generate the number mock members brighter than the magnitude cutoff that match the observations. 

\begin{figure}
\includegraphics[width=\columnwidth]{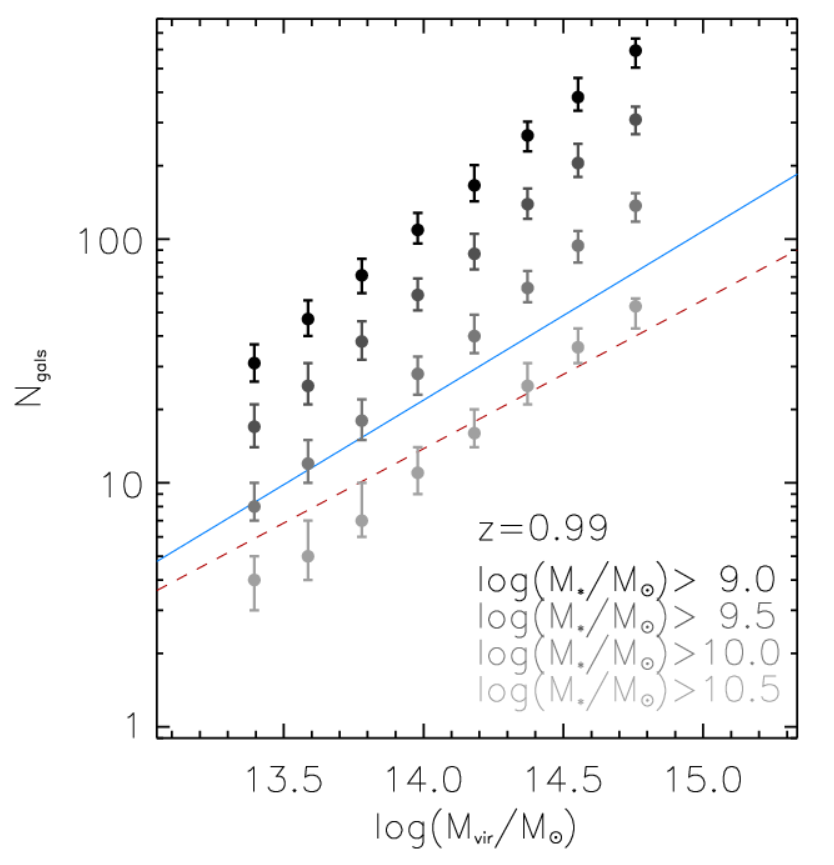}
\caption{We selected all friend-of-friend clusters and groups with masses above $\log(M_{tot}/M_{\odot}) > 13.3$ in the Millennium Run dark matter simulation embedded with the semianalytical model of \citet{delucia07} at $z$ = 0.988. Plotted is the median number of galaxies in these friend-of-friend clusters, in bins of 0.2 dex in total friend-of-friend mass. Each of the four series of points contains galaxies with stellar mass $SM$ larger than the given threshold in the labels. Error bars on the median values represent the 25th and 75th percentile of the counts distribution in each bin. The interquartile range encompasses a relatively small range in $N_{gals}$ for each given halo mass and stellar mass bin, with the largest variations reaching $\sim$20\%. The numbers of galaxies for the larger stellar mass bins are also broadly consistent with the numbers presented in Fig. \ref{fig.nmem} and estimated by Equations \ref{eq.nmem1} and \ref{eq.nmem2}.}
\label{fig.millennium}
\end{figure}

To further support our member galaxy numbers, we looked to the Millennium Run dark matter simulation embedded with the semianalytical model of \citet{delucia07} in order to try to account for the presence of poorly occupied halos that may be missing and were not included in our mocks. We took a snapshot at $z$ = 0.988 to compare with our $z\sim$1 population and selected all friend-of-friend clusters and groups with masses above $\log(M_{tot}/M_{\odot}) > 13.3$, where the friend-of-friend clusters and groups were taken from the simulation output. Fig. \ref{fig.millennium} shows the median number of galaxies in these friend-of-friend clusters in each 0.2 dex halo mass bin. For a given halo mass and a given tracer population (i.e., the stellar mass bin), we see that the interquartile range encompasses a relatively small range in $N_{gals}$, with the largest variations reaching $\sim$20\%. In addition, the numbers of galaxies appear to be broadly consistent with the numbers presented in Fig. \ref{fig.nmem} and estimated by Equations \ref{eq.nmem1} and \ref{eq.nmem2} for stellar masses larger than $\log(M_{tot}/M_{\odot}) > 10.0$ to $10.5$, which is represenative of our $z\sim$1 population. This shows that poorly occupied halos do not significantly impact the numbers of member galaxies estimated, which lends credulity to the purity and completeness estimates we will find with the mocks later on.

\subsubsection{Injecting Structures in the Mock Catalogs}

We create our mocks such that we have a total of 300 clusters and 300 groups at each redshift. We defined our mock clusters to have velocity dispersions between 580 and 1100 km s$^{-1}$, the upper bound matching the largest known structures in the ORELSE fields. Mock groups were defined as having velocity dispersions of 300 and 500 km s$^{-1}$. For each new mock generated, we used uniform random sampling of the velocity dispersions for all injected mock groups and clusters. Virial masses of each mock cluster or group are calculated directly from the imposed velocity dispersion and systemic redshift. Ignoring the uncertainty values in the fit of $M_{vir}$ versus $N_{mem}$, this corresponded to 35 to 131 members for clusters and between 9 and 25 members for groups at $z$ = 0.8. We then sample the appropriate Schechter function for the member galaxies to generate the magnitude distributions of each structure.

To test how well our detection algorithm performed when varying the velocity dispersion of the injected structures, we arranged the groups and clusters into two bins, cleanly separating their velocity dispersion ranges into lower and upper halves and effectively forming two mass bins. We injected our groups and clusters using two different arrangements for both the lower and upper mass bins. One arrangement is a single cluster and three groups and the other is two clusters alone. We stipulated our injected clusters to have their centers at least 4 Mpc apart, or 2 Mpc in the case of groups. This is done to avoid any spatial overlap between the injected structures. Such chance alignments of multiple structures in real data would lead to a higher probability of detecting a structure at their location, so our mocks cover the scenario in which detection is most difficult, meaning our purity and completeness numbers are necessarily lower limits.

For every mock at the same redshift, we used the exact same field galaxy distribution. We ran 50 mocks for each mass bin, making for a total of 150 groups and clusters for each mass bin at each of our tested redshifts. To test how well our detection algorithm performed at varying levels of spectroscopic coverage, we ran each mock through the Voronoi tessellation Monte-Carlo (VMC) four times, each at a different spectroscopic coverage level (5\%, 20\%, 50\%, and 80\%). Each time, we randomly assign that given fraction of all the galaxies in the mock spectral redshifts. We treat the remaining galaxies as only having $z_{phot}$ values. The catastrophic outlier rate is set to 6\% to match the average seen in the ORELSE photometric catalogs. The photometric redshifts $pz$ are modified by the average $z_{phot}$ error $pz_{err}$ (set to 3\% to match the average of the SC0849 and RXJ1716 fields) such that:

\begin{equation}
pz = pz + R pz_{err} (1+pz)
\end{equation}

where $R$ is a normalized Gaussian. This modification operates under the assumption that there is no change in $z_{phot}$ accuracy nor the catastrophic outlier rate as a function of magnitude to the depth of our mock data. $z_{phot}$ values that end up as catastrophic outliers are randomly set to redshifts of 0.31 and below. This reflects what we see in the real data, where the vast majority of $z_{phot}$ outliers are higher redshift galaxies that scatter to lower redshifts. After making a magnitude cut only keeping the galaxies brighter than 24.5, the VMC map is then made in the same manner as it was for the real ORELSE fields. 

\subsection{Assessing Purity and Completeness with the Mock Catalogs}\label{mock.compur}

We looked for detections in the mock VMC overdensity maps with the same parameters as in \S\ref{recovering}. In other words, we use DETECT\textunderscore THRESH = 4$\sigma$, DETECT\textunderscore MINAREA = 20 corresponding to a $\sim$0.1 Mpc$^{2}$ area, no smoothing filter, DEBLEND\textunderscore NTHRESH = 32, DEBLEND\textunderscore MINCONT = 0.01. As with the search of the real data, we discard detections in the mocks with $\sigma_{z} > 0.05$ for being unphysically large, and we exclude detections where the Gaussian peak was more than 20\% higher than the maximum flux of the fitted points to cut out detections with higher likelihoods of inaccurate total masses and redshifts. We also exclude the most poorly constrained detections by removing cases where the uncertainty on the integrated isophotal flux was larger than the integrated isophotal flux. This removal largely does not affect what we find for our purity and completeness values other than a small improvement to purity at high spectroscopic fractions. Recovered structures were identified by looking within a linking radius of 1 Mpc and a redshift window of $\Delta z$ = 0.02 for each injected structure's position and redshift, which are the identical values used for the search on the real data. 

We test whether we find the same optimal SExtractor parameters with the mocks as we did earlier with our tests on the real data. This serves as a self-consistency check and demonstrates whether the mocks are representative of the real data or not. We run SExtractor on the VMC overdensity maps of the mocks for 14 total pairs of parameters, varying the DETECT\textunderscore THRESH between 4 and 5$\sigma$ and DETECT\textunderscore MINAREA between 20, 30, 40, 50, 60, 70, and 80 square pixels. The absolute value of the DETECT\textunderscore THRESH parameter is based on the background RMS averaged from the polynomial fits of the background RMS of all the fields in the real data. The final absolute DETECT\textunderscore THRESH value is the resulting RMS value at the particular redshift multiplied by the DETECT\textunderscore THRESH $\sigma$ used. We do not vary the choice of deblending parameters, as we injected structures such that they were separated by at least 2 Mpc, so any variation in the deblending is unlikely to change our results. The detection algorithm used on these mocks hereafter is the same as the procedure we used on the real data. 

To assess the total purity and completeness values, we calculated the product of the purity and completeness numbers for all SExtractor parameter pairs across all spectroscopic coverage levels at both of our tested redshifts. The completeness is defined as only structures we recovered divided by the total number of structures we injected into the mocks. The purity is the fraction of all of our detections that we were able to match up to our injected structures. We calculate the mean purity and completeness for all mocks at each spectroscopic coverage level. We statistically quantify the uncertainties in the purity and completeness using a bootstrap method. We subsample each level of spectroscopic coverage by randomly taking 20 mocks and measuring their purity and completeness, repeating this process 1000 times. We then obtain a median purity and completeness value and their 1$\sigma$ uncertainties for a given level of spectroscopic coverage.

\begin{figure*}
\centering
\includegraphics[width=1.8\columnwidth]{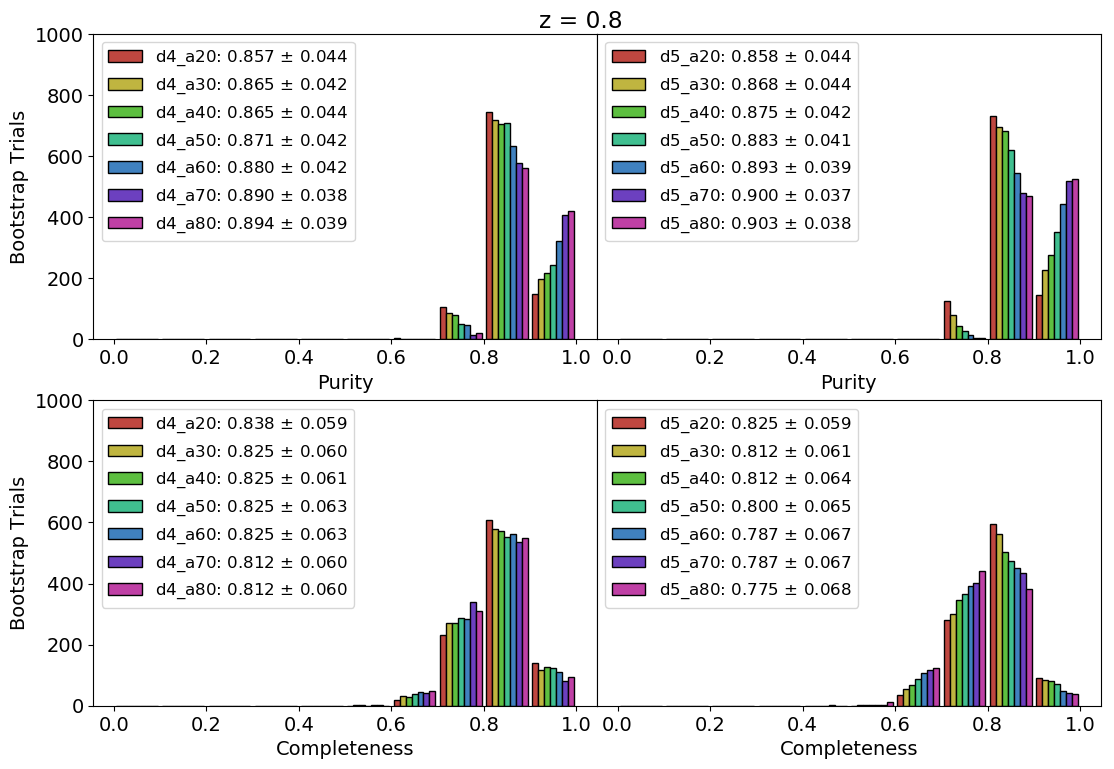}
\includegraphics[width=1.8\columnwidth]{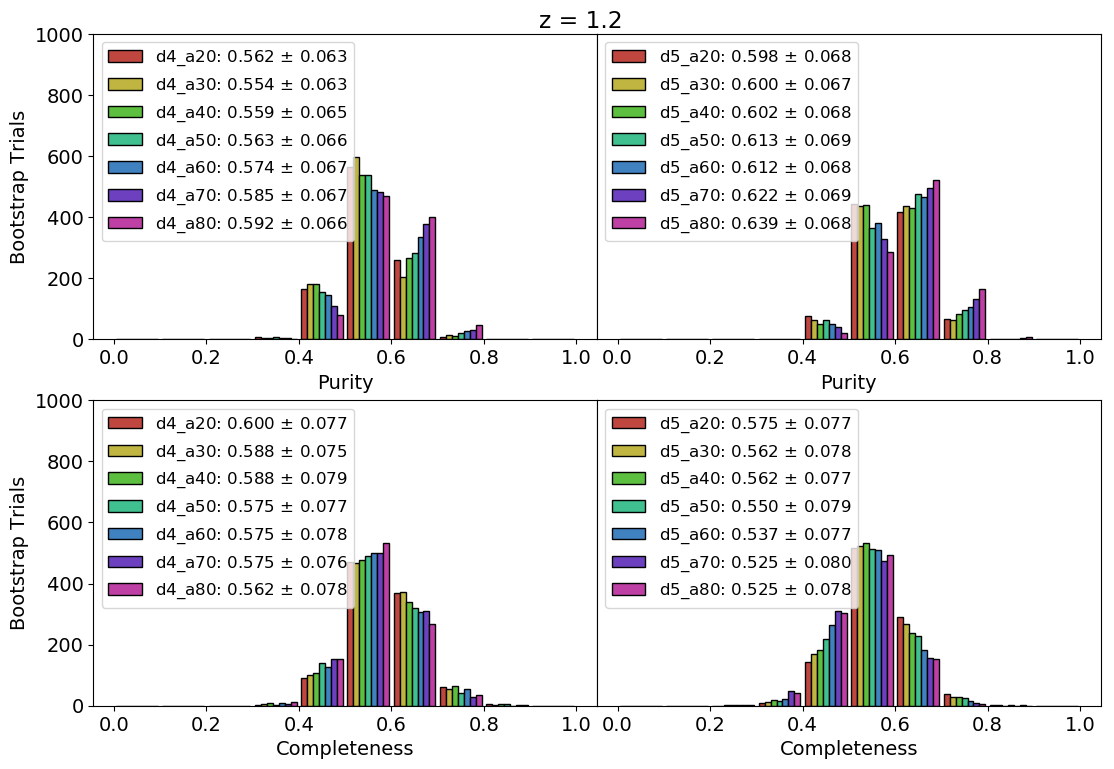}
\caption{Depicted in this figure are the purity and completeness across all mocks over all spectroscopic coverage levels. To assess the uncertainties in the purity and completeness, we used a bootstrap method where we calculated the purity and completeness for a random 20 mocks which is then repeated 1000 times. We combined our purity and completeness bootstrap routines to compute the purity and completeness values for each SExtractor parameter pair of DETECT\textunderscore THRESH of 4 and 5$\sigma$ and DETECT\textunderscore MINAREA of 20, 30, 40, 50, 60, 70, and 80 square pixels for $z=0.8$ and $z=1.2$. The DETECT\textunderscore MINAREA parameters are denoted by the color of each bar in the histogram, with the 4 and 5$\sigma$ DETECT\textunderscore THRESH parameters spilt into the left and right panels. The mean purity and completeness and their 1$\sigma$ uncertainties for each pair is given in the legends. The uncertainties in the purity and completeness values show little dependence on the choice of SExtractor parameters.} 
\label{fig.compur}
\end{figure*}

We found that there was no overall best performing set of SExtractor parameters, with each pair showing similar performance in both purity and completeness for each redshift (Fig. \ref{fig.compur}, Table \ref{tab.compur}). SExtractor was able to detect the regularly shaped structures in the mock catalogs regardless of how sensitive the detection thresholds we used were. We thus choose to use the 4$\sigma$ DETECT\textunderscore THRESH and DETECT\textunderscore MINAREA of 20 to maximize our chances of detecting an overdensity candidate, which are the same parameters we found to work best in our tests in \S\ref{recovering}. We additionally tested how our purity and completeness numbers were in the 0\% spectroscopic fraction case, finding them to be very low, with completeness under $\sim$10\% and purity at most around 50\% for both redshifts. Our purity and completeness markedly improve for even 5\% spectroscopic coverage, where our purity and completeness are, respectively, 96\% and 57\% for $z$ = 0.8, and 70\% and 29\% for $z$ = 1.2 for our choice of SExtractor parameters (Table \ref{tab.compur}). We thus choose to limit our search range to areas with spectroscopic coverage of at least 5\% when applying our algorithm to the real data.

We also assessed how the product of the purity and completeness vary when using different-sized Gaussian filters in SExtractor, as well when using no filtering at all. We found there was at most a 5 to 10\% improvement by applying some measure of filtering for our detection parameters of choice. We expect the best performance from filtering comes from recovering regularly shaped structures as we have injected into the mocks. This is not necessarily true for structures in the real data however. Since the changes were minor, we elected to use no filtering when running SExtractor on our real data so as to increase the chances of detecting all structures in our fields and to avoid biasing ourselves against detecting irregularly shaped structures.

\begin{table*}
\centering
\caption{Purity and Completeness for Different SExtractor Parameters}
\label{tab.compur}
\begin{tabular}{ccc|cccc}
\hline
Redshift & DETECT\textunderscore THRESH & DETECT\textunderscore MINAREA & 5\% $z_{spec}$ & 20\% $z_{spec}$ & 50\% $z_{spec}$ & 80\% $z_{spec}$ \\ 
\hline
\multirow{2}{*}{0.8} & \multirow{2}{*}{4} & \multirow{2}{*}{20} & P/C = 0.959/0.573 & P/C = 0.926/0.826 & P/C = 0.797/0.940 & P/C = 0.755/0.973 \\
& & & P$\cdot$C = 0.550 & P$\cdot$C = 0.765 & P$\cdot$C = 0.750 & P$\cdot$C = 0.735 \\
\hline
\multirow{2}{*}{0.8} & \multirow{2}{*}{4} & \multirow{2}{*}{30} & P/C = 0.961/0.568 & P/C = 0.930/0.823 & P/C = 0.800/0.938 & P/C = 0.764/0.973 \\
& & & P$\cdot$C = 0.546 & P$\cdot$C = 0.765 & P$\cdot$C = 0.750 & P$\cdot$C = 0.743 \\
\hline
\multirow{2}{*}{0.8} & \multirow{2}{*}{4} & \multirow{2}{*}{40} & P/C = 0.965/0.569 & P/C = 0.933/0.818 & P/C = 0.803/0.938 & P/C = 0.771/0.971 \\
& & & P$\cdot$C = 0.549 & P$\cdot$C = 0.763 & P$\cdot$C = 0.753 & P$\cdot$C = 0.749 \\
\hline
\multirow{2}{*}{0.8} & \multirow{2}{*}{4} & \multirow{2}{*}{50} & P/C = 0.968/0.562 & P/C = 0.935/0.815 & P/C = 0.811/0.935 & P/C = 0.782/0.970 \\
& & & P$\cdot$C = 0.544 & P$\cdot$C = 0.762 & P$\cdot$C = 0.758 & P$\cdot$C = 0.759 \\
\hline
\multirow{2}{*}{0.8} & \multirow{2}{*}{4} & \multirow{2}{*}{60} & P/C = 0.968/0.566 & P/C = 0.935/0.809 & P/C = 0.821/0.932 & P/C = 0.798/0.968 \\
& & & P$\cdot$C = 0.548 & P$\cdot$C = 0.756 & P$\cdot$C = 0.765 & P$\cdot$C = 0.772 \\
\hline
\multirow{2}{*}{0.8} & \multirow{2}{*}{4} & \multirow{2}{*}{70} & P/C = 0.970/0.565 & P/C = 0.939/0.805 & P/C = 0.840/0.924 & P/C = 0.807/0.961 \\
& & & P$\cdot$C = 0.548 & P$\cdot$C = 0.759 & P$\cdot$C = 0.776 & P$\cdot$C = 0.776 \\
\hline
\multirow{2}{*}{0.8} & \multirow{2}{*}{4} & \multirow{2}{*}{80} & P/C = 0.973/0.562 & P/C = 0.943/0.805 & P/C = 0.846/0.920 & P/C = 0.816/0.959 \\
& & & P$\cdot$C = 0.547 & P$\cdot$C = 0.759 & P$\cdot$C = 0.778 & P$\cdot$C = 0.783 \\
\hline
\multirow{2}{*}{0.8} & \multirow{2}{*}{5} & \multirow{2}{*}{20} & P/C = 0.983/0.565 & P/C = 0.940/0.811 & P/C = 0.799/0.924 & P/C = 0.716/0.976 \\
& & & P$\cdot$C = 0.555 & P$\cdot$C = 0.762 & P$\cdot$C = 0.738 & P$\cdot$C = 0.699 \\
\hline
\multirow{2}{*}{0.8} & \multirow{2}{*}{5} & \multirow{2}{*}{30} & P/C = 0.981/0.561 & P/C = 0.941/0.797 & P/C = 0.809/0.920 & P/C = 0.734/0.975 \\
& & & P$\cdot$C = 0.550 & P$\cdot$C = 0.750 & P$\cdot$C = 0.744 & P$\cdot$C = 0.716 \\
\hline
\multirow{2}{*}{0.8} & \multirow{2}{*}{5} & \multirow{2}{*}{40} & P/C = 0.978/0.551 & P/C = 0.941/0.785 & P/C = 0.834/0.919 & P/C = 0.751/0.974 \\
& & & P$\cdot$C = 0.539 & P$\cdot$C = 0.739 & P$\cdot$C = 0.766 & P$\cdot$C = 0.731 \\
\hline
\multirow{2}{*}{0.8} & \multirow{2}{*}{5} & \multirow{2}{*}{50} & P/C = 0.978/0.540 & P/C = 0.944/0.781 & P/C = 0.843/0.911 & P/C = 0.772/0.969 \\
& & & P$\cdot$C = 0.528 & P$\cdot$C = 0.737 & P$\cdot$C = 0.768 & P$\cdot$C = 0.748 \\
\hline
\multirow{2}{*}{0.8} & \multirow{2}{*}{5} & \multirow{2}{*}{60} & P/C = 0.979/0.531 & P/C = 0.944/0.779 & P/C = 0.848/0.901 & P/C = 0.796/0.963 \\
& & & P$\cdot$C = 0.520 & P$\cdot$C = 0.735 & P$\cdot$C = 0.764 & P$\cdot$C = 0.767 \\
\hline
\multirow{2}{*}{0.8} & \multirow{2}{*}{5} & \multirow{2}{*}{70} & P/C = 0.979/0.529 & P/C = 0.951/0.777 & P/C = 0.861/0.892 & P/C = 0.815/0.956 \\
& & & P$\cdot$C = 0.518 & P$\cdot$C = 0.739 & P$\cdot$C = 0.768 & P$\cdot$C = 0.779 \\
\hline
\multirow{2}{*}{0.8} & \multirow{2}{*}{5} & \multirow{2}{*}{80} & P/C = 0.978/0.521 & P/C = 0.953/0.766 & P/C = 0.869/0.882 & P/C = 0.824/0.948 \\
& & & P$\cdot$C = 0.510 & P$\cdot$C = 0.730 & P$\cdot$C = 0.766 & P$\cdot$C = 0.781 \\
\hline
\multirow{2}{*}{1.2} & \multirow{2}{*}{4} & \multirow{2}{*}{20} & P/C = 0.698/0.286 & P/C = 0.602/0.485 & P/C = 0.516/0.746 & P/C = 0.433/0.858 \\ 
& & & P$\cdot$C = 0.200 & P$\cdot$C = 0.292 & P$\cdot$C = 0.385 & P$\cdot$C = 0.372 \\
\hline
\multirow{2}{*}{1.2} & \multirow{2}{*}{4} & \multirow{2}{*}{30} & P/C = 0.694/0.285 & P/C = 0.598/0.472 & P/C = 0.514/0.738 & P/C = 0.440/0.861 \\
& & & P$\cdot$C = 0.198 & P$\cdot$C = 0.282 & P$\cdot$C = 0.379 & P$\cdot$C = 0.379 \\
\hline
\multirow{2}{*}{1.2} & \multirow{2}{*}{4} & \multirow{2}{*}{40} & P/C = 0.701/0.280 & P/C = 0.608/0.471 & P/C = 0.515/0.734 & P/C = 0.447/0.861 \\
& & & P$\cdot$C = 0.196 & P$\cdot$C = 0.286 & P$\cdot$C = 0.378 & P$\cdot$C = 0.385 \\
\hline
\multirow{2}{*}{1.2} & \multirow{2}{*}{4} & \multirow{2}{*}{50} & P/C = 0.712/0.276 & P/C = 0.603/0.466 & P/C = 0.517/0.730 & P/C = 0.463/0.851 \\
& & & P$\cdot$C = 0.197 & P$\cdot$C = 0.281 & P$\cdot$C = 0.377 & P$\cdot$C = 0.394 \\
\hline
\multirow{2}{*}{1.2} & \multirow{2}{*}{4} & \multirow{2}{*}{60} & P/C = 0.727/0.274 & P/C = 0.618/0.464 & P/C = 0.523/0.729 & P/C = 0.478/0.845 \\
& & & P$\cdot$C = 0.199 & P$\cdot$C = 0.287 & P$\cdot$C = 0.381 & P$\cdot$C = 0.404 \\
\hline
\multirow{2}{*}{1.2} & \multirow{2}{*}{4} & \multirow{2}{*}{70} & P/C = 0.725/0.268 & P/C = 0.626/0.458 & P/C = 0.530/0.728 & P/C = 0.495/0.843 \\
& & & P$\cdot$C = 0.194 & P$\cdot$C = 0.287 & P$\cdot$C = 0.386 & P$\cdot$C = 0.417 \\
\hline
\multirow{2}{*}{1.2} & \multirow{2}{*}{4} & \multirow{2}{*}{80} & P/C = 0.733/0.266 & P/C = 0.627/0.454 & P/C = 0.539/0.719 & P/C = 0.515/0.835 \\
& & & P$\cdot$C = 0.195 & P$\cdot$C = 0.285 & P$\cdot$C = 0.388 & P$\cdot$C = 0.430 \\
\hline
\multirow{2}{*}{1.2} & \multirow{2}{*}{5} & \multirow{2}{*}{20} & P/C = 0.747/0.228 & P/C = 0.667/0.477 & P/C = 0.542/0.706 & P/C = 0.491/0.873 \\
& & & P$\cdot$C = 0.170 & P$\cdot$C = 0.318 & P$\cdot$C = 0.383 & P$\cdot$C = 0.429 \\
\hline
\multirow{2}{*}{1.2} & \multirow{2}{*}{5} & \multirow{2}{*}{30} & P/C = 0.757/0.223 & P/C = 0.675/0.470 & P/C = 0.538/0.694 & P/C = 0.505/0.868 \\
& & & P$\cdot$C = 0.169 & P$\cdot$C = 0.317 & P$\cdot$C = 0.373 & P$\cdot$C = 0.438 \\
\hline
\multirow{2}{*}{1.2} & \multirow{2}{*}{5} & \multirow{2}{*}{40} & P/C = 0.755/0.220 & P/C = 0.683/0.459 & P/C = 0.538/0.686 & P/C = 0.511/0.858 \\
& & & P$\cdot$C = 0.166 & P$\cdot$C = 0.313 & P$\cdot$C = 0.369 & P$\cdot$C = 0.438 \\
\hline
\multirow{2}{*}{1.2} & \multirow{2}{*}{5} & \multirow{2}{*}{50} & P/C = 0.755/0.211 & P/C = 0.679/0.450 & P/C = 0.544/0.682 & P/C = 0.532/0.849 \\
& & & P$\cdot$C = 0.159 & P$\cdot$C = 0.306 & P$\cdot$C = 0.371 & P$\cdot$C = 0.452 \\
\hline
\multirow{2}{*}{1.2} & \multirow{2}{*}{5} & \multirow{2}{*}{60} & P/C = 0.753/0.206 & P/C = 0.678/0.443 & P/C = 0.553/0.672 & P/C = 0.541/0.820 \\
& & & P$\cdot$C = 0.155 & P$\cdot$C = 0.300 & P$\cdot$C = 0.372 & P$\cdot$C = 0.444 \\
\hline
\multirow{2}{*}{1.2} & \multirow{2}{*}{5} & \multirow{2}{*}{70} & P/C = 0.755/0.198 & P/C = 0.680/0.432 & P/C = 0.571/0.666 & P/C = 0.552/0.814 \\
& & & P$\cdot$C = 0.149 & P$\cdot$C = 0.294 & P$\cdot$C = 0.380 & P$\cdot$C = 0.449 \\
\hline
\multirow{2}{*}{1.2} & \multirow{2}{*}{5} & \multirow{2}{*}{80} & P/C = 0.761/0.193 & P/C = 0.680/0.429 & P/C = 0.591/0.649 & P/C = 0.580/0.806 \\
& & & P$\cdot$C = 0.147 & P$\cdot$C = 0.292 & P$\cdot$C = 0.384 & P$\cdot$C = 0.467 \\
\hline
\end{tabular}
\begin{flushleft}
Purity and completeness values for each set of SExtractor parameters we tested separated by spectroscopic coverage and redshift. There is not a strong dependence on the choice of SExtractor parameters, but we chose to use the most lenient set of parameters: DETECT\textunderscore THRESH = 4$\sigma$ and DETECT\textunderscore MINAREA = 20, to maximize our chances of detecting an overdensity candidate in the real data, which may not be regular in shape and thus more difficult to detect.
\end{flushleft}
\end{table*}

\section{Final Overdensity Candidate Catalog}\label{final}

We conducted our overdensity candidate search over a total of 15 ORELSE fields with similar depth $B$- and $V$-band imaging as well as $riz$ and some form of ground- and/or space-based NIR imaging. We typically had an average of 11 bands per field (see \S\ref{data} and Table \ref{tab.imaging} for details regarding the imaging and depth for each field). We searched for overdensity candidates with the same SExtractor parameters we found to work best overall in Sections \ref{recovering} and \ref{mocks}: DETECT\textunderscore THRESH = 4$\sigma$, DETECT\textunderscore MINAREA = 20, DEBLEND\textunderscore NTHRESH = 32, and DEBLEND\textunderscore MINCONT = 0.01. We restricted our search to overdensity candidates with mean redshifts within our total redshift range, $z$ = 0.55 to 1.37. To reduce the chance of detecting the same overdensity more than once, we looked at the largest peaks in each field and excluded detections within 0.7 Mpc and $\Delta z <$ 0.02 of those peaks, as mentioned in \S\ref{masks.linking}. 

\begin{figure*}
\centering
\includegraphics[width=2\columnwidth]{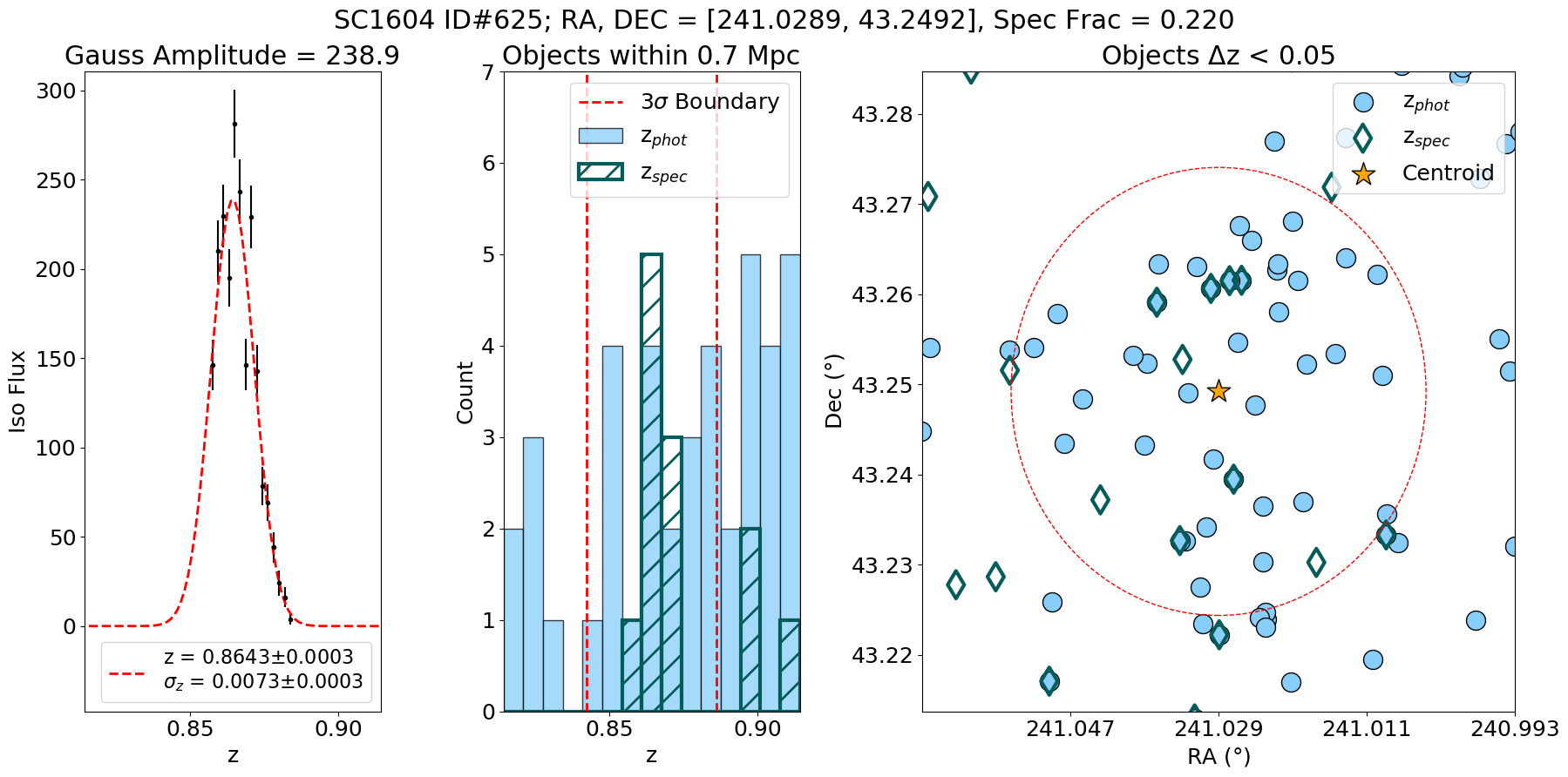}
\caption{One example of a detected overdensity candidate. The left panel shows the overdensity candidate's isophotal flux profile fitted with a Gaussian. The points in the profile come from linking individual isophotal detections in neighboring redshift slices whose flux-weighted distances were within a 1 Mpc linking radius. The associated errors in the isophotal flux is calculated by SExtractor. The center panel shows the histogram of spectroscopic and photometric members within a 0.7 Mpc radius from the centroided position of the overdensity candidate. The red dashed lines show the redshift boundary within 3$\sigma$ as determined by the Gaussian fit. The right panel plots the positions of the galaxy members within $\Delta z < 0.05$ in a square measuring 1 Mpc on a side, and the dashed red circle has a radius of 0.7 Mpc.}
\label{fig.hist}
\end{figure*}

We considered known structures identified if a detection was made within 1 Mpc of their fiducial transverse position and $\Delta z <$ 0.02 of their fiducial redshift. We discard overdensity candidates with $\sigma_{z} > 0.05$ for being unphysically large. We also exclude candidates where the Gaussian peak was more than 20\% higher than the maximum flux of the fitted points as well as candidates with larger integrated isophotal flux uncertainties than integrated isophotal fluxes, which were more likely to have inaccurate total masses and redshifts. Due to the purity and completeness numbers we saw with our tests with using reduced spectroscopic fractions in the mock catalogs in \S\ref{mock.compur}, we only included overdensity candidates in the final catalog with at least 5\% spectroscopic fractions (see later on in this section for the meaning of the spectroscopic fraction in this context). After making this cut, our sample contained 51 of the 56 previously known clusters or groups (Table \ref{tab.prev}) and 402 new overdensity candidates (Table \ref{tab.catalog}, Fig. \ref{fig.hist}, Fig. \ref{fig.histplots}).

\begin{table*}
\centering
\caption{Previously Known ORELSE Clusters and Groups}
\label{tab.prev}
\begin{tabular}{lccccccc}
\hline
Name & Redshift & RA (J2000) & Dec (J2000) & $\sigma$$^{a}$ & N$^{b}$ & log(M$_{vir}$)$^{c}$ & Recovered? \\
\hline
SC1604 Lz  & 0.5995 & 241.03282 & 43.2057    & 771.9  $\pm$ 110.0 & 21 & 14.711 $\pm$ 0.186 & Yes \\ 
SC1604 A   & 0.8984 & 241.09311 & 43.0821    & 722.4  $\pm$ 134.5 & 35 & 14.551 $\pm$ 0.243 & Yes \\ 
SC1604 B   & 0.8648 & 241.10796 & 43.2397    & 818.4  $\pm$ 74.2  & 49 & 14.722 $\pm$ 0.118 & Yes \\ 
SC1604 C   & 0.9344 & 241.03142 & 43.2679    & 453.5  $\pm$ 39.6  & 32 & 13.935 $\pm$ 0.114 & Yes \\ 
SC1604 D   & 0.9227 & 241.14094 & 43.3539    & 688.2  $\pm$ 88.1  & 70 & 14.481 $\pm$ 0.167 & Yes \\ 
SC1604 F   & 0.9331 & 241.20104 & 43.3684    & 541.9  $\pm$ 110.0 & 20 & 14.168 $\pm$ 0.265 & Yes \\ 
SC1604 G   & 0.9019 & 240.92745 & 43.4030    & 539.3  $\pm$ 124.0 & 18 & 14.169 $\pm$ 0.300 & Yes \\ 
SC1604 H   & 0.8528 & 240.89890 & 43.3669    & 287.0  $\pm$ 68.3  & 10 & 13.359 $\pm$ 0.310 & Yes \\ 
SC1604 I   & 0.9024 & 240.79746 & 43.3915    & 333.0  $\pm$ 129.4 & 7  & 13.541 $\pm$ 0.506 & Yes \\ 
SC1604 Hz  & 1.1815 & 241.07967 & 43.3215    & 661.5  $\pm$ 80.2  & 15 & 14.367 $\pm$ 0.158 & Yes \\ 
SC1324 A   & 0.7566 & 201.20129 & 30.1924    & 873.4  $\pm$ 110.8 & 43 & 14.833 $\pm$ 0.165 & Yes \\ 
SC1324 B   & 0.6971 & 201.08815 & 30.2158    & 677.1  $\pm$ 143.6 & 13 & 14.516 $\pm$ 0.276 & Yes \\ 
SC1324 C   & 0.7574 & 201.25533 & 30.4158    & 353.0  $\pm$ 182.4 & 6  & 13.652 $\pm$ 0.673 & Yes \\ 
SC1324 D   & 0.7382 & 201.00773 & 30.4164    & 205.9  $\pm$ 90.1  & 8  & 12.955 $\pm$ 0.579 & Yes \\ 
SC1324 G   & 0.6759 & 201.18736 & 30.7995    & 186.3  $\pm$ 38.8  & 10 & 12.840 $\pm$ 0.271 & Yes \\ 
SC1324 H   & 0.6990 & 201.22040 & 30.8408    & 346.4  $\pm$ 109.8 & 19 & 13.708 $\pm$ 0.393 & Yes \\ 
SC1324 I   & 0.6956 & 201.20550 & 30.9665    & 847.1  $\pm$ 96.4  & 35 & 14.808 $\pm$ 0.148 & Yes \\ 
SC1324 Hz  & 1.0979 & 201.18372 & 30.8228    & 509.3  $\pm$ 220.4 & 11 & 14.047 $\pm$ 0.564 & Yes \\ 
XLSS005    & 1.0559 & 36.773036 & -4.2972476 & 735.8  $\pm$ 108.2 & 19 & 14.536 $\pm$ 0.191 & Yes \\ 
SG0023 A   & 0.8396 & 6.02560   & 4.3590     & 412.8  $\pm$ 119.2 & 14 & 13.836 $\pm$ 0.376 & Yes \\ 
SG0023 B1  & 0.8290 & 5.97570   & 4.3884     & 176.3  $\pm$ 29.6  & 23 & 12.730 $\pm$ 0.219 & Yes \\ 
SG0023 B2  & 0.8453 & 5.96970   & 4.3820     & 277.8  $\pm$ 41.0  & 38 & 13.319 $\pm$ 0.192 & Yes \\ 
SG0023 C   & 0.8466 & 5.92470   & 4.3807     & 385.3  $\pm$ 54.3  & 45 & 13.744 $\pm$ 0.184 & No* \\ 
SG0023 M   & 0.8472 & 5.96740   & 4.3199     & 418.8  $\pm$ 68.9  & 14 & 13.853 $\pm$ 0.214 & Yes \\ 
SG0023 Hz  & 0.9799 & 5.99464   & 4.3570     & 218.2  $\pm$ 62.6  & 9  & 12.970 $\pm$ 0.374 & Yes \\ 
RXJ1757    & 0.6931 & 269.33196 & 66.525991  & 862.3  $\pm$ 107.9 & 34 & 14.832 $\pm$ 0.250 & Yes \\ 
RXJ1757 Hz & 0.9456 & 269.20697 & 66.593766  & 290.5  $\pm$ 92.6  & 8  & 13.352 $\pm$ 0.415 & Yes \\ 
RXJ1821    & 0.8168 & 275.38451 & 68.465768  & 1119.6 $\pm$ 99.6  & 52 & 15.142 $\pm$ 0.116 & Yes \\ 
RXJ1821 Hz & 0.9189 & 275.24066 & 68.437054  & 684.8  $\pm$ 97.4  & 19 & 14.476 $\pm$ 0.185 & Yes \\ 
SC0910 LzA & 0.7600 & 137.68681 & 54.3436    & 524.0  $\pm$ 159.4 & 11 & 14.166 $\pm$ 0.396 & No  \\ 
SC0910 LzB & 0.7859 & 137.60865 & 54.3897    & 165.5  $\pm$ 30.5  & 10 & 12.658 $\pm$ 0.240 & No  \\ 
SC0910 A   & 1.1034 & 137.51280 & 54.3099    & 840.4  $\pm$ 244.0 & 23 & 14.698 $\pm$ 0.378 & Yes \\ 
SC0910 B   & 1.1007 & 137.68489 & 54.3725    & 724.7  $\pm$ 151.4 & 25 & 14.506 $\pm$ 0.272 & Yes \\ 
Cl1429     & 0.9871 & 217.28141 & 42.6826    & 911.1  $\pm$ 84.2  & 38 & 14.831 $\pm$ 0.185 & Yes \\ 
Cl1137     & 0.9553 & 174.39786 & 30.008930  & 534.6  $\pm$ 81.1  & 28 & 14.144 $\pm$ 0.197 & Yes \\ 
RXJ1053    & 1.1285 & 163.43097 & 57.591476  & 898.0  $\pm$ 142.0 & 28 & 14.778 $\pm$ 0.206 & Yes \\ 
RXJ1053 Hz & 1.2049 & 163.20387 & 57.584000  & 916.3  $\pm$ 194.8 & 11 & 14.786 $\pm$ 0.277 & Yes \\ 
RXJ1221 A  & 0.7017 & 185.53798 & 49.2329    & 426.6  $\pm$ 71.3  & 18 & 13.923 $\pm$ 0.218 & Yes \\ 
RXJ1221 B  & 0.7000 & 185.34103 & 49.3138    & 753.2  $\pm$ 122.5 & 36 & 14.654 $\pm$ 0.222 & Yes \\ 
RXJ1716 A  & 0.8158 & 259.10074 & 67.085108  & 624.1  $\pm$ 136.1 & 40 & 14.380 $\pm$ 0.284 & Yes \\ 
RXJ1716 B  & 0.8092 & 259.21686 & 67.139647  & 1120.6 $\pm$ 101.5 & 83 & 15.145 $\pm$ 0.118 & Yes \\ 
RXJ1716 C  & 0.8146 & 259.25725 & 67.152497  & 678.4  $\pm$ 57.8  & 39 & 14.489 $\pm$ 0.111 & No  \\ 
RXJ1716 Hz & 0.8531 & 259.30007 & 67.183591  & 757.4  $\pm$ 99.2  & 16 & 14.623 $\pm$ 0.171 & Yes \\ 
Cl1350 A   & 0.8012 & 207.88457 & 60.0371    & 351.2  $\pm$ 92.4  & 9  & 13.802 $\pm$ 0.343 & Yes \\ 
Cl1350 B   & 0.8017 & 207.53970 & 60.1034    & 300.0  $\pm$ 118.5 & 10 & 13.429 $\pm$ 0.515 & Yes \\ 
Cl1350 C   & 0.7996 & 207.71545 & 60.1148    & 802.4  $\pm$ 83.9  & 43 & 14.712 $\pm$ 0.149 & Yes \\ 
RCS0224 A  & 0.7780 & 36.15714  & -0.0949    & 825.4  $\pm$ 193.2 & 34 & 14.754 $\pm$ 0.305 & Yes \\ 
RCS0224 B  & 0.7781 & 36.14123  & -0.0394    & 710.7  $\pm$ 58.8  & 52 & 14.559 $\pm$ 0.108 & Yes \\ 
RCS0224 Hz & 0.8454 & 36.32021  & -0.0928    & 437.7  $\pm$ 115.8 & 15 & 13.911 $\pm$ 0.345 & Yes \\ 
SC0849 A   & 1.2637 & 132.23463 & 44.761780  & 714.4  $\pm$ 171.6 & 13 & 14.448 $\pm$ 0.313 & Yes \\ 
SC0849 B   & 1.2639 & 132.29977 & 44.865903  & 261.5  $\pm$ 62.6  & 12 & 13.139 $\pm$ 0.312 & Yes \\ 
SC0849 C   & 1.2609 & 132.24443 & 44.866012  & 839.2  $\pm$ 111.8 & 25 & 14.659 $\pm$ 0.174 & Yes \\ 
SC0849 D   & 1.2703 & 132.14184 & 44.896338  & 697.2  $\pm$ 111.2 & 23 & 14.415 $\pm$ 0.208 & Yes \\ 
SC0849 E   & 1.2601 & 132.27496 & 44.959253  & 445.1  $\pm$ 71.9  & 14 & 13.833 $\pm$ 0.210 & Yes \\ 
SC0849 LzB & 0.5678 & 132.24487 & 44.896184  & 495.2  $\pm$ 169.3 & 8  & 14.140 $\pm$ 0.445 & No  \\ 
SC0849 LzA & 1.1394 & 132.28743 & 44.855804  & 176.3  $\pm$ 107.1 & 6  & 12.655 $\pm$ 0.791 & Yes \\ [0.7ex]
\hline
\end{tabular}
\begin{flushleft}
$a$: 1 Mpc velocity dispersion in km s$^{-1}$, determined with a biweight estimator.

$b$: Number of galaxies used for the velocity dispersion calculation.

$c$: Virial mass in units of solar mass, calculated from the formula given in \citet{Lemaux12}.

All 56 previously known clusters and groups in the 15 ORELSE fields we conducted our search over, of which we recovered 51 with our VMC overdensity detection method. RA and Dec centers were calculated from an $i$-band weighted mean of galaxies within 1 Mpc (or $z$-band for redshifts greater than 0.95). Because of how we cut out detections within 0.7 Mpc of more significant peaks, we miss the detection of SG0023 C as we find it within 0.38 Mpc of the larger SG0023 B2 detection. Dropping the separation threshold to include SG0023 C however results in more duplicate detections of other known structures and thus likely more duplicate detections of the overdensity candidates. We thus decide that the 0.7 Mpc separation is more appropriate for the general use case here.
\end{flushleft}
\end{table*}

For one of the known structures that was not detected, SG0023 C, our minimum separation threshold of 0.7 Mpc between detections results in it being dropped from the detected sample as its detection is within 0.38 Mpc of where we find the more significant SG0023 B2 detection. If we were to drop the separation threshold so as to include SG0023 C, we find 14 total duplicate detections of other known structures, whereas keeping the 0.7 Mpc separation only results in 5 duplicate detections. As the supercluster in SG0023 is an especially difficult to separate system, we elect to use the 0.7 Mpc separation for its advantages in the more general case scenario and reduce our probability of having duplicate detections of the same overdensity candidates.

\begin{figure*}
\centering
\includegraphics[width=2\columnwidth]{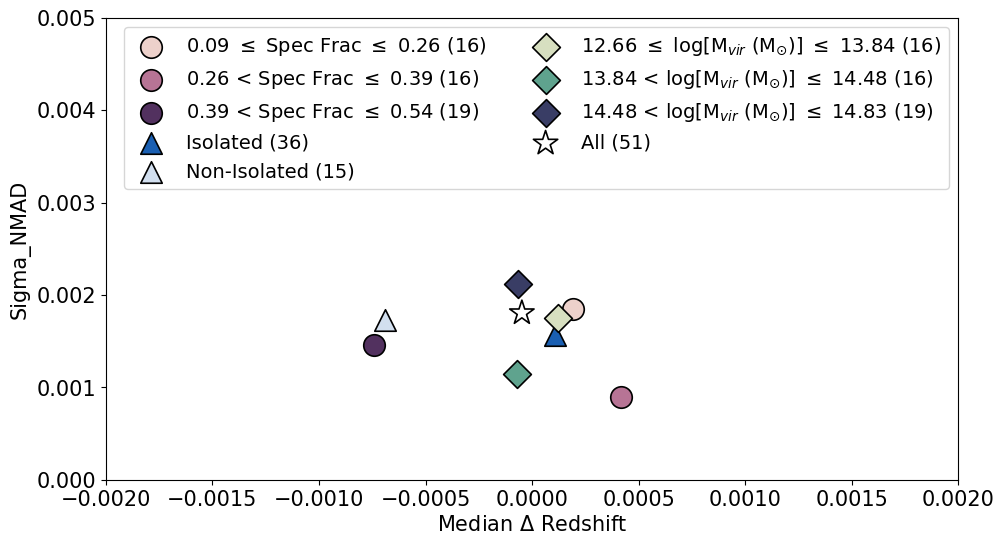}
\caption{Median redshift offsets and their median absolute deviations for all recovered known structures between what was found with our Gaussian fits and the fiducial values. These were additionally binned into three sets of sub-samples based on their spectroscopic fraction, mass, and whether they were isolated from other structures or not. In each case, the differences between the bins were minor, and the absolute scale on both axes is on the order of tens of thousandths at most or $<$100 km s$^{-1}$ at these redshifts.}
\label{fig.nmad}
\end{figure*}

We checked sub-samples of the recovered known structures based on their spectroscopic fraction, mass, and isolation from other structures. We calculated the redshift offsets between what we find with our Gaussian fits and the fiducial values, as well as the median absolute deviation of these offsets. The variations in the median redshift offset and absolute deviations were tens of thousandths at most, or $<$100 km s$^{-1}$ at these redshifts, for each sub-sample, showing strong promise for our ability to find structures even at small spectroscopic fractions and low masses (Fig. \ref{fig.nmad}).

We approximate the spectroscopic fraction of an overdensity candidate by calculating the fraction of $z_{phot}$ galaxies within 0.7 Mpc of the overdensity candidate's barycenter and $\Delta z < 0.05$ of its redshift with a high-quality $z_{spec}$ counterpart. We use 0.7 Mpc as an average of the typical sizes of the known ORELSE groups and clusters. Because the redshifts of $z_{phot}$ galaxies are more uncertain than $z_{spec}$ galaxies, they may be scattered in or out of a given redshift range. The true spectroscopic fraction $Q_{true}$ involves estimating the true number of $z_{phot}$ galaxies in the redshift range with the following:

\begin{equation}
N_{true} = N_{phot} \times \frac{P_{gal}}{C_{gal}}
\end{equation}

\begin{equation}
P_{gal} = \frac{N_{conf,A}}{N_{conf,A}+N_{conf,out}}
\end{equation}

\begin{equation}
C_{gal} = \frac{N_{conf,A}}{N_{conf,A}+N_{conf,B}}
\end{equation}

\begin{equation}
Q_{true} = \frac{N_{conf,A}+N_{conf,B}}{N_{true}}
\end{equation}

$P_{gal}$ and $C_{gal}$ are the purity and completeness for the member galaxies respectively. Note that these purity and completeness values have no relation to the calculated purity and completeness of the detections in the mocks. $N_{true}$ is the true number of galaxies in the redshift range. $N_{phot}$ is the number of $z_{phot}$ galaxies in the redshift range. $N_{conf,A}$ and $N_{conf,B}$ are the $z_{spec}$ galaxies in the redshift range with a match to a $z_{phot}$ galaxy with a match inside or outside the redshift range respectively. $N_{conf,out}$ are the $z_{phot}$ galaxies with a $z_{spec}$ match outside the redshift range. Mathematically, $Q_{true}$ simplifies to:

\begin{equation}
\begin{split}
Q_{true} & = \frac{N_{conf,A}+N_{conf,B}}{N_{phot}} \frac{C}{P} \\
         & = \frac{N_{conf,A}+N_{conf,B}}{N_{phot}} \frac{\frac{N_{conf,A}}{N_{conf,A}+N_{conf,B}}} {\frac{N_{conf,A}+N_{conf,out}}{N_{conf,A}}} \\
         & = \frac{N_{conf,A}+N_{conf,out}}{N_{phot}}
\end{split}
\end{equation}

\subsection{Estimation of the Total Masses of the Candidates}
To convert the isophotal flux from SExtractor into a physically meaningful value, we devised a new method of estimation by fitting the integrated Gaussian isophotal flux of our isolated known structures to their virial mass, where as in Section \ref{recovering}, we define a structure as isolated if no other known structures with a redshift $\Delta z < 0.02$ are within a radial distance of 2.5 Mpc. While we make some assumption on the dynamical state of the structures used to calibrate this mass estimate, we selected a large number of isolated structures for this calibration both to limit the possible dynamical perturbation of these structures by surrounding structure and to average over variation in dynamical states. Further, as we will show in \S\ref{masscomp}, this new method of mass estimation correlates extremely well with similar overdensity-based mass estimates meaning its accuracy has limited dependence on the type of structure or its galaxy population. Further, by virtue of the fact that virial mass estimates essentially match all other total mass measurements in ORELSE for those structures where independent mass measurements are available, our mass estimation also correlates well with X-ray, lensing, and Sunyaev-Zel'dovich masses of ORELSE structures.

We excluded four isolated structures from our fit: SC1604 Lz, SC1324 A, SC1324 I, and RXJ1053 Hz. SC1604 Lz is located near the redshift limit of our spectroscopic coverage. Its extended irregular structure and high mass additionally imply that it is not dynamically relaxed. SC1324 A and I are located at the most southern and northern edges of the field respectively and therefore partially off of the photometric footprint in that field. These structures in the VMC map are clipped by the edges of the image, and thus their integrated isophotal fluxes are artificially low. RXJ1053 Hz suffers from sparse sampling of its underlying velocity distribution and only has a small number of members.

\begin{figure*}
\centering
\includegraphics[width=2\columnwidth]{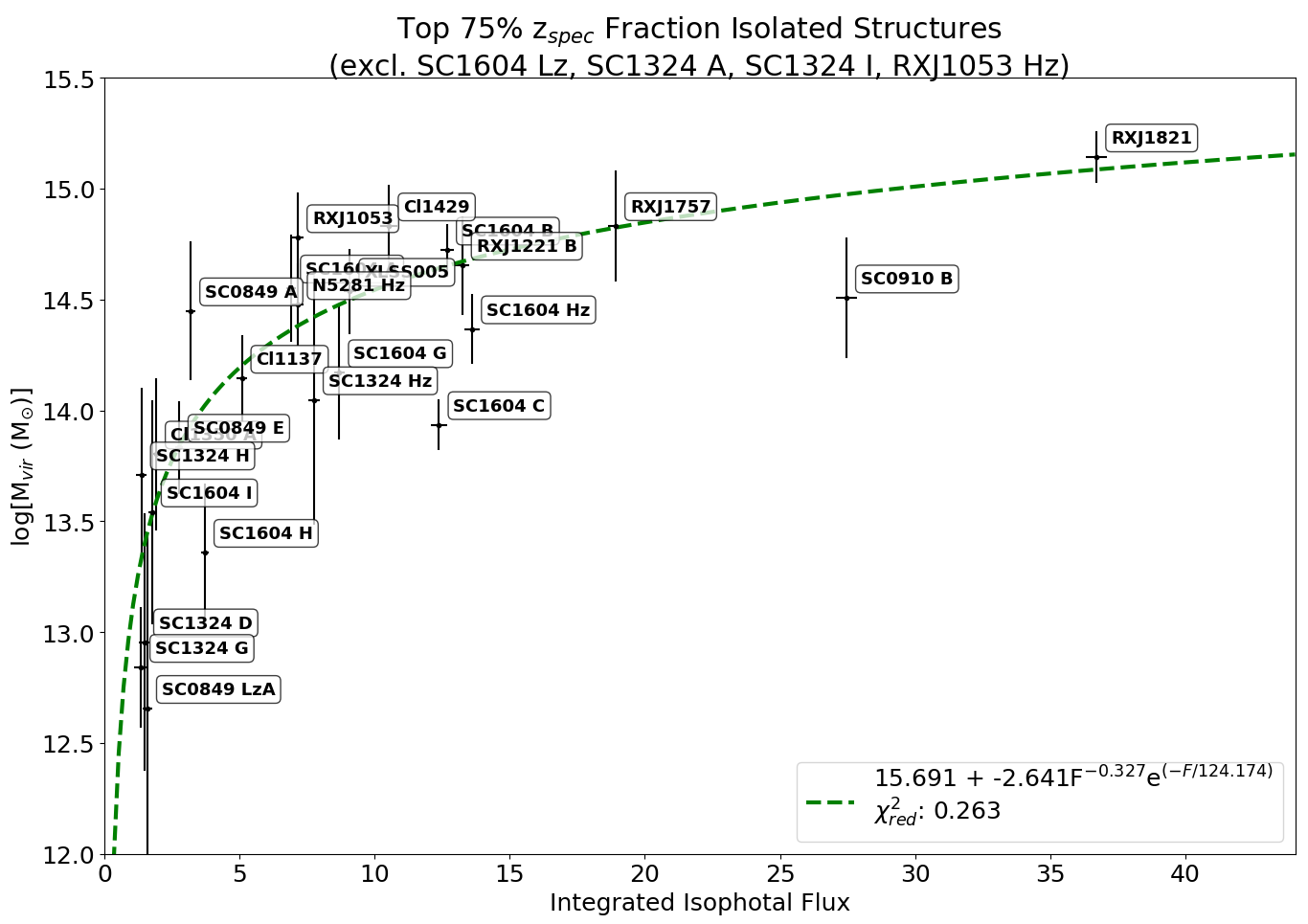}
\caption{The integrated isophotal flux F of the Gaussian fits for selected isolated known structures, plotted against their virial masses. These structures were chosen based on removing the bottom 25\% of all recovered known structures in spectroscopic fraction and selecting the isolated structures among the remaining. Four structures, SC1604 Lz, SC1324 A, SC1324 I, and RXJ1053 Hz, are treated as outliers in the fitting. In the fitting process, we only included the errors on the virial masses as they dominated over the errors on the integrated isophotal fluxes, though including both sets of errors did not meaningfully change the shapes of the fits. Of the four models we fitted for the virial mass, the pseudo-Schechter function had the smallest scatter and reduced $\chi^{2}$, where the terms are $a$ = 15.691 $\pm$ 0.010, $b$ = -2.641 $\pm$ 0.033, $c$ = -0.327 $\pm$ 0.039, and $d$ = 124.174 $\pm$ 0.740, with an associated reduced $\chi^{2}$ of 0.263.}
\label{fig.mass}
\end{figure*}

We fitted four different models to the virial mass and integrated isophotal flux relation: a linear fit, a quadratic fit, an exponential fit, and a pseudo-Schechter function fit of the form log(M$_{vir}$) = $a + b F^{c} e^{-(F/d)}$, where M$_{vir}$ is the virial mass in units of solar mass and F is the integrated isophotal flux. We find that there is a large scatter in masses at the low integrated isophotal flux end. To investigate this, we applied our fits to several different subgroups of our recovered isolated known structures cutting on their spectroscopic fractions (the top 75\% or 50\%) and number of members (requiring at least 10, 15, or 20 members). The spectroscopic fraction cuts were made on all recovered known structures, and the fits were applied to the isolated structures among them. Because the virial mass errors dominated, we only included them in the fitting and $\chi^{2}$ calculations, though the fits did not change meaningfully when we also included the integrated isophotal flux errors in the fitting process. We settled on using the pseudo-Schechter function fit from the isolated structures with the top 75\% of spectroscopic fractions as it produced the smallest reduced $\chi^{2}$ of all the fits we attempted. We take the pseudo-Schechter model to use as our scaling relation between the integrated isophotal flux and virial mass with the parameters, $a$ = 15.691 $\pm$ 0.010, $b$ = -2.641 $\pm$ 0.033, $c$ = -0.327 $\pm$ 0.039, and $d$ = 124.174 $\pm$ 0.740, with an associated reduced $\chi^{2}$ of 0.263 (Fig. \ref{fig.mass}).

\subsubsection{Purity and Completeness by Mass}

With a relation between mass and isophotal flux now in hand, we can calculate the purity and completeness numbers by mass bin in the mock catalogs. As a proxy for the ``true'' masses of all of the structures we injected into the mocks, we compute their virial masses based on their inputted virial radius, using Equation \ref{eq.vir} in \S\ref{recovering}. When we compare these virial masses to the fitted masses based on their isophotal flux, we find the median difference between the two to be close to 0 for all spectroscopic fractions and both redshifts.

\begin{figure*}
\centering
\includegraphics[width=1.5\columnwidth]{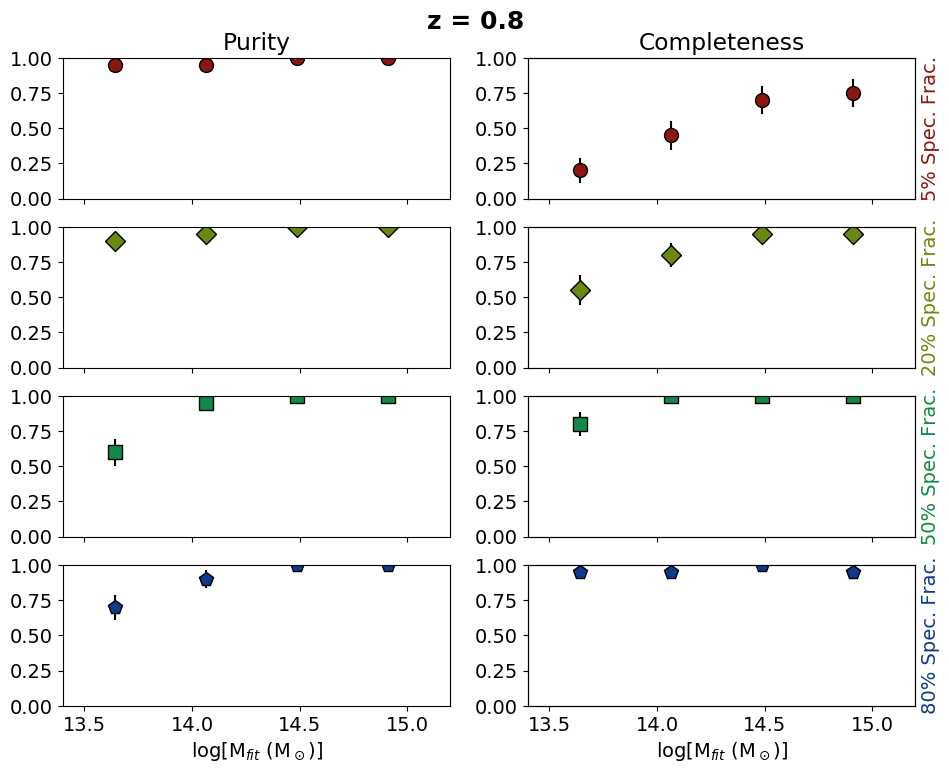}
\includegraphics[width=1.5\columnwidth]{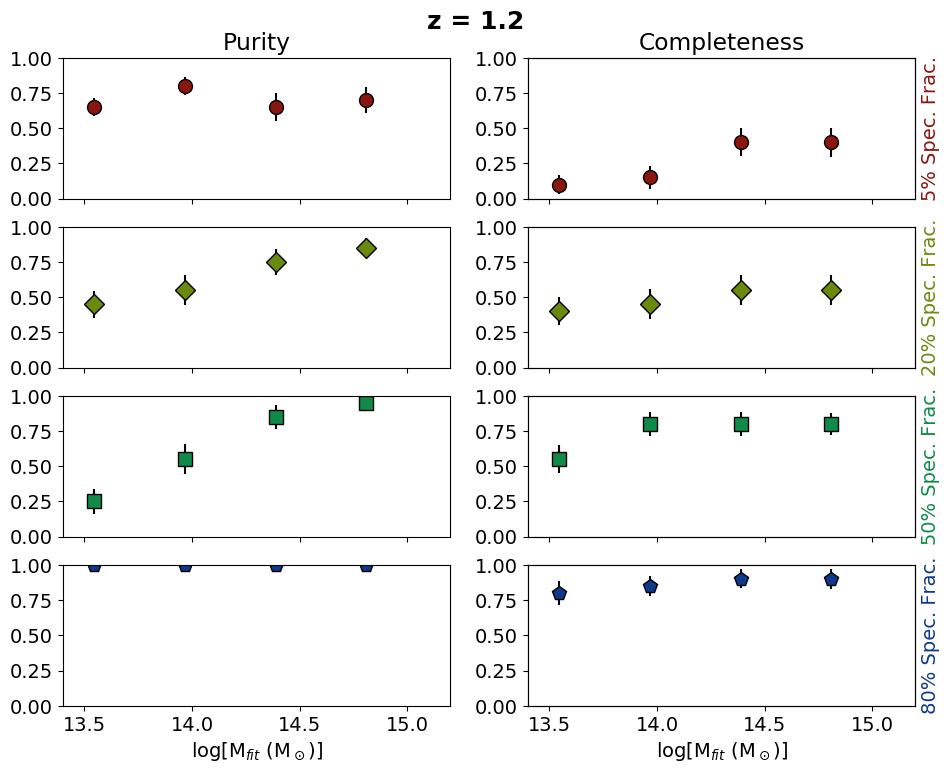}
\caption{Purity and completeness numbers for the mock catalogs divided into four mass bins. The points plotted are the median values and their 1$\sigma$ uncertainty over 1000 bootstrap trials picking 20 random detections at a time. In some cases, the size of the marker was larger than the range of the error bar. Note that the mass bins are slightly different between the $z$ = 0.8 and $z$ = 1.2 plots due to small differences in the mass ranges for each redshift. We found that the purity tended to be higher at lower spectroscopic fractions, which is likely due to the many photometric objects washing out the small overdensities that arise from field fluctuations. It is possible these overdensities are real structure in the field, as we take the field from real data, so our purity numbers here are functionally lower limits.}
\label{fig.compur_mass}
\end{figure*}

For the purity and completeness calculations, we divide the masses into four equal bins, which span from 13.2 $< \log(M/M_{\odot}) <$ 15.0. We use the virial mass for the recovered injected structures and the fitted masses for the spurious detections. For each mass bin at each spectroscopic fraction at each redshift, we compute the purity and completeness through a bootstrap method, similar to the process in \S\ref{mock.compur}. We subsample and compute the purity and completeness of a random 20 detections, then repeat the process 1000 times (Fig. \ref{fig.compur_mass}).

At low masses, it becomes difficult to separate overdensities that arise from real structure versus those that are nothing more than field fluctuations. In these cases, lower spectroscopic fractions which wash out small overdensities to the point of non-detection are more resistant to contamination by field fluctuations, and so we see lower purity numbers at higher spectroscopic fractions. However, we do not know if the low mass spurious detections are actually present in the field or not, as the galaxies are taken from real data, so we caution readers to treat the purity numbers as lower limits.  As we only have purity and completeness estimates for masses above $\log(M/M_{\odot}) = 13.2$, the lower mass candidates we found must similarly be taken with a grain of salt.

\subsubsection{Comparison with Overdensity Mass Estimation}
\label{masscomp}

As an additional check on our mass estimation, we compared our fitted mass estimation $M_{fit}$ to the method similar to the approach that is used in \citet{Cucciati18}. To make this comparison we first compiled the overdensity masses for the isolated structures which lay in the top half of all isolated structures in terms of spectral fraction, i.e., a similar population to that used in the fit described above. We then used the formalism of \citet{Cucciati18} to measure the overdensity masses at a variety of different equivalent spherical radii by varying the isodensity threshold that defines the structure. The equivalent spherical radius is given as $(3/4*V)^{1/3}$, where $V$ is the volume of the isolated structure above a given density threshold. This overdensity mass was calculated at five different radii, running from $R_{vir}$ to $5R_{vir}$ in steps of one $R_{vir}$. At each step, the overdensity mass is calculated as $M_{tot}=\rho_{m}V(1+\delta_m)$, where $\rho_m$ is the comoving matter density, $V$ is the volume of the isolated structure, and $\delta_m$ is the average matter overdensity within that volume. The average matter overdensity was calculated by the average $\delta_{gal}$ via $\delta_m = \delta_{gal}/b$, where $b$ is the bias factor of the galaxies comprising the overdensity measurement. While \citet{Cucciati18} adopted a single bias factor, appropriate for their analysis at a single redshift, here we are required to adopt a bias factor that varies for each structure. This varying bias factor was estimated in the following manner. For the entire ORELSE spectroscopic sample subject to the criteria discussed in \S\ref{data}, we measured the median specific star-formation rate (SSFR) as a function of redshift as estimated by the \texttt{FAST} SED fitting described in \citet{Tomczak17}. The entire ORELSE sample was used instead of each individual field to smooth out any decrease in the average SSFR of galaxies caused by an individual structure \citep[e.g.,][]{tomczak19}. For an isolated structure at a given redshift, the average SSFR and the structure redshift was used to estimate $b$ from \citet{Coil17}. The adopted bias factor ranged from $b=1.12$ at $z=0.7$ to $b=1.27$ at $z=1.2$. 

In order to determine the appropriate equivalent spherical radius for comparison to $M_{fit}$, we additionally calculated the elongation of each of the isolated structures at each step using the formalism of \citet{Cucciati18}. Briefly, elongation is defined as the average effective radius in the two transverse dimensions relative to the effective radius in the line-of-sight dimension. The effective radius is defined as $R_e=\sqrt{(\Sigma_{i}w_{i}(x_{i}-x_{peak})^2/\Sigma_{i}(w_i))}$, where $i$ is the $i_{th}$ pixel of the VMC map within the volume bounded by the structure along a given dimension, $w_i$ is the $\delta_{gal}$ value of that pixel, and $x_i$ is the location of the pixel relative to the barycenter location, $x_{peak}$, along the right ascension, declination, or redshift axes. Elongations, $E_{z/xy}$, were estimated in the range 6-30, with a mean value of $\langle E_{z/xy}\rangle=15.2$, i.e., at a given overdensity, the average isolated structure was 15.2$\times$ larger in the line-of-sight dimension than the transverse dimensions. Given the logic presented in \S\ref{mockmakeup}, we choose the equivalent spherical radius that encompasses a volume similar to $4/3\pi$$1.5R_{vir}^3 E_{z/xy}$, as this radius contains a large fraction of the true members of a system while minimizing interlopers. This equivalent spherical radius is $4R_{vir}$. Overdensity masses for each isolated structure were measured in volumes with equivalent spherical radii of $4R_{vir}$ and compared to $M_{fit}$. This comparison yielded a median offset of $\langle \tilde{\Delta}_{ \log(M_{tot}) - \log(M_{fit})} \rangle = -0.221$ and a $\sigma_{NMAD} = 0.391$ over $\sim$2.5 orders of magnitude in structure masses ($12.7 \la \log(M_{tot}/M_{\odot}) \la 15.2$). We note here that these values do not change meaningfully if we adopt a slightly smaller or larger equivalent spherical radius. Given the large number of assumptions in each method, this level of concordance between the two mass estimates is impressive, and we conclude that the $M_{fit}$ values measured here are broadly consistent with the $M_{tot}$ methodology of \citet{Cucciati18}. 

We also compare these $M_{tot}$ values with the $M_{vir}$ of each respective isolated structures. By design, since $M_{fit}$ is anchored to $M_{vir}$ and because we see a large degree of concordance between $M_{tot}$ and $M_{fit}$, these values should be broadly similar. We verify that expectation here, finding a median offset of $\langle \tilde{\Delta}_{M_{tot} - M_{vir}}) \rangle = -0.220$ and a $\sigma_{NMAD} = 0.155$ again over $\sim$2.5 orders of magnitude in structure masses ($12.7 \la \log(M_{tot}/M_{\odot}) \la 15.2$). Again, given the different assumptions associated with the two methods, disavowing knowledge of the previous comparison, the level of concordance is striking. Finally, from our tests with the mock catalogs, we recall that our purity and completeness tend to drop off at lower masses, especially at higher redshifts. Because of this, we make no attempt to push the analysis in this paper to structure masses lower than the minimum mass range probed by the mocks, or $\log(M_{tot}/M_{\odot})\sim13.5$.

\section{Conclusions}

Searching for galaxy overdensities is challenging, and many established methods depend on using assumptions on their shapes and other physical properties to detect them. In this work, we have presented a powerful new technique, Voronoi tessellation Monte-Carlo (VMC) mapping, and applied it to the rich ORELSE data set cut at 18 mag $\leq i/z \leq$ 24.5 mag (the equivalent of $10^9 - 10^{10} M_{\odot}$ stellar mass-limited sample), recovering 51 of the 56 known structures and finding 402 new overdensity candidates. Though we've applied the VMC method to one particular data set in this work, it can similarly be used for any photometric and spectroscopic data set.

How many overdensities the VMC method will find is tied heavily to the choice of its various parameters related to the mapping and detection, all of which we extensively tested to find what gave the best performance overall. We tested how varying the Source Extractor (SExtractor) detection parameters, DETECT\textunderscore THRESH and DETECT\textunderscore MINAREA, as well as the deblending parameters, DEBLEND\textunderscore NTHRESH and DEBLEND\textunderscore MINCONT, would affect how many known ORELSE structures we could recover. We also tested the SExtractor detection parameters on constructed mock catalogs with injected groups and clusters, where unlike with the real ORELSE data, we have the advantage of having full knowledge of what structures are present. From these tests, we concluded that the best parameters to use were DETECT\textunderscore THRESH = 4$\sigma$, DETECT\textunderscore MINAREA = 20, DEBLEND\textunderscore NTHRESH = 32, and DEBLEND\textunderscore MINCONT = 0.01. We also saw that the VMC method in its broad application can find many overdensities across a wide redshift range. However, it has some difficulty separating close or blended systems while avoiding splitting individual structures elsewhere. In this work, we elected to compromise on parameters that were not able to completely split systems like the superclusters in SG0023 or RXJ1716 as they performed best for all of the fields overall.

With the mock catalogs, we confirmed the best DETECT\textunderscore THRESH and DETECT\textunderscore MINAREA parameters we found with testing a subset of known structures by assessing our purity and completeness numbers across a range of simulated spectroscopic fractions. We found impressively high purity and completeness rates even at relatively small spectroscopic fractions, such as 0.92/0.83 and 0.60/0.49 for purity/completeness at $z = 0.8$ and $z = 1.2$ respectively for spectroscopic fractions of $\sim$20\%.

\begin{figure*}
\centering
\includegraphics[width=2\columnwidth]{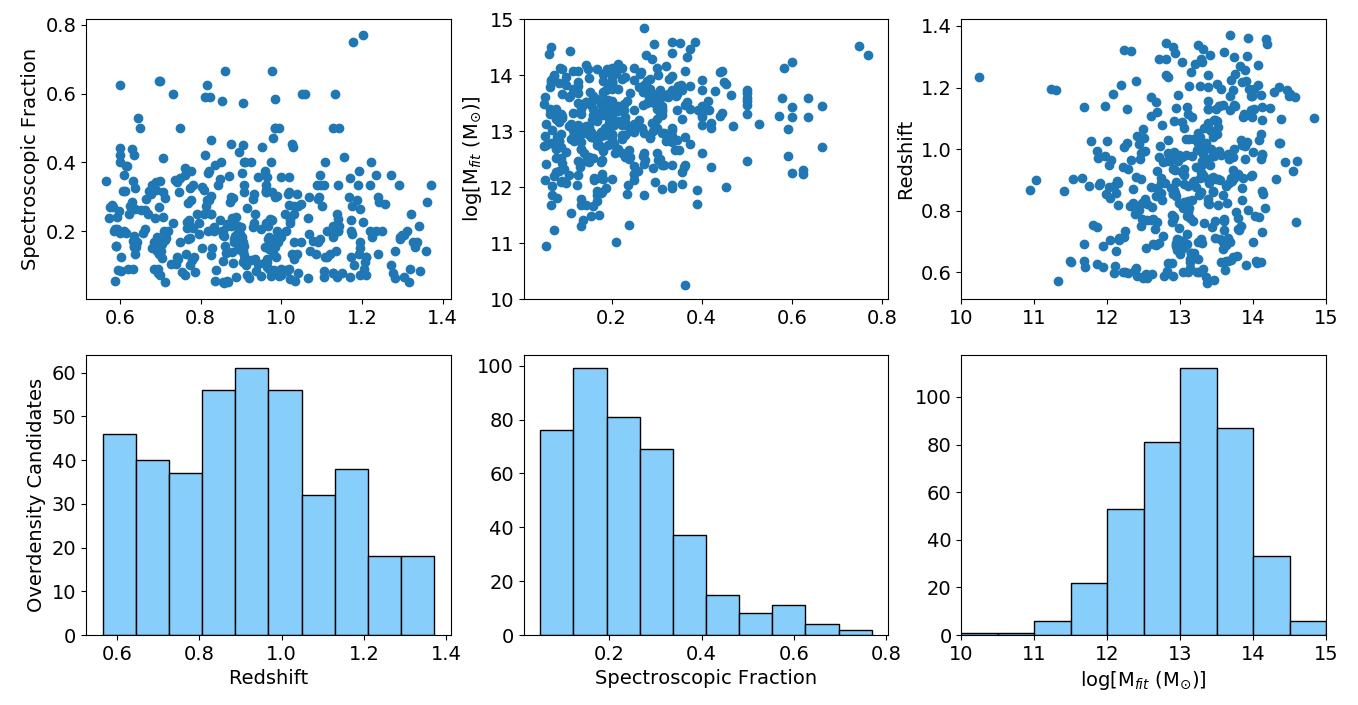}
\caption{Scatter plots (top) and histograms (bottom) of the redshifts, spectroscopic fractions, and masses of the 402 new overdensity candidates listed in Table \ref{tab.catalog}. The same x-axis is shared between the top and bottom panels for each column. The redshift range spanned between 0.565 $< z <$ 1.371. We set the spectroscopic fraction floor at 5\%, and it reached a maximum of 76.9\%. The estimated masses were between 10.2 $<\log(M_{fit}/M_{\odot})<$ 14.8. Our sensitivity to low-mass overdensities as well as the total number of overdensities detected tend to decline with greater redshift, which is the same behavior we saw in our mock catalog tests.}
\label{fig.histplots}
\end{figure*}

Table \ref{tab.catalog} lists our 402 new overdensity candidates with their redshifts, transverse positions, fitted masses, and spectroscopic fractions. The total redshift range spanned by our candidates was 0.565 $< z <$ 1.371. We fixed the spectroscopic fraction floor to be no lower than 5\%, and highest value was 76.9\%. The estimated masses were between 10.2 $<\log(M_{fit}/M_{\odot})<$ 14.8 (Fig. \ref{fig.histplots}). From purity and completeness tests with the mock catalogs, we found that the purity and completeness tend to be uniformly high at around $\log(M_{fit}/M_{\odot} \simgt$ 14.5, especially at high spectroscopic coverage. We can also derive a total mass function from the overdensity candidates we detected, corrected using the purity and completeness numbers from the mocks and interpolated based on the three dimensions of redshift, mass, and spectroscopic coverage. We will be including such analysis in full in an upcoming paper. Note that we only have purity and completeness estimates for masses above $\log(M_{fit}/M_{\odot}) = 13.2$, so the lower mass candidates we find here must be taken with a grain of salt. However, we wished to include everything the algorithm found, and these low mass candidates make for good follow-up targets for confirmation in future work. Access to the code and maps we used in this work is also available upon request.

With the ORELSE dataset, we have demonstrated the ability of the VMC method to measure precise systemic redshifts, provide an estimate of the total gravitating mass, and maintain high levels of purity and completeness for both groups and clusters at intermediate redshifts ($z\sim1$) even in the case of only moderate levels of spectroscopy. These factors speak to the value of the VMC method in applications for both current and future imaging surveys that either contain or overlap with some spectroscopic component. One such survey that has already been undertaken is the VIMOS Ultra Deep Survey \citep[VUDS;][]{lefevre15}, a spectroscopic redshift survey of $\sim$5000 faint galaxies at redshifts of $2<z\simeq$ 6 over 1 square degree of the sky in well-known extragalactic fields with deep, high-quality imaging. While the VMC method has already demonstrated the ability to detect several overdensities in VUDS \citep[e.g.,][]{Cucciati18,Lemaux18}, we are currently adapting the methods presented in this paper to perform a systematic search for forming groups, clusters, and superclusters within VUDS in the high-redshift Universe. However, this survey will eventually be eclipsed in size and in depth by ongoing or near-term future surveys, e.g., the Hyper Suprime-Cam Subaru Strategic Program with follow-up observations from the Prime Focus Spectrograph, as well as future surveys envisioned for the coming decade following the rise of thirty meter class telescopes, dedicated 10-$m$ class telescopes such as the Maunakea Spectroscopic Explorer, and the next generation of space observatories that will enable imaging and spectroscopic surveys across an incredible redshift baseline for large swaths of the sky. With regards to the redshift regime covered by this paper, the new Anglo-Australian Telescope (AAT) Deep Extragalactic VIsible Legacy Survey \citep[DEVILS;][]{Davies18} is a spectroscopic campaign designed for high completeness that will help to overcome many of the issues with using photometric redshifts at $0.3<z<1.0$. The Wide Area VISTA Extra-galactic Survey-Deep \citep[WAVES-Deep;][]{Driver16} is a 4MOST Consortium Design Reference Survey aiming to obtain roughly 1.2 million spectroscopic galaxy redshifts over a 100 square degree area at z$\sim$1. Tools such as the VMC method presented here and its variants will be powerful in utilizing these data most effectively to find overdensities of galaxies of differing types from the local universe to the very highest of redshifts. 

\section*{Acknowledgements}

{\footnotesize
This material is based upon work supported by the National Science Foundation under Grant No. 1411943. Part of the work presented herein is supported by NASA Grant Number NNX15AK92G. This work was additionally supported by the France-Berkeley Fund, a joint venture between UC Berkeley, UC Davis, and le Centre National de la Recherche Scientifique de France promoting lasting institutional and intellectual cooperation between France and the United States. DH would like to thank her advisor David Tholen for being able to allocate time to working on a study in an outside field. PFW acknowledges funding through the H2020 ERC Consolidator Grant 683184 and the support of an EACOA Fellowship from the East Asian Core Observatories Association. BCL gratefully acknowledges Gianni Zamorani, Sandro Bardelli, and Elena Zucca for discussions helpful in developing the VMC technique. We also thank the anonymous referee for their helpful contribution. This study is based, in part, on data collected at the Subaru Telescope and obtained from the SMOKA, which is operated by the Astronomy Data Center, National Astronomical Observatory of Japan. This work is based, in part, on observations made with the Spitzer Space Telescope, which is operated by the Jet Propulsion Laboratory, California Institute of Technology under a contract with NASA. UKIRT is supported by NASA and operated under an agreement among the University of Hawaii, the University of Arizona, and Lockheed Martin Advanced Technology Center; operations are enabled through the cooperation of the East Asian Observatory. When the data reported here were acquired, UKIRT was operated by the Joint Astronomy Centre on behalf of the Science and Technology Facilities Council of the U.K. This study is also based, in part, on observations obtained with WIRCam, a joint project of CFHT, Taiwan, Korea, Canada, France, and the Canada-France-Hawaii Telescope which is operated by the National Research Council (NRC) of Canada, the Institut National des Sciences de l'Univers of the Centre National de la Recherche Scientifique of France, and the University of Hawai'i. Some portion of the spectrographic data presented herein was based on observations obtained with the European Southern Observatory Very Large Telescope, Paranal, Chile, under Large Programs 070.A-9007 and 177.A-0837. The remainder of the spectrographic data presented herein were obtained at the W.M. Keck Observatory, which is operated as a scientific partnership among the California Institute of Technology, the University of California, and the National Aeronautics and Space Administration. The Observatory was made possible by the generous financial support of the W.M. Keck Foundation. We thank the indigenous Hawaiian community for allowing us to be guests on their sacred mountain, a privilege, without which, this work would not have been possible. We are most fortunate to be able to conduct observations from this site.}

\bibliographystyle{mnras}
\bibliography{newtech} 

\begin{table*}
\centering
\caption{New Overdensity Candidates}
\label{tab.catalog}
\begin{tabular}{lccccccc}
\hline
Candidate ID & Points$^{a}$ & Redshift & $\sigma_{z}^{b}$ & RA (J2000) & Dec (J2000) & log(M$_{fit}$)$^{c}$ & Spec. Fraction \\
\hline
SC1604 6 & 6 & 0.5727 $\pm$ 0.0014 & 0.0048 $\pm$ 0.0020 & 241.00285 & 43.35538 & 11.32$^{+0.38}_{-0.58}$ & 0.238 \\     [0.7ex]
SC1604 18 & 11 & 0.6014 $\pm$ 0.0004 & 0.0068 $\pm$ 0.0006 & 241.10267 & 43.13988 & 13.23$^{+0.15}_{-0.14}$ & 0.194 \\   [0.7ex]
SC1604 19 & 9 & 0.5998 $\pm$ 0.0005 & 0.0051 $\pm$ 0.0007 & 241.13713 & 43.37485 & 12.24$^{+0.13}_{-0.12}$ & 0.625 \\    [0.7ex]
SC1604 20 & 9 & 0.5988 $\pm$ 0.0006 & 0.0053 $\pm$ 0.0008 & 241.07052 & 43.36816 & 12.28$^{+0.14}_{-0.14}$ & 0.259 \\    [0.7ex]
SC1604 22 & 11 & 0.6013 $\pm$ 0.0002 & 0.0043 $\pm$ 0.0002 & 241.06520 & 43.30805 & 13.10$^{+0.11}_{-0.10}$ & 0.400 \\   [0.7ex]
SC1604 46 & 9 & 0.6000 $\pm$ 0.0006 & 0.0061 $\pm$ 0.0006 & 241.05988 & 43.27654 & 13.26$^{+0.16}_{-0.16}$ & 0.441 \\    [0.7ex]
SC1604 202 & 7 & 0.6927 $\pm$ 0.0009 & 0.0052 $\pm$ 0.0014 & 241.15393 & 43.28733 & 11.68$^{+0.24}_{-0.28}$ & 0.143 \\   [0.7ex]
SC1604 206 & 10 & 0.6975 $\pm$ 0.0006 & 0.0082 $\pm$ 0.0010 & 241.14705 & 43.38570 & 13.59$^{+0.18}_{-0.19}$ & 0.636 \\  [0.7ex]
SC1604 223 & 11 & 0.7008 $\pm$ 0.0005 & 0.0065 $\pm$ 0.0005 & 240.86598 & 43.41580 & 13.04$^{+0.14}_{-0.13}$ & 0.143 \\  [0.7ex]
SC1604 266 & 8 & 0.7184 $\pm$ 0.0012 & 0.0069 $\pm$ 0.0016 & 241.12362 & 43.42163 & 12.77$^{+0.24}_{-0.26}$ & 0.200 \\   [0.7ex]
SC1604 290 & 46 & 0.7307 $\pm$ 0.0001 & 0.0061 $\pm$ 0.0001 & 240.86678 & 43.39457 & 14.12$^{+0.13}_{-0.14}$ & 0.222 \\  [0.7ex]
SC1604 337 & 6 & 0.7324 $\pm$ 0.0022 & 0.0054 $\pm$ 0.0020 & 241.12159 & 43.25468 & 12.25$^{+0.34}_{-0.45}$ & 0.600 \\   [0.7ex]
SC1604 343 & 36 & 0.7818 $\pm$ 0.0002 & 0.0079 $\pm$ 0.0002 & 240.87332 & 43.39353 & 14.03$^{+0.14}_{-0.14}$ & 0.209 \\  [0.7ex]
SC1604 355 & 33 & 0.7785 $\pm$ 0.0005 & 0.0145 $\pm$ 0.0005 & 240.92604 & 43.40497 & 14.08$^{+0.14}_{-0.15}$ & 0.115 \\  [0.7ex]
SC1604 356 & 10 & 0.7480 $\pm$ 0.0011 & 0.0070 $\pm$ 0.0015 & 240.85166 & 43.36172 & 11.85$^{+0.20}_{-0.22}$ & 0.167 \\  [0.7ex]
SC1604 406 & 9 & 0.7703 $\pm$ 0.0005 & 0.0053 $\pm$ 0.0005 & 241.17687 & 43.34504 & 12.54$^{+0.13}_{-0.12}$ & 0.231 \\   [0.7ex]
SC1604 415 & 11 & 0.7769 $\pm$ 0.0003 & 0.0071 $\pm$ 0.0004 & 241.02976 & 43.20301 & 13.80$^{+0.15}_{-0.15}$ & 0.375 \\  [0.7ex]
SC1604 498 & 12 & 0.8086 $\pm$ 0.0005 & 0.0066 $\pm$ 0.0005 & 241.03156 & 43.24164 & 13.12$^{+0.14}_{-0.13}$ & 0.250 \\  [0.7ex]
SC1604 499 & 17 & 0.8100 $\pm$ 0.0005 & 0.0077 $\pm$ 0.0006 & 241.10693 & 43.33826 & 13.04$^{+0.13}_{-0.12}$ & 0.591 \\  [0.7ex]
SC1604 523 & 10 & 0.8258 $\pm$ 0.0023 & 0.0102 $\pm$ 0.0035 & 241.09897 & 43.19012 & 12.48$^{+0.32}_{-0.41}$ & 0.303 \\  [0.7ex]
SC1604 525 & 11 & 0.8214 $\pm$ 0.0008 & 0.0072 $\pm$ 0.0008 & 241.19860 & 43.36250 & 13.01$^{+0.15}_{-0.15}$ & 0.292 \\  [0.7ex]
SC1604 538 & 9 & 0.8225 $\pm$ 0.0045 & 0.0113 $\pm$ 0.0065 & 241.16770 & 43.38293 & 12.55$^{+0.48}_{-0.90}$ & 0.591 \\   [0.7ex]
SC1604 552 & 9 & 0.8318 $\pm$ 0.0008 & 0.0071 $\pm$ 0.0009 & 241.09338 & 43.05800 & 13.11$^{+0.17}_{-0.17}$ & 0.206 \\   [0.7ex]
SC1604 625 & 15 & 0.8643 $\pm$ 0.0003 & 0.0073 $\pm$ 0.0003 & 241.02888 & 43.24921 & 14.12$^{+0.14}_{-0.15}$ & 0.220 \\  [0.7ex]
SC1604 633 & 6 & 0.8657 $\pm$ 0.0011 & 0.0039 $\pm$ 0.0012 & 240.94707 & 43.27677 & 10.95$^{+0.29}_{-0.42}$ & 0.056 \\   [0.7ex]
SC1604 637 & 10 & 0.8797 $\pm$ 0.0088 & 0.0142 $\pm$ 0.0065 & 241.02850 & 43.43176 & 13.63$^{+0.37}_{-0.57}$ & 0.107 \\  [0.7ex]
SC1604 690 & 6 & 0.8768 $\pm$ 0.0007 & 0.0033 $\pm$ 0.0005 & 241.00935 & 43.22017 & 12.03$^{+0.17}_{-0.17}$ & 0.237 \\   [0.7ex]
SC1604 700 & 18 & 0.9028 $\pm$ 0.0003 & 0.0073 $\pm$ 0.0002 & 241.03047 & 43.21775 & 14.32$^{+0.14}_{-0.15}$ & 0.368 \\  [0.7ex]
SC1604 767 & 7 & 0.9010 $\pm$ 0.0021 & 0.0062 $\pm$ 0.0029 & 241.18364 & 43.13483 & 11.02$^{+0.43}_{-0.75}$ & 0.211 \\   [0.7ex]
SC1604 811 & 5 & 0.9038 $\pm$ 0.0019 & 0.0054 $\pm$ 0.0035 & 241.19251 & 43.21757 & 11.53$^{+0.58}_{-1.34}$ & 0.111 \\   [0.7ex]
SC1604 912 & 11 & 0.9370 $\pm$ 0.0015 & 0.0086 $\pm$ 0.0018 & 241.03441 & 43.44029 & 12.29$^{+0.22}_{-0.23}$ & 0.083 \\  [0.7ex]
SC1604 933 & 8 & 0.9318 $\pm$ 0.0013 & 0.0062 $\pm$ 0.0007 & 241.12834 & 43.26951 & 13.73$^{+0.19}_{-0.20}$ & 0.308 \\   [0.7ex]
SC1604 960 & 18 & 0.9691 $\pm$ 0.0037 & 0.0193 $\pm$ 0.0030 & 240.78757 & 43.37800 & 14.10$^{+0.20}_{-0.22}$ & 0.148 \\  [0.7ex]
SC1604 997 & 52 & 0.9812 $\pm$ 0.0010 & 0.0269 $\pm$ 0.0010 & 240.77694 & 43.35182 & 14.12$^{+0.15}_{-0.15}$ & 0.086 \\  [0.7ex]
SC1604 1032 & 49 & 0.9575 $\pm$ 0.0054 & 0.0411 $\pm$ 0.0032 & 240.77708 & 43.35194 & 14.43$^{+0.16}_{-0.18}$ & 0.108 \\ [0.7ex]
SC1604 1050 & 21 & 0.9693 $\pm$ 0.0006 & 0.0106 $\pm$ 0.0006 & 240.86665 & 43.36175 & 13.67$^{+0.15}_{-0.15}$ & 0.091 \\ [0.7ex]
SC1604 1102 & 10 & 0.9794 $\pm$ 0.0018 & 0.0129 $\pm$ 0.0037 & 241.02810 & 43.42602 & 13.56$^{+0.28}_{-0.34}$ & 0.158 \\ [0.7ex]
SC1604 1103 & 10 & 0.9798 $\pm$ 0.0019 & 0.0118 $\pm$ 0.0033 & 241.12359 & 43.38821 & 13.09$^{+0.28}_{-0.33}$ & 0.471 \\ [0.7ex]
SC1604 1112 & 10 & 0.9816 $\pm$ 0.0022 & 0.0119 $\pm$ 0.0043 & 241.09527 & 43.44931 & 12.99$^{+0.33}_{-0.43}$ & 0.176 \\ [0.7ex]
SC1604 1242 & 18 & 1.0333 $\pm$ 0.0007 & 0.0093 $\pm$ 0.0009 & 240.94486 & 43.26764 & 13.12$^{+0.17}_{-0.16}$ & 0.056 \\ [0.7ex]
SC1604 1247 & 10 & 1.0317 $\pm$ 0.0017 & 0.0107 $\pm$ 0.0021 & 241.05053 & 43.30065 & 13.18$^{+0.22}_{-0.24}$ & 0.167 \\ [0.7ex]
SC1604 1269 & 10 & 1.0377 $\pm$ 0.0005 & 0.0054 $\pm$ 0.0006 & 241.01870 & 43.17909 & 12.50$^{+0.14}_{-0.13}$ & 0.308 \\ [0.7ex]
SC1604 1329 & 11 & 1.0861 $\pm$ 0.0002 & 0.0044 $\pm$ 0.0002 & 241.01972 & 43.09836 & 13.39$^{+0.12}_{-0.12}$ & 0.133 \\ [0.7ex]
SC1604 1339 & 10 & 1.0996 $\pm$ 0.0044 & 0.0225 $\pm$ 0.0066 & 241.04940 & 43.25374 & 14.84$^{+0.21}_{-0.28}$ & 0.273 \\ [0.7ex]
SC1604 1366 & 10 & 1.1287 $\pm$ 0.0116 & 0.0180 $\pm$ 0.0107 & 241.20819 & 43.38557 & 13.49$^{+0.44}_{-0.82}$ & 0.200 \\ [0.7ex]
SC1604 1385 & 10 & 1.1738 $\pm$ 0.0004 & 0.0076 $\pm$ 0.0004 & 240.99754 & 43.37170 & 13.98$^{+0.15}_{-0.16}$ & 0.200 \\ [0.7ex]
SC1604 1386 & 12 & 1.1718 $\pm$ 0.0003 & 0.0071 $\pm$ 0.0002 & 240.91559 & 43.33163 & 13.97$^{+0.14}_{-0.14}$ & 0.231 \\ [0.7ex]
\hline
\end{tabular}
\begin{flushleft}
$a$: Number of points used in the Gaussian fit of the isophotal flux.

$b$: 1$\sigma$ redshift dispersion showing width of the Gaussian.

$c$: Total mass in units of solar mass, calculated from the fit given in Fig. \ref{fig.mass}.

The full table of our 402 newly found overdensity candidates. Many of our candidates had redshift precisions smaller than $z < 0.01$ and several candidates occupy the low mass overdensity regime of $\sim 5 \times 10^{13} M_{\odot}$. We caution readers to take our lower mass candidates with a grain of salt, as we only have purity and completeness estimates for candidates with masses $\log(M_{fit}/M_{\odot}) \geq 13.2$.
\end{flushleft}
\end{table*}

\begin{table*}
\centering
\contcaption{New Overdensity Candidates}
\begin{tabular}{lccccccc}
\hline
Candidate ID & Points$^{a}$ & Redshift & $\sigma_{z}^{b}$ & RA (J2000) & Dec (J2000) & log(M$_{fit}$)$^{c}$ & Spec. Fraction \\
\hline
SC1604 1398 & 16 & 1.1781 $\pm$ 0.0002 & 0.0101 $\pm$ 0.0003 & 241.15552 & 43.35774 & 14.51$^{+0.14}_{-0.15}$ & 0.750 \\ [0.7ex]
SC1604 1440 & 16 & 1.2023 $\pm$ 0.0005 & 0.0141 $\pm$ 0.0007 & 241.14045 & 43.37799 & 14.35$^{+0.15}_{-0.16}$ & 0.769 \\ [0.7ex]
SC1604 1493 & 10 & 1.2396 $\pm$ 0.0008 & 0.0078 $\pm$ 0.0008 & 241.01789 & 43.17570 & 13.39$^{+0.16}_{-0.16}$ & 0.300 \\ [0.7ex]
SC1604 1496 & 5 & 1.2340 $\pm$ 0.0017 & 0.0047 $\pm$ 0.0028 & 241.07546 & 43.29556 & 10.25$^{+0.59}_{-1.49}$ & 0.364 \\  [0.7ex]
SC1604 1513 & 16 & 1.2728 $\pm$ 0.0014 & 0.0128 $\pm$ 0.0009 & 241.15944 & 43.37114 & 14.07$^{+0.16}_{-0.17}$ & 0.364 \\ [0.7ex]
SC1604 1537 & 9 & 1.2923 $\pm$ 0.0008 & 0.0077 $\pm$ 0.0010 & 241.11096 & 43.20679 & 12.80$^{+0.17}_{-0.16}$ & 0.333 \\  [0.7ex]
SC1324 11 & 10 & 0.6437 $\pm$ 0.0006 & 0.0073 $\pm$ 0.0010 & 201.20014 & 30.21687 & 13.13$^{+0.18}_{-0.17}$ & 0.528 \\
\hline
\end{tabular}
\end{table*}

\begin{table*}
\centering
\contcaption{New Overdensity Candidates}
\begin{tabular}{lccccccc}
\hline
Candidate ID & Points$^{a}$ & Redshift & $\sigma_{z}^{b}$ & RA (J2000) & Dec (J2000) & log(M$_{fit}$)$^{c}$ & Spec. Fraction \\
\hline
XLSS005 346 & 13 & 0.6299 $\pm$ 0.0003 & 0.0052 $\pm$ 0.0003 & 36.64844 & -4.47962 & 12.89$^{+0.11}_{-0.10}$ & 0.286 \\  [0.7ex]
XLSS005 358 & 11 & 0.6346 $\pm$ 0.0009 & 0.0080 $\pm$ 0.0010 & 36.71511 & -4.54406 & 13.14$^{+0.17}_{-0.17}$ & 0.421 \\  [0.7ex]
XLSS005 362 & 10 & 0.6315 $\pm$ 0.0004 & 0.0050 $\pm$ 0.0004 & 36.51317 & -4.39425 & 12.83$^{+0.13}_{-0.12}$ & 0.346 \\  [0.7ex]
XLSS005 379 & 12 & 0.6516 $\pm$ 0.0183 & 0.0163 $\pm$ 0.0088 & 36.47950 & -4.46707 & 13.78$^{+0.46}_{-1.19}$ & 0.262 \\  [0.7ex]
XLSS005 395 & 13 & 0.6378 $\pm$ 0.0002 & 0.0050 $\pm$ 0.0002 & 36.76602 & -4.53178 & 13.66$^{+0.13}_{-0.13}$ & 0.324 \\  [0.7ex]
XLSS005 399 & 9 & 0.6363 $\pm$ 0.0003 & 0.0029 $\pm$ 0.0003 & 36.84608 & -4.58340 & 11.68$^{+0.06}_{-0.05}$ & 0.133 \\   [0.7ex]
XLSS005 437 & 5 & 0.6332 $\pm$ 0.0004 & 0.0026 $\pm$ 0.0005 & 36.69795 & -4.47262 & 11.51$^{+0.16}_{-0.17}$ & 0.172 \\
\hline
\end{tabular}
\end{table*}

\begin{table*}
\centering
\contcaption{New Overdensity Candidates}
\begin{tabular}{lccccccc}
\hline
Candidate ID & Points$^{a}$ & Redshift & $\sigma_{z}^{b}$ & RA (J2000) & Dec (J2000) & log(M$_{fit}$)$^{c}$ & Spec. Fraction \\
\hline
XLSS005 1653 & 31 & 0.9162 $\pm$ 0.0002 & 0.0061 $\pm$ 0.0001 & 36.42077 & -4.40508 & 13.98$^{+0.14}_{-0.14}$ & 0.071 \\ [0.7ex]
XLSS005 1654 & 11 & 0.8813 $\pm$ 0.0007 & 0.0080 $\pm$ 0.0010 & 36.49487 & -4.49830 & 13.10$^{+0.17}_{-0.17}$ & 0.231 \\ [0.7ex]
XLSS005 1656 & 10 & 0.8864 $\pm$ 0.0012 & 0.0084 $\pm$ 0.0015 & 36.44889 & -4.61527 & 13.05$^{+0.21}_{-0.22}$ & 0.185 \\ [0.7ex]
XLSS005 1671 & 11 & 0.8886 $\pm$ 0.0007 & 0.0093 $\pm$ 0.0009 & 36.80858 & -4.30730 & 13.67$^{+0.17}_{-0.17}$ & 0.240 \\ [0.7ex]
XLSS005 1695 & 7 & 0.8823 $\pm$ 0.0007 & 0.0047 $\pm$ 0.0009 & 36.47430 & -4.19802 & 11.89$^{+0.18}_{-0.18}$ & 0.176 \\  [0.7ex]
XLSS005 1714 & 10 & 0.8756 $\pm$ 0.0161 & 0.0156 $\pm$ 0.0102 & 36.84403 & -4.40244 & 13.22$^{+0.52}_{-1.31}$ & 0.300 \\ [0.7ex]
XLSS005 1725 & 10 & 0.8891 $\pm$ 0.0006 & 0.0057 $\pm$ 0.0008 & 36.50960 & -4.65077 & 12.15$^{+0.14}_{-0.13}$ & 0.182 \\
\hline
\end{tabular}
\end{table*}

\begin{table*}
\centering
\contcaption{New Overdensity Candidates}
\begin{tabular}{lccccccc}
\hline
Candidate ID & Points$^{a}$ & Redshift & $\sigma_{z}^{b}$ & RA (J2000) & Dec (J2000) & log(M$_{fit}$)$^{c}$ & Spec. Fraction \\
\hline
XLSS005 2828 & 10 & 1.1056 $\pm$ 0.0035 & 0.0117 $\pm$ 0.0035 & 36.79601 & -4.14645 & 13.16$^{+0.29}_{-0.35}$ & 0.083 \\ [0.7ex]
XLSS005 2834 & 9 & 1.1129 $\pm$ 0.0014 & 0.0104 $\pm$ 0.0028 & 36.86683 & -4.28207 & 13.11$^{+0.27}_{-0.32}$ & 0.125 \\  [0.7ex]
XLSS005 2839 & 10 & 1.1125 $\pm$ 0.0009 & 0.0107 $\pm$ 0.0013 & 36.86673 & -4.34315 & 13.88$^{+0.18}_{-0.19}$ & 0.200 \\ [0.7ex]
XLSS005 2844 & 6 & 1.1072 $\pm$ 0.0070 & 0.0083 $\pm$ 0.0050 & 36.69562 & -4.24084 & 12.60$^{+0.52}_{-1.09}$ & 0.400 \\  [0.7ex]
XLSS005 2848 & 11 & 1.1290 $\pm$ 0.0016 & 0.0108 $\pm$ 0.0011 & 36.90042 & -4.18909 & 14.01$^{+0.18}_{-0.19}$ & 0.182 \\ [0.7ex]
XLSS005 2852 & 9 & 1.1204 $\pm$ 0.0005 & 0.0063 $\pm$ 0.0006 & 36.58711 & -4.56302 & 12.68$^{+0.13}_{-0.12}$ & 0.133 \\  [0.7ex]
XLSS005 2861 & 15 & 1.1266 $\pm$ 0.0003 & 0.0100 $\pm$ 0.0004 & 36.57163 & -4.43365 & 13.97$^{+0.15}_{-0.15}$ & 0.200 \\
\hline
\end{tabular}
\end{table*}

\begin{table*}
\centering
\contcaption{New Overdensity Candidates}
\begin{tabular}{lccccccc}
\hline
Candidate ID & Points$^{a}$ & Redshift & $\sigma_{z}^{b}$ & RA (J2000) & Dec (J2000) & log(M$_{fit}$)$^{c}$ & Spec. Fraction \\
\hline
SG0023 288 & 9 & 0.7154 $\pm$ 0.0005 & 0.0044 $\pm$ 0.0005 & 5.97599 & 4.27707 & 12.22$^{+0.10}_{-0.09}$ & 0.179 \\      [0.7ex]
SG0023 328 & 9 & 0.7491 $\pm$ 0.0004 & 0.0046 $\pm$ 0.0004 & 5.90218 & 4.37707 & 12.46$^{+0.12}_{-0.10}$ & 0.500 \\      [0.7ex]
SG0023 387 & 17 & 0.8341 $\pm$ 0.0004 & 0.0075 $\pm$ 0.0004 & 5.81936 & 4.41367 & 13.17$^{+0.12}_{-0.11}$ & 0.212 \\     [0.7ex]
SG0023 443 & 11 & 0.8475 $\pm$ 0.0005 & 0.0070 $\pm$ 0.0006 & 5.87266 & 4.40156 & 12.94$^{+0.14}_{-0.13}$ & 0.231 \\     [0.7ex]
SG0023 528 & 10 & 0.9033 $\pm$ 0.0185 & 0.0165 $\pm$ 0.0117 & 5.85472 & 4.49141 & 13.39$^{+0.53}_{-1.57}$ & 0.158 \\     [0.7ex]
SG0023 542 & 13 & 0.9058 $\pm$ 0.0014 & 0.0112 $\pm$ 0.0023 & 5.84260 & 4.41348 & 12.74$^{+0.22}_{-0.24}$ & 0.167 \\     [0.7ex]
SG0023 576 & 9 & 0.9293 $\pm$ 0.0009 & 0.0077 $\pm$ 0.0015 & 5.94700 & 4.31904 & 12.68$^{+0.21}_{-0.22}$ & 0.212 \\
\hline
\end{tabular}
\end{table*}

\begin{table*}
\centering
\contcaption{New Overdensity Candidates}
\begin{tabular}{lccccccc}
\hline
Candidate ID & Points$^{a}$ & Redshift & $\sigma_{z}^{b}$ & RA (J2000) & Dec (J2000) & log(M$_{fit}$)$^{c}$ & Spec. Fraction \\
\hline
Cl1429 158 & 10 & 0.9331 $\pm$ 0.0008 & 0.0075 $\pm$ 0.0010 & 217.15585 & 42.72360 & 12.93$^{+0.18}_{-0.18}$ & 0.222 \\  [0.7ex]
Cl1429 171 & 10 & 0.9855 $\pm$ 0.0009 & 0.0095 $\pm$ 0.0012 & 217.43174 & 42.77845 & 13.65$^{+0.18}_{-0.19}$ & 0.300 \\  [0.7ex]
Cl1429 176 & 10 & 0.9855 $\pm$ 0.0001 & 0.0044 $\pm$ 0.0001 & 217.34975 & 42.72999 & 14.12$^{+0.14}_{-0.15}$ & 0.583 \\  [0.7ex]
Cl1429 188 & 11 & 0.9834 $\pm$ 0.0009 & 0.0084 $\pm$ 0.0010 & 217.39123 & 42.59193 & 13.31$^{+0.17}_{-0.17}$ & 0.500 \\  [0.7ex]
Cl1429 207 & 10 & 0.9854 $\pm$ 0.0008 & 0.0071 $\pm$ 0.0006 & 217.20412 & 42.66691 & 13.46$^{+0.16}_{-0.16}$ & 0.500 \\  [0.7ex]
Cl1429 214 & 6 & 0.9868 $\pm$ 0.0022 & 0.0074 $\pm$ 0.0043 & 217.32585 & 42.58052 & 12.46$^{+0.49}_{-0.93}$ & 0.125 \\   [0.7ex]
Cl1429 257 & 8 & 0.9953 $\pm$ 0.0002 & 0.0043 $\pm$ 0.0002 & 217.41620 & 42.73773 & 13.53$^{+0.14}_{-0.14}$ & 0.500 \\
\hline
\end{tabular}
\end{table*}

\begin{table*}
\centering
\contcaption{New Overdensity Candidates}
\begin{tabular}{lccccccc}
\hline
Candidate ID & Points$^{a}$ & Redshift & $\sigma_{z}^{b}$ & RA (J2000) & Dec (J2000) & log(M$_{fit}$)$^{c}$ & Spec. Fraction \\
\hline
RXJ1716 199 & 11 & 0.8131 $\pm$ 0.0009 & 0.0076 $\pm$ 0.0012 & 259.38881 & 67.18763 & 12.61$^{+0.19}_{-0.19}$ & 0.273 \\ [0.7ex]
RXJ1716 202 & 9 & 0.8160 $\pm$ 0.0011 & 0.0060 $\pm$ 0.0012 & 259.22600 & 67.24113 & 12.31$^{+0.21}_{-0.23}$ & 0.625 \\  [0.7ex]
RXJ1716 319 & 10 & 0.8524 $\pm$ 0.0004 & 0.0069 $\pm$ 0.0005 & 259.21004 & 67.23587 & 13.58$^{+0.15}_{-0.15}$ & 0.579 \\ [0.7ex]
RXJ1716 364 & 12 & 0.8741 $\pm$ 0.0006 & 0.0089 $\pm$ 0.0006 & 259.18415 & 67.17011 & 13.84$^{+0.16}_{-0.16}$ & 0.455 \\ [0.7ex]
RXJ1716 420 & 10 & 0.9694 $\pm$ 0.0035 & 0.0151 $\pm$ 0.0050 & 259.47834 & 67.22165 & 13.89$^{+0.29}_{-0.37}$ & 0.182 \\ [0.7ex]
RXJ1716 430 & 10 & 1.1086 $\pm$ 0.0007 & 0.0094 $\pm$ 0.0009 & 259.31087 & 67.21130 & 13.88$^{+0.17}_{-0.18}$ & 0.125 \\ [0.7ex]
Cl1350 1 & 9 & 0.7407 $\pm$ 0.0141 & 0.0165 $\pm$ 0.0123 & 207.75349 & 60.12973 & 13.80$^{+0.47}_{-1.22}$ & 0.125 \\
\hline
\end{tabular}
\end{table*}

\begin{table}
\caption{ORELSE Imaging Data}
\label{tab.imaging}
\begin{tabular}{l|lll}
\hline
Field & Band & Telescope/Instrument & Depth$^{a}$ \\
\hline
SG0023 & $B$ & Subaru/Suprime-Cam & 26.4 \\
 & $V$ & Subaru/Suprime-Cam & 26.2 \\
 & $R^{+}$ & Subaru/Suprime-Cam & 25.3 \\
 & $r^{\prime}$ & Palomar200/LFC & 25.7 \\
 & $I^{+}$ & Subaru/Suprime-Cam & 25.2 \\
 & $i^{\prime}$ & Palomar200/LFC & 25.2 \\
 & $z^{\prime}$ & Palomar200/LFC & 23.8 \\
\hline
\end{tabular}
\begin{flushleft}
$a$: Magnitude depth at 80\% completeness.
\end{flushleft}
\end{table}

\begin{table}
\contcaption{ORELSE Imaging Data}
\begin{tabular}{l|lll}
\hline
Field & Band & Telescope/Instrument & Depth$^{a}$ \\
\hline
RXJ1053 & $u$ & CFHT/MegaCam & 24.8 \\
 & $B$ & Subaru/Suprime-Cam & 26.1 \\
 & $V$ & Subaru/Suprime-Cam & 26.1 \\
 & $g$ & CFHT/MegaCam & 25.7 \\
 & $R_{C}$ & Subaru/Suprime-Cam & 25.2 \\
 & $R^{+}$ & Subaru/Suprime-Cam & 26.4 \\
 & $r$ & CFHT/MegaCam & 24.5 \\
\hline
\end{tabular}
\end{table}

\begin{table}
\contcaption{ORELSE Imaging Data}
\begin{tabular}{l|lll}
\hline
Field & Band & Telescope/Instrument & Depth$^{a}$ \\
\hline
Cl1429 & $B$ & Subaru/Suprime-Cam & 26.7 \\ 
 & $V$ & Subaru/Suprime-Cam & 26.2 \\ 
 & $r^{\prime}$ & Palomar200/LFC & 24.2 \\ 
 & $i^{\prime}$ & Palomar200/LFC & 23.5 \\ 
 & $z^{\prime}$ & Palomar200/LFC & 22.7 \\ 
 & $Y$ & Subaru/Suprime-Cam & 23.2 \\ 
 & $J$ & UKIRT/WFCam & 21.9 \\
\hline
\end{tabular}
\end{table}

\begin{table}
\contcaption{ORELSE Imaging Data}
\begin{tabular}{l|lll}
\hline
Field & Band & Telescope/Instrument & Depth$^{a}$ \\
\hline
RXJ1821 & $B$ & Subaru/Suprime-Cam & 26.0 \\
 & $V$ & Subaru/Suprime-Cam & 26.0 \\
 & $r^{\prime}$ & Palomar200/LFC & 24.4 \\
 & $i^{\prime}$ & Palomar200/LFC & 24.3 \\
 & $z^{\prime}$ & Palomar200/LFC & 23.3 \\
 & $Y$ & Subaru/Suprime-Cam & 23.4 \\
 & $J$ & CFHT/WIRCam & 21.4 \\
 & $K_s$ & CFHT/WIRCam & 21.7 \\
 & $[3.6]$ & {\it Spitzer}/IRAC & 23.9 \\
 & $[4.5]$ & {\it Spitzer}/IRAC & 23.8 \\ [0.5ex]
\hline
SC0910 & $B$ & Subaru/Suprime-Cam & 24.4 \\
 & $V$ & Subaru/Suprime-Cam & 25.6 \\
 & $R_{C}$ & Subaru/Suprime-Cam & 26.4 \\
 & $I^{+}$ & Subaru/Suprime-Cam & 25.8 \\
 & $Z^{+}$ & Subaru/Suprime-Cam &  24.8 \\
 & $J$ & UKIRT/WFCam & 22.1 \\
 & $K$ & UKIRT/WFCam & 21.7 \\
 & $[3.6]$ & {\it Spitzer}/IRAC & 23.2 \\
 & $[4.5]$ & {\it Spitzer}/IRAC & 23.2 \\ [0.5ex]
\hline
\end{tabular}
\end{table}

\bsp	
\label{lastpage}
\end{document}